\documentclass[aps,color,twocolumn,pdflatex,citeautoscript,superscriptaddress,notitlepage]{revtex4-2}
\usepackage[T1]{fontenc}
\usepackage{graphicx}
\usepackage{dcolumn}
\usepackage{comment}
\usepackage{amsmath,amssymb,bbold,bm}
\usepackage{amsfonts}
\usepackage{listings}
\usepackage{braket}
\usepackage{float,latexsym}
\usepackage{multirow}
\usepackage{tikz}
\usepackage{color}
\usepackage{xr}
\usepackage[normalem]{ulem}
\usepackage{pifont}
\usepackage[unicode=true,
 bookmarks=true,bookmarksnumbered=true,bookmarksopen=false,citecolor={blue},urlcolor={magenta}, breaklinks=true,pdfborder={0 0 0},backref=false,colorlinks=true]{hyperref}

\usepackage{cleveref}
\crefname{appendix}{Appendix}{Appendices}
\crefname{equation}{Eq.}{Eqs.}
\crefname{figure}{Fig.}{Figs.}
\crefname{table}{Table}{Tables}
\crefname{section}{Section}{Sections}
\crefname{mythe}{Theorem}{Theorems}
\crefname{mydef}{Definition}{Definitions}

\def\ie{{\it i.e.},\ }

\def\nono{\nonumber}
\allowdisplaybreaks
\def\kk{\mathbf{k}}

\def\QQ{\mathbf{Q}}
\def\qq{\mathbf{q}}
\def\bsigma{\boldsymbol{\sigma}}
\def\aa{\mathbf{a}}
\def\bb{\mathbf{b}}
\def\RR{\mathbf{R}}
\def\rr{\mathbf{r}}
\def\tt{\mathbf{t}}
\def\PH{\mathcal{P}}
\def\calH{\mathcal{H}}
\def\calD{\mathcal{D}}

\begin{document}
\title{Higher-Order Topological Superconductivity in Twisted Bilayer Graphene}
\begin{abstract}
We show that introducing spin-singlet or spin-triplet superconductivity into twisted bilayer graphene induces higher-order topological superconductivity.
$C_{2z}T$-protected corner states of Majorana Kramers pairs appear at the boundary between domains with opposite signs of pairing, and zero modes materialize in Abrikosov vortices.
The topology of the superconducting phase originates from the anomaly \cite{Song:2020b}--—the absence of a lattice support---of the single-valley band structure of twisted bilayer graphene, which is protected by $C_{2z}T$ and the particle-hole symmetry ${\cal P}$.
We prove that any pairing (spin-singlet or spin-triplet) term preserving valley-U(1), spin-SU(2), time-reversal, $C_{2z}T$, and ${\cal P}$ \emph{must} drive the system into a higher-order topological superconductor phase.
Here spin-SU(2) is the global spin-SU(2) for the singlet pairing and a combination of two SU(2)'s in the two valleys for the triplet pairing.
Using a Dirac Hamiltonian, we demonstrate the existence of corner modes and confirm this with numerical calculations.
These corner states are stable even if the approximate particle-hole symmetry $\mathcal{P}$ is weakly broken, which is true in  experimental setups.
Finally, we suggest an experiment to detect the topological superconductivity: by observing the fractional Josephson effect in a TBG-TSC Josephson system.
\end{abstract}
\author{Aaron Chew}
\affiliation{Department of Physics, Princeton University, Princeton, New Jersey 08544, USA}
\author{Yijie Wang}
\affiliation{International Center for Quantum Materials, School of Physics, Peking University, Beijing 100871, China}
\author{B. Andrei Bernevig}
\affiliation{Department of Physics, Princeton University, Princeton, New Jersey 08544, USA}
\author{Zhi-Da Song}
\email{zhidas@princeton.edu}
\affiliation{Department of Physics, Princeton University, Princeton, New Jersey 08544, USA}

\maketitle
{\bf \emph{Introduction.}}~Twisted bilayer graphene (TBG) plays host to a plethora of exciting physics, including superconductivity, correlated insulators, the quantum anomalous Hall effect, and ferromagnetism \cite{Bistritzer:2011, Cao:2018a, Cao:2018b, Lu:2019, Yankowitz:2019, Sharpe:2019, Tarnopolsky:2019, Saito:2020a, Stepanov:2020, Arora:2020, Serlin:2019, Cao:2020, Polshyn:2019, Xie:2019, Jiang:2019, Choi:2019, Zondiner:2020, Wong:2020, Saito:2020b, Das:2020, Saito:2020c, Nuckolls:2020, Wu:2020a, Xu:2018a, Koshino:2018, Ochi:2018, Zou:2018, Xu:2018b, Guinea:2018, Venderbos:2018, You:2019, Fu:2018, Wu:2020b, Lian:2019, Wu:2018, Isobe:2018, Liu:2018, Bultinck:2020a, Zhang:2019, Liu:2019a, Liu:2019b, Dodaro:2018, Efimkin:2018, Gonzales:2019, Yuan:2018, Kang:2019, Bultinck:2020b, Khalaf:2020, Kang:2018, Po:2018, Song:2019, Po:2019, Ahn:2019, Bouhon:2019,Lian:2020, Xie:2020,Kang:2020,Julku:2020,Lu:2020,Soejima:2020,Pixley:2019,Konig:2020,Padhi:2020, Christos:2020, Hejazi:2020, Wang:2020, Liao:2020, Onari:2020, Chen:2020, Khalaf:2020, He:2021, pierce2021unconventional,zhang2021spin,Fernandes_2021,jian2021charge4e,Sharpe_2021,lewandowski2021does,lin2021spinorbit,Wang_2021b,gonzalez2021magnetic,Kim_2021,cha2021strange,goodwin2021flat,zhang2021fractional,liu2021gig,sheffer2021chiral,phong2021band,shavit2021theory,chaudhary2021shiftcurrent,xie2021fractional,Thomson_2021,Kwan_2021,kwan2020domain,Wang_2021,Chichinadze_2020b,Chichinadze_2020a}.
The richness stems from several remarkable properties: the nearly flat bands that emerge at the magic angle, which allow for interactions to dominate the physics \cite{Bistritzer:2011}; the (previously thought) fragile topology of these bands, whereupon adding additional trivial bands renders the system trivial \cite{Ahn:2019, Po:2019, Song:2019}; and the effective symmetries that appear in certain limits of TBG, including a unitary particle-hole symmetry $P$ that appears at charge neutrality of the single-particle bands \cite{Song:2019, Song:2020b}.
It will be more convenient to use the anti-unitary particle-hole symmetry ${\cal P}= P C_{2z}T$, which is local in real space.

In Ref. \cite{Song:2020b}, some authors of the present work showed that the Bistritzer-MacDonald model of single-valley TBG is anomalous: it cannot be realized in a lattice model that preserves the $C_{2z}T$ and ${\cal P}$ symmetries.
It is well-known (e.g., \cite{Fu:2008}) that an anomalous band structure plus a pairing term that respects the protecting symmetries can yield a topological superconductor (TSC).
We prove that TBG plus a pairing term will lead to a TSC phase, which we term TBG-TSC, as long as the pairing preserves spin-SU(2), valley-U(1), time-reversal, $C_{2z}T$, and ${\cal P}$ symmetries.
Spin and valley remain good quantum numbers in the superconducting phase.

We use a Dirac theory to demonstrate the topology of TBG-TSC.
Without pairing, TBG has eight Dirac cones: one per combination of Moir\'e valley (at $K_M$ and $K_M'$), graphene valley $\eta = \pm 1$, and spin $s = \pm 1$.
We introduce spin-singlet or spin-triplet superconductivity into TBG at charge neutrality to gap the Dirac cones.
Since no superconductivity  has yet been observed at charge neutrality in experiment, it must be introduced via proximity to a superconductor.
We demonstrate that the gapped Dirac theory yields a \emph{higher-order} symmetry-protected TSC, and zero modes appear bound to corners of the system.
Higher-order topological phases exhibit gapless modes not necessarily in $d - 1$ dimensions, as is usually expected from the bulk-boundary correspondence, but also in $d - 2$ dimensions or lower \cite{Benalcazar:2017a, Benalcazar:2017b, Schindler:2018, Song:2017, Slager:2015, Peng:2017, Matsugatani:2018, Trifunovic:2019, Zhang:2020a, Zhang:2020b, Watanabe:2020, Parameswaran:2017, Tiwari:2020, Vu:2020, Khalaf:2018, Geier:2018}.
When pairing is present the system may realize Majorana corner modes, studied for example in Refs.~\onlinecite{Yan:2017,Yan:2018,Wang:2018, Teo:2013, Xu:2014, Khalaf:2018, Benalcazar:2014, Zhu:2018, Dwivedi:2018, Wang:2018b, Liu:2018b, Wu:2019, Zhu:2019, Pan:2019, Zeng:2019, Kheirkhah:2020, Franca:2019, Bomantara:2020, Hsu:2020, Varjas:2019, Pahomi:2020, Laubscher:2019, Wu:2020c, Plekhanov:2019, Zhang:2019b, Wu:2020d, Roy:2019, Yan:2019, De:2020, Wu:2020e, Wu_2020f, Bomantara:2020, Roy:2020, Wu:2020f, Tiwari:2020, Eschmann:2020, Wu:2020g, Li:2020, Ghosh:2020, Huang:2020}.
Each valley yields four gapped Dirac cones in the \emph{non-redundant} BdG basis.
Within a  single valley, domain walls (in the $C_{2x}$-invariant direction) capture two helical modes,
corners bind two complex fermion zero modes (or four Majoranas) per valley, which can be labeled by Moir\'e valley.
The four total fermionic corner modes are globally protected by $\text{valley-U(1)}, C_{2z}T, {\cal P}, $and a chiral symmetry $S$ that emerges as a result of time-reversal symmetry.
The corner states are pinned to $C_{3z}^i C_{2x} C_{3z}^{-i}$-invariant ($i=0,1,2$) corners of the system.

We verify the corner modes numerically.
We also demonstrate at the level of free fermions that $C_{2x}$-symmetric edges are gapless.
We then consider interactions using bosonization, and argue that even in the presence of translation symmetry along a $C_{2x}$-protected edge, strong interactions can gap out the edge.
We also examine the fate of the zero modes under interactions, and conclude with an experimental setup to detect TBG-TSC, namely, creating a Josephson junction using TBG-TSC and observing the effect of winding superconducting phase.

{\bf \emph{Dirac theory.}}~
TBG obeys spin-SU(2) and valley-U(1) symmetries \footnote{TBG actually obeys a higher symmetry, $U(2) \times U(2)$, one for each valley.  We do not require this full symmetry for our analysis.}.
The first originates from the negligible spin-orbit coupling of graphene and the second emerges at small twist angle of TBG.
The valley-U(1) symmetry splits the Hamiltonian into two sectors, denoted by $\eta = \pm 1$ \cite{Bistritzer:2011}.
Due to the length scale difference between graphene lattice and Moir\'e lattice, a scattering process between the two graphene valleys involves momentum transfer far larger than the Moir\'e reciprocal vectors and is suppressed.
Because our pairing is intervalley, valley-U(1) is still preserved and we can still divide the superconducting Hamiltonian into two independent sectors.
Valley-U(1) symmetry is critical; without it TBG is not anomalous.
We discuss the fate of U(1)-symmetry breaking along domain walls and edges in Appendix~\ref{app:U1-breaking}; we expect that if the domain wall is smooth over the length scale of the graphene lattice but sharp over the Moir\'e lattice, valley-U(1) is still a good symmetry.

The low energy physics of TBG can be described by four Dirac points for each spin $s$:
\begin{equation}
H_0^{{(s)}}(\kk) = \mu_0 (k_x \tau_z \sigma_x + k_y {\tau_0} \sigma_y)
\end{equation}
where $\tau_{0,z}$ are Pauli matrices representing the two valleys and $ \mu_0, \sigma_{x,z}$ are Pauli matrices denoting Moir\'e valley ($K_M$ and $K_M'$) and sublattice, respectively.
Enforcing spin rotation forces $H_0^{(\uparrow)} = H_0^{(\downarrow)}$, so we drop the spin index.
This Hamiltonian respects the discrete symmetries: $T$ (spinless time-reversal), $\mathcal{P}$ (approximate anti-unitary particle-hole), $C_{2z}$, $C_{3z}$, and $C_{2x}$, where $T$ and $\cal P$ are anti-unitary and satisfy $T^2=1$, ${\cal P}^2=-1$.
The representations of the discrete symmetries for this Dirac theory are summarized in Table \ref{SymmetryTable}.
As explained in Ref.~\onlinecite{Song:2019,Song:2020b}, a unitary particle-hole symmetry $P$ emerges when the twist angle $\theta$ is small ($\sim 1^\circ$).
Here we have defined ${\cal P}$ as the combined operation $P C_{2z}T$, which is local in real space \cite{Song:2020b}.
Readers can refer to Appendix \ref{app:BdG-TBG} for the microscopic definitions of these symmetries.
All the discrete symmetries commute with the valley-U(1) and spin-SU(2) rotations and they also commute with each other except for $\{\mathcal{P},C_{2x}\}=0$, $C_{3z} C_{2x} = C_{2x} C_{3z}^{-1}$.
Each valley and spin sector has a magnetic space group generated by $C_{2z}T$, $C_{3z}$, $C_{2x}$, and ${\cal P}$ \cite{Song:2019,Song:2020b}.
The anomaly of the single-valley Hamiltonian $H^{(\eta)}_0$, defined as the block of $H_0$ with $\tau_z=\eta$, is reflected as the fact that one cannot gap $H^{(\eta)}_0$ by adding terms preserving $C_{2z}T$ and $\cal P$ symmetries.
Breaking the valley-U(1) symmetry will remove this anomaly.
For example, the valley-mixing term $\mu_z\tau_x\sigma_x$ satisfies both $C_{2z}T$ and $\PH$, and opens a gap in the Dirac Hamiltonian.

\begin{table}
\begin{tabular}{c|c|c|c|c|c|c}
Symmetry & Action on $H$ &  Action on $\cal{H}$ & $\kk \rightarrow$ & $K_M$ & $\Gamma_M$ & $M_M$  \\
\hline
$C_{2x}$ & $\mu_x\sigma_x$ & $\mu_x\sigma_x$ & $C_{2x} \kk$ & $K_M'$ & $\Gamma_M$ & $M_M$   \\
$C_{3z}$ & $e^{\frac{i2\pi}{3} \tau_z \sigma_z}$ & $e^{\frac{i2\pi}{3} \tau_z \sigma_z}$   & $C_{3z}\kk$  & $K_M$ & $\Gamma_M$ & $C_{3z} M_M$ \\
$C_{2z}T$ & $\sigma_x K$ & $\sigma_x K$ & $\kk$  & $K_M$ & $\Gamma_M$ & $M_M$ \\
$T$ & $\tau_x \mu_x K$ & $\tau_x \mu_x K$ & $-\kk$  & $K_M'$ & $\Gamma_M$ & $M_M$ \\
${\cal P}$ & $i\mu_y\sigma_x K$ & $i\xi_z\mu_y\sigma_x K$ & $-\kk$  & $K_M'$ & $\Gamma_M$ & $M_M$ \\
$S$ & - & $\xi_y$ & $\kk$  & $K_M$ & $\Gamma_M$ & $M_M$
\end{tabular}
\caption{Table of symmetries in TBG along with their actions on $H$ (the free Hamiltonian of TBG) and $\calH$ (the BdG Hamiltonian of TBG for both valleys) low-energy degrees of freedom.
The $\PH$ symmetry only emerges at charge neutrality and will disappear if the chemical potential moved away, and $S$ only exists for the BdG Hamiltonian.
}
\label{SymmetryTable}
\end{table}

It is worth mentioning that $\PH$ corresponds to a charge-conjugation symmetry $\PH_c$ of the many-body flat-band Hamiltonian of TBG \cite{TBG-III}.
$\PH_c$ has the same representation matrix as $\PH$ except that it is unitary and transforms annihilation operators to creation operators (and vise versa):
\begin{equation} \label{eq:Pc-maintext}
\PH_c c_{\kk,\eta,\nu,\alpha,s} \PH_c^{-1} = \sum_{\nu' \beta} c_{-\kk,\eta,\nu',\beta,s}^\dagger [i\mu_y]_{\nu'\nu} [\sigma_x]_{\beta \alpha},
\end{equation}
where $\nu,\nu'$ represent the Moir\'e valley, $\alpha,\beta$ represent the sublattice, and $s$ represents spin.
In this work, we regard $\PH_c$ as a physical symmetry and $\{\PH,H\}=0$ as a constraint satisfied by the single particle Hamiltonian imposed by $\PH_c$.
(See \cref{app:sym-continuous-model} and Ref. \onlinecite{TBG-III} for detailed discussions on the relation between $\PH$ and $\PH_c$.)

\begin{figure}[t!]
\includegraphics[width=\columnwidth]{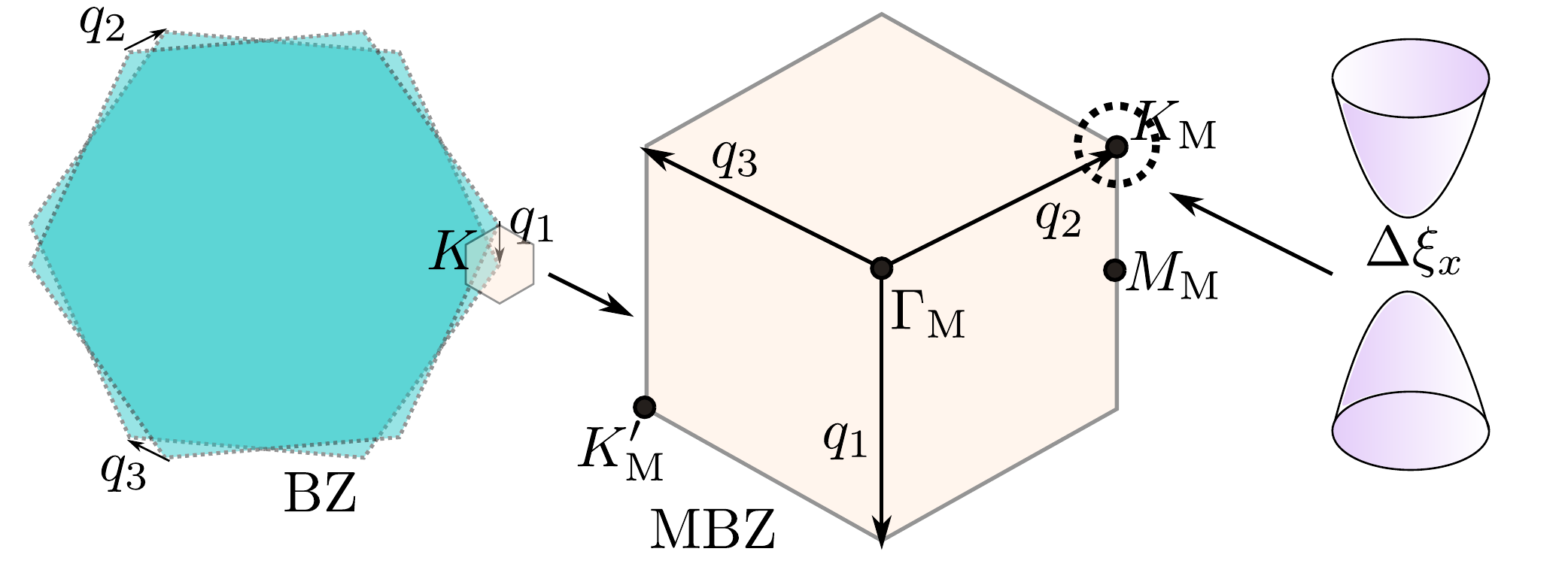}
\caption{TBG possesses four Dirac cones per graphene valley (eight total), which we hybridize with spin-singlet (and later spin-triplet) pairing.  The four cones per graphene valley can be labeled by spin and Moir\'e valley.  Pairing between opposite spin, valley, and Moir\'e valley drives us into the topological phase.
}
\label{fig:DiracCones}
\end{figure}

\begin{figure*}
\includegraphics[width=\textwidth]{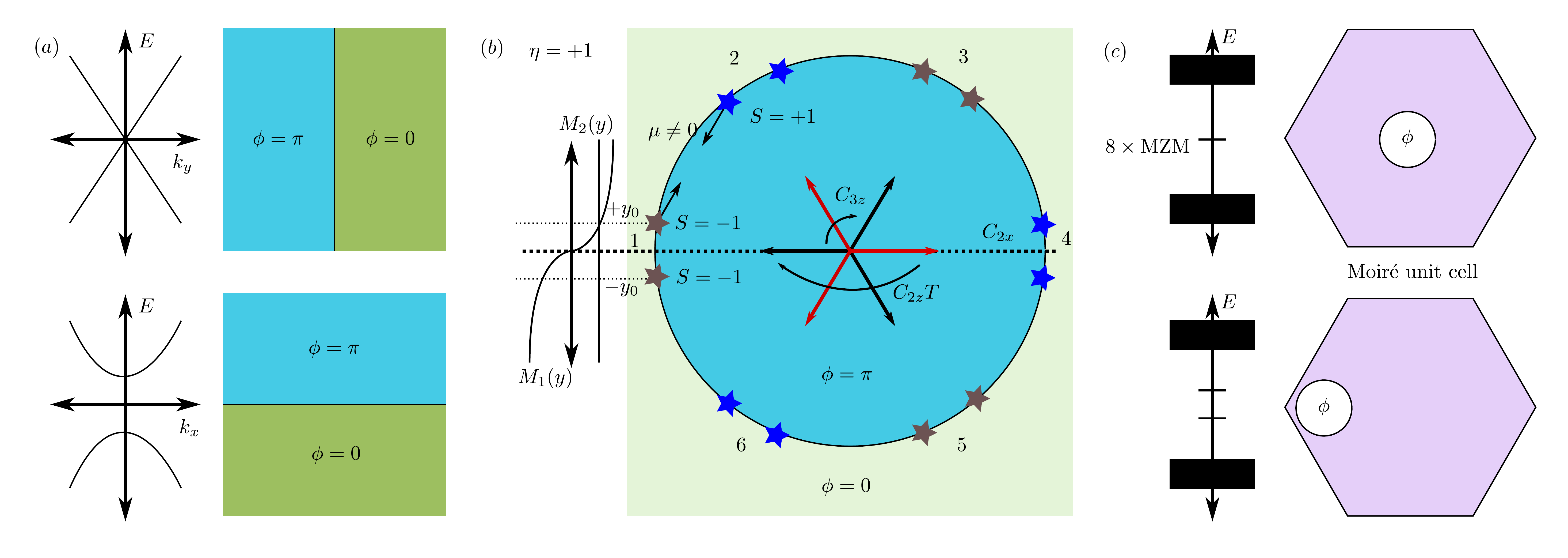}
\caption{(a) Edge states in TBG-TSC, as discussed in Appendix~\ref{app:edge-state-Dirac}. The top panel is an edge along the $y$ direction.  If we assume translation symmetry, then $C_{2x}$ will conspire to keep the band structure gapless (neglecting interactions).  On the other hand, a domain wall along the $x$ direction will be generically be gapped even with free fermions.  (b) Schematic of TBG-TSC.  Corner modes (indicated by stars) are protected by TRS, and are captured along domain walls between regions separating superconductors with phases $\phi = 0, \pi$.  In the presence of reflection symmetry these corner modes are pinned to $C_{2x}$-invariant points, and are mapped onto other zero modes via $C_{3z}$ and $C_{2z}T$.  For weak ${\cal P}$-breaking perturbations (i.e. adding $V \xi_z$) these corner modes shift to points $\pm y_0$; for strong ${\cal P}$-breaking perturbations they annihilate with their partners. (c) Magnetic flux piercing TBG-TSC.  If the flux pierces the $C_{2z}$ center of the system we find $4$ Majorana zero modes {per valley} (8 total) bound to the vortex.  Shifting away from the center causes splitting between the modes.  See Appendix~\ref{app:vortex} for detailed calculations.
}
\label{HexagonFig}
\end{figure*}

We now show that the BdG Hamiltonian of TBG in each valley sector is in Altland-Zirnbauer symmetry class CII, which is equipped with a chiral symmetry $S$, a particle-hole symmetry ${\cal P}$, and an emergent ``time-reversal'' $\tilde{T}=S{\cal P}$ satisfying ${\cal P}^2=\tilde{T}^2=-1$.
Intervalley spin-singlet pairing, which creates one fermion in each valley and thus preserves the total valley number, takes the form
\begin{align}
\Delta^{(\eta)}_{\nu\alpha; \nu'\beta}(\kk) c_{\kk,\eta,\nu,\alpha,\uparrow}^\dagger c_{-\kk,-\eta,-\nu',\beta,\downarrow}^\dagger +  h.c..
\label{PairingTerm}
\end{align}
The pairing term pairs opposite Moir\'e valley.  Switching into the non-redundant BdG basis:
\begin{align}
(
c_{\kk,\eta,\nu,\alpha,\uparrow} \   \cdots  \   c^\dagger_{-\kk,-\eta,-\nu',\alpha',\downarrow} \  \cdots
)^T,
\label{BdGBasis}
\end{align} yields the BdG Hamiltonian
{\small
\begin{align}
{\cal H}^{(\eta)}(\kk) = \begin{bmatrix}
H_0^{(\eta)}(\kk) - E_F & \Delta^{(\eta)}(\kk) \\
\Delta^{(\eta)\dagger}(\kk) & -\mu_x H_0^{(-\eta)T}(-\kk) \mu_x + E_F
\end{bmatrix},
\label{BdGHam}
\end{align} }%
with $H_0^{(\eta)}(\kk)$ being the hopping Hamiltonian projected into the valley $\tau_z = \eta$ of TBG (not spin!) and $E_F$ the chemical potential.
Then $T=\tau_x\mu_x K$ and spin-SU(2) constrains the form of the pairing and hopping Hamiltonians to satisfy:
\begin{equation}
H_0^{(\eta)}(\kk) = \mu_x H_0^{(-\eta)*}(-\kk) \mu_x,\quad
\Delta^{(\eta)}(\kk) = \Delta^{(\eta) \dagger}(\kk),
\label{SymmetriesHam}
\end{equation}
which yields the BdG Hamiltonian
\begin{align} \label{eq:BdG-Dirac}
{\cal H}^{(\eta)}(\kk) &= (H_0^{(\eta)}(\kk) - E_F)\xi_z + \Delta^{(\eta)}(\kk)\xi_x.
\end{align}
We use the Pauli matrices $\xi_{z,x}$ for particle-hole space.
Note that in the non-redundant BdG basis, the opposite spin is included in the annihilation operators.
There are two copies of this Hamiltonian, one for each valley; we focus on the positive valley $\eta = +1$.
In the BdG basis Eq.~\ref{BdGBasis}, spinful time-reversal ${\cal T} = i\hat{s}_y T$ (with $\hat{s}_y$ the spin operator corresponding to $y$) transforms the BdG spinor as
\begin{align}
  (
c_{\kk,\eta,\nu,\alpha,\uparrow} \   \cdots  \   c^\dagger_{-\kk,-\eta,-\nu',\alpha',\downarrow} \  \cdots
)^T \rightarrow \nonumber \\
(
c_{-\kk,-\eta,-\nu,\alpha,\downarrow} \   \cdots  \   -c^\dagger_{\kk,\eta,\nu',\alpha',\uparrow} \  \cdots
)^T \nonumber \\
=i\xi_y (
c_{\kk,\eta,\nu,\alpha,\uparrow}^\dagger \   \cdots  \   c_{-\kk,-\eta,-\nu',\alpha',\downarrow} \  \cdots
)^T,
\end{align} which corresponds to a unitary operator $i\xi_y$ accompanied by a particle-hole exchange.  This is the chiral symmetry $S = \xi_y$, and it anti-commutes with the Hamiltonian.  (The $i$ can be gauged away as typically chiral symmetry is chosen to square to $+1$.)
See Appendix~\ref{app:BdG-symmetry} for a microscopic derivation of the chiral symmetry $S$.

We consider the simplest spin-singlet pairing: $\Delta_{\nu\alpha;\nu'\beta}^{(+)}(\kk) = \Delta \delta_{\nu,\nu'} \delta_{\alpha\beta}$, $\Delta$ real, \ie
\begin{align}
{\cal H}^{(+)}(k) &= \xi_z {\mu_0} (k_x \sigma_x + k_y \sigma_y) \nonumber \\
&- E_F \xi_z {\mu_0\sigma_0} + \Delta \xi_x {\mu_0\sigma_0}.
\label{BulkHamiltonian}
\end{align}
As detailed in Appendix \ref{app:uniform-s}, such a pairing term corresponds to a homogeneous on-site spin-singlet pairing introduced at each carbon atom in TBG.
This spin-singlet pairing commutes with the symmetry operators $T$, $C_{2z}T$, $C_{3z}$, $C_{2x}$.
One can show that the pairing term is also invariant under the charge conjugation $\PH_c$ (\cref{eq:Pc-maintext}).
In the BdG formalism, the $\PH_c$ symmetry leads to the constraint $\PH \calH^{(\eta)}(\kk) \PH^{-1} = - \calH^{(\eta)}(-\kk)$ with $\PH=i \xi_z \mu_y\sigma_x K$.
The $\xi_z$ matrix in $\PH$ comes from the fact that the particle part and the hole part of the BdG basis consist of opposite Moir\'e valley (\cref{BdGBasis}) and the $\PH_c$ operator gives the two Moir\'e valleys the opposite signs (\cref{eq:Pc-maintext}).
(See \cref{app:BdG-symmetry} for detailed discussions on the form of $\PH$ in the BdG formalism.)
We hence identify the equivalent symmetry class of the BdG Hamiltonian in each valley as CII because in the BdG formalism $\PH^2=-1$ and $\tilde{T}^2=(S\PH)^2=-1$.
The symmetries of Eq. (\ref{BulkHamiltonian}) are summarized in Table \ref{SymmetryTable}.

Eq. (\ref{BulkHamiltonian}) is fully gapped in the bulk, and there is a symmetric copy of this Hamiltonian in the other valley ($\eta=-$), which can be obtained by applying spinless $T$ to Eq. (\ref{BulkHamiltonian}).
It is worth noting that there are no independent copies in the other spin sector  ($s=\downarrow$) because we have already included them in the non-redundant BdG basis.

{\bf \emph{Edge Hamiltonian and Corner States.}}~In this section we explicitly demonstrate the existence of edge states and corner states bound to domain walls of pairing terms with phase difference $\pi$.
We restrict ourselves to the valley sector $\eta = +$.
We first consider a domain wall perpendicular to the $x$ axis: $\Delta(x) \xi_x $, where $\Delta(x)=\Delta_0$ for $x>0$ and $-\Delta_0$ for $x<0$, {see Fig.~\ref{HexagonFig}(a)}.
As $k_y$ is a good quantum number we expect the states localized in the $x$ direction and propagating along $y$.
This calculation is carried out in Appendix~\ref{app:edge-state-Dirac}.
As expected for Dirac fermions under a mass change, we find four gapless edge modes per valley (two chiral and two anti-chiral); their particle-hole, Moir\'e valley, and sublattice indices are given by $\{\xi_y,\mu_z,\sigma_x\}=\{1,1,1\}$, $\{1,-1,1\}$, $\{-1,1,-1\}$, $\{-1,-1,-1\}$, respectively.
The projected Hamiltonian (performed in Appendix~\ref{app:edge-state-Dirac}) on the edge modes is
\begin{align}
H^{\text{edge}}(k_y) = k_y \xi_y' \mu_0.
\label{ProjHam}
\end{align}
Here $\xi_y'$ and $\mu_0$ are Pauli matrices defined on the domain wall.
The Pauli matrix $\xi_y'$ is not to be confused with the original Pauli matrix $\xi_y$, though it turns out that the Pauli $\mu_0$ coincides for both edge and bulk.
The projected chiral, particle-hole, and $C_{2x}$ symmetries are $S^{\rm edge}=\xi_z' {\mu_0}$, ${\cal P}^{\rm edge} = i \xi_z' \mu_y K$, $C_{2x}^{\rm edge}=\xi_z'\mu_x$, respectively.
No homogeneous gap term is allowed by these symmetries.
To be specific, the only {homogeneous} gap term {(anti-commuting with the Hamiltonian)} that is allowed by $S$ and ${\cal P}$ is { $\xi_x'\mu_0$}; however, this breaks $C_{2x}$.
Therefore, the TBG-TSC has protected gapless edge states on the pairing domain walls in the $y$-direction.
The number of edge solutions is doubled due to the other valley ($\eta=-1$).
Because we have used the \emph{non-redundant} BdG basis, the zero mode solutions for the corner states are not Majoranas, but complex fermions.

TBG-TSC also manifests itself as unavoidable corner states.
To see this, we consider a circular region with positive pairing term $\Delta_0$ surrounded by negative pairing term $-\Delta_0$ (Fig. \ref{HexagonFig}(b)).
The 1st high symmetry corner, \ie the corner in the $x$-direction, has the same symmetries as the edge discussed in the last paragraph except it does not have the translation symmetry along $y$.
Hence the $S$- and ${\cal P}$-symmetric gap term $M_1(y) \xi_x'\mu_0$ is now allowed but must change sign under $y\to -y$ to preserve $C_{2x}$, $M_1(y) = -M_1(-y)$.
The zero of $M_1(y)$ at $y=0$ leads to two Jackiw-Rebbi complex fermion zero modes (per valley), as derived in Appendix~\ref{app:Dirac}.
We find that the two complex fermion zero modes in each valley have the same chiral eigenvalue $+1$ and thus are robust against arbitrary perturbations respecting the chiral symmetry \footnote{Unlike unitary symmetries, a chiral symmetry disallows hoppings between states of the same chiral eigenvalue.}.
The two zero modes must be located at the same position in real space because of ${\cal P}$ --- since ${\cal P}$ is a local operator and satisfies ${\cal P}^{2}=-1$, due to Kramers theorem, it must transform one fermionic zero mode to another at the same position.
We call such a pair of fermionic zero modes a Kramers doublet.
Due to the $C_{3z}$ and $C_{2z}T$ symmetries, zero modes also appear at other points in the sample, as shown in Fig. (\ref{HexagonFig}).
Since $[C_{3z},S]=0$ and $\{C_{2z}T,S\}=0$, zero modes at the 3rd and 5th corners, which are respectively rotated from the 1st corner by $C_{3z}$ and $C_{3z}^{-1}$, have the chiral eigenvalue $+1$; whereas zero modes at the 2nd, 4th and 6th corners, which are respectively rotated from the 5th, 1st, 3rd corners by $C_{2z}T$, have the chiral eigenvalue $-1$.
The other valley has opposite chiral eigenvalues. In short, both ${\cal P}$ and $S$ protect the zero modes from splitting, while $S$ also keeps the zero modes at $0$ energy.

In fact, $C_{3z}$ and $C_{2x}$ are not required for the corner modes to exist, but their presence will bind the the zero modes at the $C_{2x}$-invariant corners \cite{Langbehn:2017} and their $C_{3z}$-symmetric locations.
Breaking of $C_{3z}$ and $C_{2x}$ in a practical setup may change the locations of corner states but cannot remove them.
In order for two Kramers doublets to annihilate they must carry opposite chiral eigenvalues; thus we require a doublet of $S = +1$ and another doublet of $S = -1$ to come together.
$C_{2z}T$ will reflect these doublets to the opposite corner, giving a total of four doublets per valley required to annihilate completely.
Since our system has six doublets per valley, we cannot fully annihilate every corner while preserving $C_{2z}T$, and thus our system is topological.

${\cal P}$ is an approximate symmetry in TBG at charge neutrality with small twist angle; either much larger twist angle or finite doping will break the ${\cal P}$ symmetry.
Thus the robustness of corner modes upon ${\cal P}$-breaking perturbations is crucial for them to be experimentally relevant.
Since the two zero modes at a same corner have same chiral eigenvalue, they must be locally stable against ${\cal P}$-breaking perturbations.
By explicit calculations given in Appendix~\ref{app:Dirac}, we find that the ${\cal P}$-breaking terms will split the two zero modes in real space but leave their energies unchanged.

We have numerically confirmed the existence of edge states and corner states using the BM model of TBG plus a spin-singlet pairing (\cref{app:num_TBG}).
The evolution of the corner modes (\cref{HexagonFig}) under the $\PH$ breaking term, which is chosen as the chemical potential in the calculation, is also observed in \cref{app:num-corner}.

Ref. \onlinecite{Christos:2020} found that only four types of pairings in TBG lead to full gaps in the Bogoliubov bands: the singlet pairing forming the representation $A_1$ (uniform spin-singlet), the singlet pairing forming the representation $A_2$, the triplet pairing forming the representation $B_1$, and the triplet pairing forming the
representation $B_2$.
Here $A_{1,2}$ and $B_{1,2}$ are notations of the irreducible representations of the point group $D_6$.
In the above we have discussed the $A_1$ pairing.
In \cref{app:other-pairing} we analyze all four cases and find that $B_1$ is also topological whereas $A_2$ and $B_2$ are trivial.
As discussed in the end of \cref{app:B1B2-pairing}, due to the separate spin-SU(2) symmetries in the two valleys, there is no real difference between spin-singlet and spin-triplet in TBG, and $A_1$ and $B_1$ are related by a combination of two different spin-charge rotations in the two valleys.

{\bf \emph{Interactions.}}~We now consider the fate of TBG-TSC in the presence of strong interactions.
As typical bandwidths of TBG near the magic angle are very small, we expect interactions to play a comparatively large role.
The low-energy form of the edge Hamiltonian Eq.~\ref{ProjHam} suggests a bosonization treatment.
Along a $C_{2x}$-invariant edge, we know that the free-fermion Hamiltonian is gapless.
We carry out the bosonization in Appendix~\ref{app:bosonization}: there are $4$ pairs of helical Dirac modes total, as each of the eight Dirac cones yields a chiral edge state.
Each of the eight chiral states may be labeled by a combination of chirality, graphene valley, Moir\'e valley, or spin, with one label being redundant.
For cleaner notation, we opt to use spin in place of graphene valley, and combine Moir\'e valley and chirality into a new label we call flavor.
\begin{align}
\psi_{\alpha n s} \sim e^{i(\varphi_{\alpha s} + n \theta_{\alpha s})},
\end{align}
where $\alpha,n,s$  denote flavor $1,2$, chirality $R,L$, and spin $\uparrow, \downarrow$ respectively.
$n$ in the exponent is $\pm 1$ for chirality $R/L$, respectively.
We find that it is possible to gap out the edge degrees of freedom even for a $C_{2x}$-symmetric edge.

One may also find explicit expressions for the zero modes bound to the corners, as performed in Appendix~\ref{app:corner-dirac}.
We denote the corner state operators by $\Phi_{as}$, with $a$ denoting flavor and $s$ spin.
We may also perform an analysis of what gap terms are allowed by symmetry for the zero modes, and show that even if all symmetries are present the corner modes may completely gap out with interactions, as shown in Appendix~\ref{app:corner-interactions}.

{\bf \emph{Valley Symmetry Breaking.}}~Our realization of topological edges and Majorana corner modes requires that valley-U(1) symmetry be preserved, so the valleys do not couple.
In fact, we prove in Appendix \ref{app:U1-breaking} that along a hard edge connecting TBG with the vacuum, either U(1) valley symmetry or $\PH$ is strongly broken, and hence has no chance of realizing the zero modes.
This problem can be circumvented by instead considering a domain wall between superconductors of phase difference $\pi$ instead of a hard edge.
While the domain wall still introduces scattering that may link the valleys, the breaking of valley-U(1) is no longer enforced on general grounds.
We may consider, then, a slowly varying domain wall where the jump in superconducting phase is slow on the scale of the graphene unit cell, but rapid on the scale the Moir\'e unit cell.

{\bf \emph{Experimental Detection.}}~TBG-TSC, as a higher-order topological superconductor, should yield a fascinating assortment of experimental signatures.
At the free fermion level, these are readily apparent:  altering the chemical potential (e.g. by adjusting gate voltage) will shift the location of the corner states because of the breaking of $\PH$.
The corner modes remain at zero energy over a finite range of voltages until they annihilate one another, as shown in \cref{HexagonFig}(b) and proven in \cref{app:corner-split}.
Along $C_{2x}$-invariant edges, the edge states remain gapless and thus offer transport signatures.
As illustrated in \cref{HexagonFig}(c) and detailed in \cref{app:vortex}, zero modes also appear at the center of vortexes of the pairing order parameter.  However, interactions complicate the detection of the gapless states and edges; we show in Appendix~\ref{app:bosonization},\ref{app:corner-interactions} that symmetry-preserving interactions can gap the corner states and gapless edges.

\begin{figure}
    \centering
    \includegraphics[width=\columnwidth]{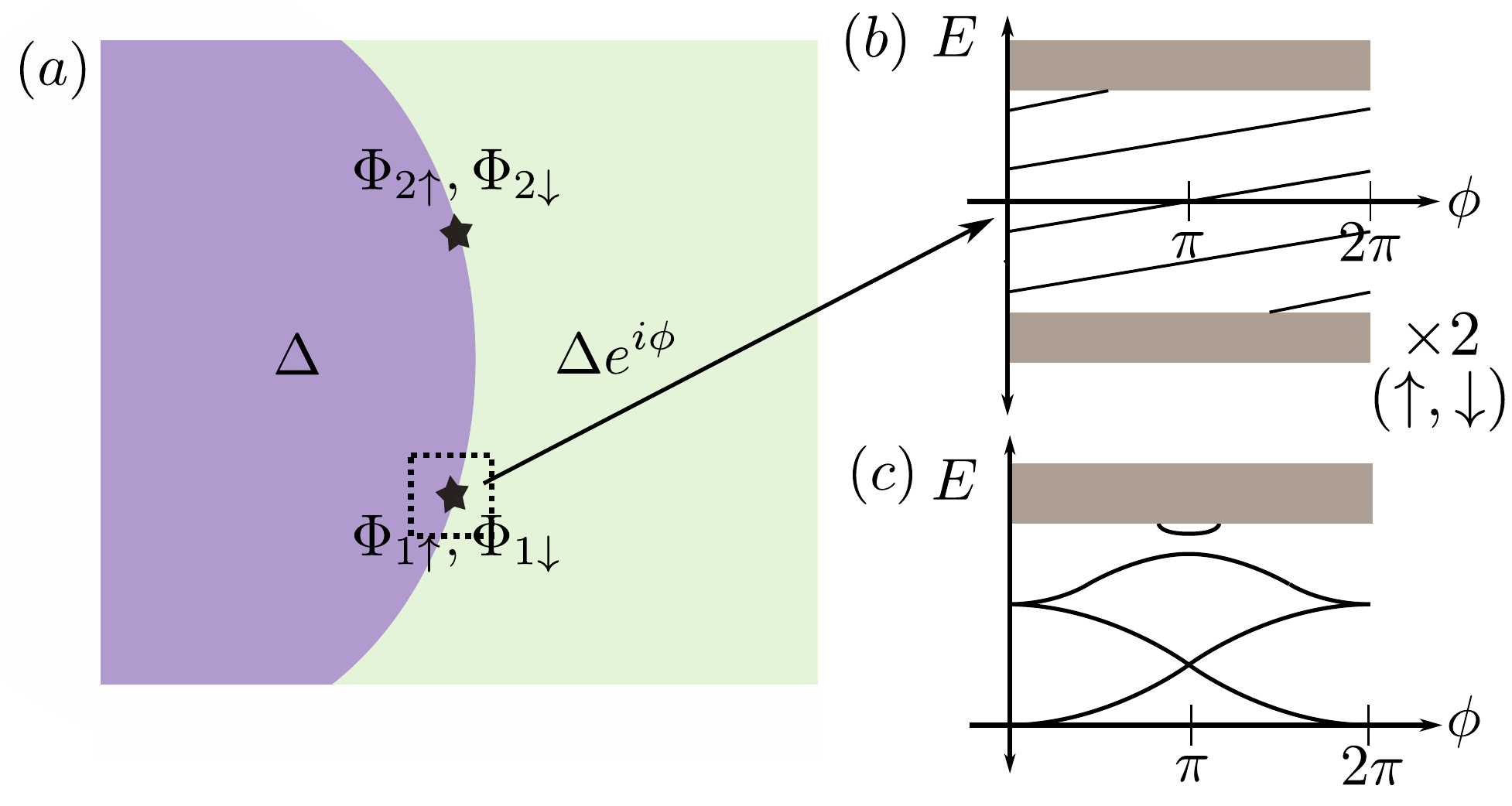}
    \caption{Josephson junction for TBG-TSC.  In (a), we apply a chemical potential via gate to shift the {pairs of} corner modes $1,2$ away from one another.  At $\phi = \pi$, 4 total complex fermion corner modes (at the free fermion level) are pinned to zero energy, and breaking $\PH$ shifts the two pairs of two complex fermions in space.  Varying the superconducting phase away from $\pi$ will allow the corner modes to shift from zero energy, as depicted in (b), the single-particle spectrum (at finite chemical potential) for $\Phi_{1\uparrow}$.  There is an identical copy of the spectrum for the opposite spin sector $\Phi_{1\downarrow}$, and doubled again for the other pair of in-gap modes $\Phi_{2s}$, though since that is separated in space we will not consider it.  In (c), the multi-particle states involved in the ground state evolution are plotted.  See Appendix~\ref{app:josephson} for an in-depth discussion.}
    \label{fig:josephson}
\end{figure}

We propose a further setup to detect the higher-order topology with interactions: the fractional Josephson effect.
Fig.~\ref{fig:josephson} shows a sheet of TBG-TSC hybridized to a Josephson junction between two superconductors.
At phase difference $\phi = \pi$, the {4 complex fermion} corner modes $\Phi_{{as}}$ exist at zero energy (at the level of free fermions).
As with the helical edges, we opt for spin $s$ instead of valley to label the corner states,  and the remaining two states are labeled by a second index $a$.
Changing the chemical potential will shift them in location, but so long as the symmetry breaking is not too large the zero modes are stable.
Changing the phase difference away from $\phi = \pi$ allows the corner states to shift away from zero energy.  We will denote these states as $\Phi_{as}[\phi]$ and they are in-gap states; close to $\phi = \pi$ they are smaller than the gap but they are not pinned at $E = 0$.  Consider the Josephson junction Hamiltonian
\begin{align}
    H[\phi] &= \xi_z \mu_0 (-i\partial_x \tau_z \sigma_x + -i\partial_y \sigma_y) + \Delta \Theta(-x) \xi_x \nonumber \\
    &+ \Delta \Theta(x) [\cos(\phi) \xi_x + \sin(\phi) \xi_y].
\end{align} This Josephson Hamiltonian breaks the chiral symmetry $S = \xi_y$, but the symmetry is restored if we also take $\phi \rightarrow -\phi$.  Hence the spectra of the in-gap states is symmetric about $\phi = \pi: E[\phi] = -E[-\phi]$.

What is the fate of the four in-gap states as $\phi \neq \pi$?  The pairs $\Phi_{1s}[\phi]$ and $\Phi_{2s}[\phi]$ are related by $C_{2x}$ and yield identical spectra.  The spin-up and spin-down states are related by spin rotation and thus \emph{also} have identical spectra.  In addition, at $\phi = 0, 2\pi$ the Josephson Hamiltonian becomes a homogeneous sheet of TBG-TSC, and thus the in-gap states will flow into the continuum at $\phi = 0, 2\pi$ (though using a soft instead of a hard domain wall will allow for in-gap states even at $\phi = 0, 2\pi$).  Hence the single-particle spectrum appears as Fig.~\ref{fig:josephson}(b).  Each in-gap mode $\Phi_{as}[\phi]$ carries valley number $+1$.  A pumping cycle that winds $\phi$ by $2\pi$ will begin in the ground state, with all negative energies unoccupied, and end up occupying positive energy states which all carry valley number $+1$.  At the level of free fermions, valley-$U(1)$ is a good symmetry, and so no matter how many times we wind $\phi$, the valley number continues to increase with no chance of mapping back to the ground state.  The Josephson junction is aperiodic.

We now add interactions, and show this aperiodic pumping cycle becomes $6\pi$ periodic, or a $Z_3$ fractional Josephson effect \cite{Kitaev:2001,Kwon:2003, Fu:2009, Zhang:2013, Liu:2014, Mellars:2016,  Camjayi:2017, Haim:2018}.
Begin in the ground state at $\phi = 0$.  In Fig.~\ref{fig:josephson}(b), the single-particle spectrum, this corresponds to all negative energy states occupied and positive energy states unoccupied.
Tracing the states as $\phi$ winds by $2\pi$, we end in an excited state where the first positive energy state for both spins is occupied; the valley number has changed by $+2$.
So long as valley-U(1) is conserved, there is no way for the multi-particle ground state to return to its original form, as each winding changes the valley number.
However, as argued in next paragraph, in many-body language, the valley-U(1) is reduced to a $Z_3$ symmetry, that allows for six-body pairing terms.
If such terms are allowed, then the many-body spectrum avoids as in Fig.~\ref{fig:josephson}(c), resulting in a $Z_3$ fractional Josephson effect.
(See Appendix~\ref{app:josephson} for more details.)

The presence of valley symmetry originates from the separation of the $K, K'$ points in the graphene BZ.
Because the graphene valleys are far apart, scattering between the valleys is suppressed, yielding number conservation within each valley.
With interactions, however, processes involving multiple fermions are possible.
Because $3K = 0$ modulo a reciprocal lattice vector in the graphene BZ, six-body terms giving a net momentum transfer of $6K$ are technically allowed by graphene translation, breaking valley-U(1) to $Z_3$.

{\bf \emph{Conclusion and Discussion.}}~We have shown that proximitizing twisted-bilayer graphene with the $A_1$ spin-singlet (or $B_1$ spin-triplet) superconductivity \emph{must} yield a higher-order topological superconductor.
Multiple zero modes are bound to corners of the system.
We have explicitly demonstrated the topological phase and proved its existence with the Wilson loop formalism (\cref{app:WilsonLoop}).
We have also examined the fate of the gapless edge states and zero modes under symmetric interactions, and concluded with possible experimental signatures of the zero modes (\cref{app:josephson}).

Our work harnesses the unusual bulk topological properties of TBG in a special superconducting heterostructure.
By combining a relatively simple phase (a spin-singlet superconductor) with TBG, we engineer a higher-order topological superconductor with corner modes bound to domain walls.
Our work begs the question if other heterostructures can exploit the anomalous structure of TBG to yield even more exotic topological phases; for example, by using ferromagnets or quantum Hall systems.

We also conjecture that introducing spin-singlet or spin-triplet superconductivity into the recently realized mirror symmetric magic-angle twisted trilayer graphene (MATTG) \cite{park2020TTG} also leads to topological superconductor because in MATTG a single valley has an odd number of Dirac points protected by the $C_{2z}T$ symmetry, which is also anomalous and usually only appears as the surface state of the axion insulator.

{\bf \emph{Acknowledgements.}}~A.C. was supported by the Gordon and Betty Moore Foundation through Grant GBMF8685 towards the Princeton theory program.
Further funding for this work came from the DOE Grant No. DE-SC0016239, the Schmidt Fund for Innovative Research, Simons Investigator Grant No. 404513, and the Packard Foundation.
Further support was provided by the NSF-EAGER No. DMR 1643312, NSF- MRSEC No. DMR-1420541 and DMR-2011750, ONR No. N00014-20-1-2303, BSF Israel US foundation No. 2018226, and the Princeton Global Network Funds.

\bibliography{TBGTSCRefs}
\appendix
\tableofcontents
\section{The BdG formalism of superconductor phase in twisted bilayer graphene} \label{app:BdG-TBG}

In this section, we derive the BdG Hamiltonian from the continuous two-valley model of TBG described in Ref. \cite{Song:2019} and explain the symmetries.

\subsection{A brief review of the continuous two-valley model of TBG}
\begin{figure}[t]
\centering
\includegraphics[width=1.0\linewidth]{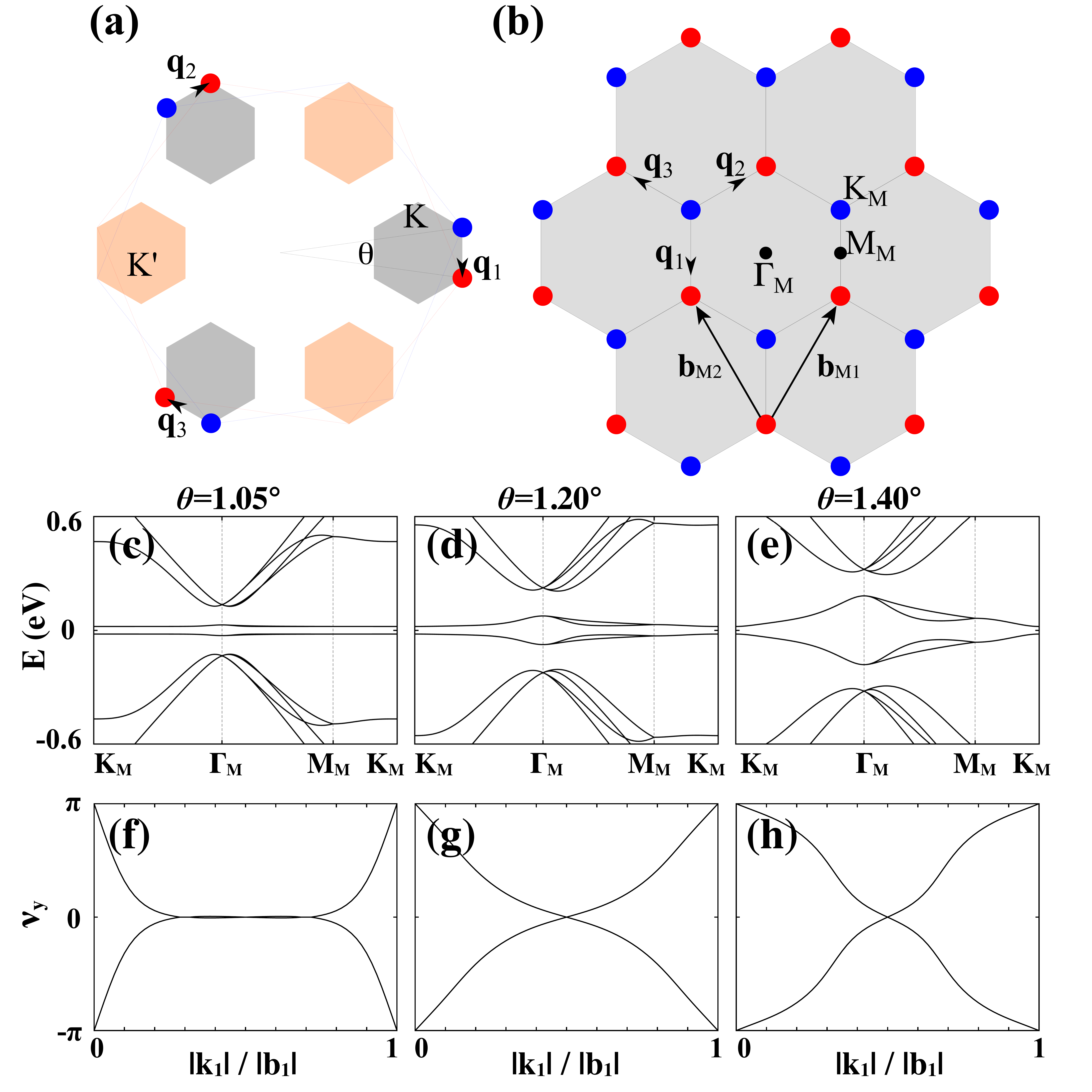}
\caption{\label{fig:MB_WL} (a) The graphene Brillouin zones of the two layers. The blue (red) hexagon represents the top (bottom) layer. With a twist angle $\theta$, two Moir\'e Brillouin zones are formed (small hexagons) for the two graphene valleys. (b) The Moir\'e Brillouin zone. Blue and red points denote Dirac cones from the top and bottom layers, respectively. $\mathbf{b}_{M1,M2}$ are the reciprocal lattice vectors.
(c)-(e) Bogoliubov Moir\'e bands with the $s$-wave pairing, solved at $\theta=1.05^\circ,1.20^\circ,1.40^\circ$, respectively.
(f)-(h) Wilson loops of the -1 and -2 Bogoliubov bands, solved at $\theta=1.05^\circ,1.20^\circ,1.40^\circ$, respectively. All the calculations are carried out at $w_1=110$meV, $w_0=0.8w_1$, $v_F=5.944{\rm eV\cdot \mathring{A}}$, and $|\Delta_0|=20$meV.
}
\end{figure}

The two-valley Hamiltonian for TBG reads
\begin{align} \label{eq:H0-second-quantized}
    \hat{H}_0 = \sum_{\kk \QQ \QQ'\eta s \alpha \beta} H^{(\eta)}_{\QQ\alpha,\QQ'\beta}(\kk) c^\dagger_{\kk \QQ \eta \alpha s} c_{\kk \QQ' \eta \beta s},
\end{align}
where $\eta=\pm$, $\alpha(\beta)=1,2$, and $s=\uparrow\downarrow$ are the valley, sublattice, and spin indices, respectively.
$\QQ$-vectors form a honeycomb lattice generated by $\qq_1=k_D(0,-1)$, $\qq_2=k_D(\frac{\sqrt3}2,\frac12)$, $\qq_3=k_D(-\frac{\sqrt3}2,\frac12)$, with $k_D=2|K|\sin\frac{\theta}2$ being the  distance between the Dirac point $K_+$ in the top layer and the Dirac point $K_-$ in the bottom layer (\cref{fig:MB_WL}(a)).
The single-particle Hamiltonian is given by
\begin{align} \label{eq:A2-Hk}
    H_{\QQ\alpha,\QQ'\beta}^{(+)}(\kk) = & v_F k_D  \delta_{\QQ,\QQ'} [(\kk - \QQ) \cdot \bsigma]_{\alpha\beta} \nono \\
+ & \sum_{j = 1}^3 (\delta_{\QQ' - \QQ,\qq_j} + \delta_{\QQ-\QQ',\qq_j}) T^j_{\alpha \beta},
\end{align}
\begin{align}
    H_{\QQ\eta,\QQ'\beta}^{(-)} (\kk) = & - v_F k_D  \delta_{\QQ,\QQ'} [(\kk - \QQ) \cdot \bsigma^*]_{\alpha\beta} \nono \\
+ & \sum_{j = 1}^3 (\delta_{\QQ' - \QQ,\qq_j} + \delta_{\QQ-\QQ',\qq_j}) T^{j*}_{\alpha \beta},
\end{align}
and
{\small
\begin{align}
    T^j = w_0 \sigma_0 + w_1 (\cos(\frac{2\pi}{3}(j-1)) \sigma_x + \sin(\frac{2\pi}{3}(j-1)) \sigma_y).
\end{align}}%
Here $\bsigma=(\sigma_x,\sigma_y)$, $\bsigma^*=(\sigma_x,-\sigma_y)$ are Pauli matrices for the sublattice degree of freedom, $v_F$  is the Fermi velocity, and $w_0$ and $w_1$ are the inter-layer couplings dominated by the AA-stacking and AB-stacking regions, respectively.
Due to the corrugation effect in the $z$-direction, $w_0$ is usually smaller than $w_1$.
The Moir\'e Brillouin zone (MBZ) is spanned by $\bb_{M1}=\frac{2\pi}{a_{M0}} (\frac{1}{\sqrt3},1)$, $\bb_{M_2} = \frac{2\pi}{a_{M0}} (-\frac{1}{\sqrt3},1)$.
The high symmetry points in the MBZ are shown in \cref{fig:MB_WL}(b).
Correspondingly, the lattice vectors of the Moir\'e unit cell are $\aa_{M1} = a_{M0} (\frac{\sqrt3}2,\frac12)$, $\aa_{M2} = a_{M0} (-\frac{\sqrt3}2,\frac12)$, where $a_{M0}=\frac{4\pi}{3k_D}$ is the Moir\'e lattice constant.
In this work, we choose the parameters as $v_F=5.944 \mathrm{eV\cdot \mathring{A}}$, $|K|=1.703\mathring{A}^{-1}$, $w_1=110$meV, $w_0=0.8w_1$.

Each electron operator $c^\dagger_{\kk \QQ \eta \alpha s}$ is a Fourier transformation of the electron operators of all the atomic orbitals in the two layers of graphene.
In next subsection, we will use this definition to obtain the form of singlet pairing in momentum space.
We first split the $\QQ$-lattice into two sublattices:
\begin{equation}
\mathcal{Q}_+=\{ \kk_{K_M} +n \bb_{M1} + m\bb_{m_2}\, |\, n,m\in \mathbb{Z} \},
\end{equation}
\begin{equation}
\mathcal{Q}_-=\{ \kk_{K_M'} +n \bb_{M1} + m\bb_{m_2}\, |\, n,m\in \mathbb{Z} \},
\end{equation}
where $\kk_{K_M}$ and $\kk_{K_M'}$ are the high symmetry momenta $K_M$ and $K_M'$ in the MBZ, respectively.
In \cref{fig:MB_WL}(b), $\mathcal{Q}_+$ and $\mathcal{Q}_-$ are colored as blue and red, respectively.
Then we define the index
\begin{equation}
\zeta_\QQ = \begin{cases}
1,\qquad & \QQ \in \mathcal{Q}_+ \\
-1,\qquad & \QQ \in \mathcal{Q}_- \\
\end{cases}
\end{equation}
to indicate the sublattice $\QQ$ belonging to.
As explained in Ref. \cite{Song:2019}, the operator $c_{\kk,\QQ,\eta,\alpha,s}$ is a Fourier transformation of the orbitals in the layer $l = \eta\zeta_\QQ $, with $l=\pm $ representing the top and bottom layers.
To be specific, we have
\begin{equation}
c_{\kk,\QQ,\eta,\alpha,s}^\dagger = \frac{1}{\sqrt{N_M N_0}} \sum_{\RR \in l}
e^{i (\eta\mathbf{K}_l + \kk-\QQ)\cdot (\RR+\tt_\alpha) } c_{\RR,l,\alpha,s}^\dagger ,
\end{equation}
where $N_M$ is the number of Moir\'e unit cells in the system, $N_0$ is the number of graphene unit cells in each Moir\'e unit cell,  $l=\eta\zeta_\QQ$ is the layer index, $\RR$ indexes all the graphene unit cells in the layer $l$, and $c_{\RR,l,\alpha,s}^\dagger$ is the creation operator of the orbital at sublattice $\tt_\alpha$ in the unit cell $\RR$ in the layer $l$.
Accordingly, the inverse transformation is
\begin{align}
c_{\RR,l,\alpha,s}^\dagger = &  \frac{1}{\sqrt{N_M N_0}} \sum_{\eta} \sum_{\kk \in \mathrm{MBZ}} \sum_{\QQ\in \mathcal{Q}_{\eta l }} e^{-i (\eta\mathbf{K}_l + \kk-\QQ)\cdot (\RR+\tt_\alpha) } \nono\\
& \times c_{\kk,\QQ,\eta,\alpha,s}^\dagger . \label{eq:cR-ck}
\end{align}

\subsection{Symmetries of the continuous two-valley model} \label{app:sym-continuous-model}
Here we review the symmetries of the two-valley model.
Within each valley there is spin and charge conservation, but pairing between valleys reduces the symmetry group to (global) spin-SU(2) and valley-U(1) ($e^{i\eta \theta}$), as explained in App.~\ref{app:uniform-s}.
Second, each valley of the model respects the symmetry of the magnetic space group $P6'2'2$ (\# 177.151 in the BNS setting \cite{Song:2019}, which is generated by $C_{3z}$, $C_{2x}$, and $C_{2z}T$.
Third, the two valleys are related by the time-reversal symmetry $T$.
For unitary and anti-unitary ($C_{2z}T$, $T$) symmetries, the actions of these symmetries on the Hamiltonian are
\begin{equation}
H(g\kk) = D(g) H(\kk) D^\dagger(g),\;
H(g\kk) = D(g) H^*(\kk) D^\dagger(g),
\end{equation}
respectively, with
\begin{equation} \label{eq:C3z-D}
D_{\mathbf{Q}^{\prime},\mathbf{Q}}\left(C_{3z}\right)=
    \delta_{\mathbf{Q}^{\prime},C_{3z}\mathbf{Q}}
    \cdot e^{i\frac{2\pi}{3}\tau_z\cdot \sigma_{z}},
\end{equation}
\begin{equation} \label{eq:C2x-D}
D_{\mathbf{Q}^{\prime},\mathbf{Q}}\left(C_{2x}\right)=\delta_{\mathbf{Q}^{\prime},C_{2x}\mathbf{Q}}
    \cdot \tau_0 \cdot \sigma_{x},
\end{equation}
\begin{equation}
D_{\mathbf{Q}^{\prime},\mathbf{Q}}\left(C_{2z}T\right)=\delta_{\mathbf{Q}^{\prime},\mathbf{Q}}
    \cdot \tau_0 \cdot \sigma_{x},
\end{equation}
\begin{equation}\label{eq:TRS-D}
D_{\mathbf{Q}^{\prime},\mathbf{Q}}\left(T\right)=\delta_{\mathbf{Q}^{\prime},-\mathbf{Q}}
    \cdot \tau_x \cdot \sigma_{0},
\end{equation}
Here $\tau_{x}$ ($\tau_0$) is the Pauli (identity) matrix for the valley index, and $\sigma_{x,z}$ ($\sigma_0$) are the Pauli (identity) matrices for the sublattice index.

At small twist angle ($\theta\sim 1^\circ$), the Hamiltonian has an emergent unitary particle-hole symmetry:
\begin{equation}
H(-\kk) = - D(P) H(\kk) D^\dagger(P)
\end{equation}
with
\begin{equation}
D_{\QQ',\QQ}(P) = \delta_{\QQ',-\QQ} \zeta_\QQ \cdot \tau_0 \cdot \sigma_0.
\end{equation}
Then we can define an anti-unitary particle-hole symmetry as $\mathcal{P} = C_{2z}T P$, whose action is given by
\begin{equation} \label{eq:PH-D}
D_{\QQ',\QQ} (\PH) = \delta_{\QQ',-\QQ} \zeta_\QQ \cdot \tau_0 \cdot \sigma_x \ .
\end{equation}
For the single-particle Hamiltonian, we can define the symmetry operators as $g=D(g)$ for unitary symmetries and $g=D(g)K$ for anti-unitary symmetries, where $K$ is the complex conjugation.
The commutation or anti-commutation relations between $\PH$ and the other symmetries are
\begin{align}
& [C_{3z},\PH] = 0,\quad
\{C_{2x},\PH\}=0,\quad
[C_{2z}T,\PH]=0,\nono\\
& \{T,\PH\}=0.
\end{align}

The symmetry operators can also be written in a second quantized form.
For $g=C_{3z}$, $C_{2x}$, $C_{2z}T$, and $T$, the action on electron operators is
\begin{equation} \label{eq:g-action}
g c_{\kk,\QQ,\eta,\alpha,s}^\dagger g^{-1} = \sum_{\QQ',\eta',\alpha'} c_{g\kk,\QQ',\eta',\alpha',s}^\dagger D_{\QQ',\eta',\alpha', \QQ,\eta,\alpha} (g) .
\end{equation}
Notice that $C_{2z}T$ and $T$ are still anti-unitary in the second quantized form.
On the other hand, the particle-hole symmetry $\PH$, which anti-commutes with $H$, corresponds to a {\it unitary} charge conjugation symmetry that commutes with the second quantized Hamiltonian $\hat{H}_0$ (but changes the filling).
We denote the charge conjugation as $\PH_c$.
It transforms the creation operator and the annihilation operator to each other as
{\small
\begin{equation} \label{eq:Pc-def}
\begin{aligned}
\PH_c c^{\dagger}_{\kk,\QQ,\eta,\alpha,s} \PH_c^{-1} &=  \sum_{\QQ',\eta',\alpha'} c_{-\kk,\QQ',\eta',\alpha',s} D_{\QQ',\eta',\alpha', \QQ,\eta,\alpha} (\PH) \\
\PH_c c_{\kk,\QQ,\eta,\alpha,s} \PH_c^{-1} &=  \sum_{\QQ',\eta',\alpha'} c^\dagger_{-\kk,\QQ',\eta',\alpha',s} D^*_{\QQ',\eta',\alpha', \QQ,\eta,\alpha} (\PH)
\end{aligned}
\end{equation} }
One can verify that \cref{eq:H0-second-quantized} is invariant under $\PH_c$ \cite{TBG-III}.

\subsection{The effective symmetries of the BdG Hamiltonian} \label{app:BdG-symmetry}

We first assume spin-singlet pairing between the two valleys.
Hence we introduce the BdG basis:
\begin{equation} \label{eq:BdG-basis}
\psi_{\kk,\QQ,\eta,\alpha,s} =
\begin{cases}
c_{\kk,\QQ,\eta,\alpha,\uparrow},\qquad & \text{if}\ s=\uparrow \\
c_{-\kk,-\QQ,-\eta,\alpha,\downarrow}^\dagger,\qquad & \text{if}\ s=\downarrow
\end{cases}\
\end{equation}
and write the Hamiltonian with pairing as
\begin{equation} \label{eq:HBdG-second-quantized}
\hat{H} = \sum_{ \substack{\kk,\QQ,\QQ' \\ \eta\alpha\beta ss' }} \mathcal{H}^{(\eta)}_{\QQ  \alpha s \, ;\, \QQ' \beta s'}(\kk) \psi^\dagger_{\kk,\QQ,\eta,\alpha,s} \psi_{\kk,\QQ',\eta,\beta,s'} \ .
\end{equation}
We refer to $\mathcal{H}^{(\eta)}$ as the BdG Hamiltonian.
Here the spin index also indicates the particle ($s=\uparrow$) and hole ($s=\downarrow$) degree of freedom.
We have assumed that the pairing happens only between different valleys such that the valley-U(1) is preserved and the BdG Hamiltonian matrix is diagonal in the valley index.
The full symmetry SU(2) $\times$ SU(2), with spin conservation in each valley, would force pairing between fermions of the same valley, yielding a pair-density wave at finite momentum.
We do not wish to use this form of pairing, and so work with a reduced symmetry group.
In the singlet pairing case, only total spin-SU(2) is conserved.
Now we derive the explicit form of the free part of the BdG Hamiltonian, \ie the BdG Hamiltonian without pairing:
\begin{align}
\hat{H}_0 = & \sum_{\substack{\kk\QQ\QQ'\\ \eta \alpha\beta}} \Big( H^{(\eta)}_{\QQ\alpha, \QQ'\beta}(\kk) \psi_{\kk,\QQ,\eta,\alpha,\uparrow}^\dagger \psi_{\kk,\QQ',\eta,\beta,\uparrow} \nono\\
& + H^{(-\eta)}_{-\QQ\alpha, -\QQ'\beta}(-\kk) \psi_{\kk,\QQ,\eta,\alpha,\downarrow} \psi_{\kk,\QQ',\eta,\beta,\downarrow}^\dagger \Big) .
\end{align}
By commuting $\psi_{\kk,\QQ,\eta,\alpha,\downarrow}$ and $\psi_{\kk,\QQ',\eta,\beta,\downarrow}^\dagger$ in the second term, we have
\begin{equation}
\hat{H}_0 = \sum_{\substack{\kk\QQ\QQ'\\ \eta \alpha\beta}} \calH^{(\eta,0)}_{\QQ\alpha s, \QQ'\beta s'}(\kk) \psi_{\kk,\QQ,\eta,\alpha,s}^\dagger \psi_{\kk,\QQ',\eta,\beta,s'} + const,
\end{equation}
where
\begin{equation}\small
\calH^{(\eta,0)}_{\QQ\alpha s, \QQ'\beta s'}(\kk) = \delta_{ss'} \big( \delta_{s\uparrow} H^{(\eta)}_{\QQ\alpha, \QQ'\beta}(\kk) - \delta_{s\downarrow} H^{(-\eta)T}_{-\QQ\alpha, -\QQ'\beta}(-\kk) \big).
\end{equation}
Due to the {spinless} time-reversal symmetry \cref{eq:TRS-D}, the free Hamiltonian {without pairing} satisfies $H^{(\eta)}_{\QQ',\QQ}(\kk) = H^{(-\eta)*}_{-\QQ',-\QQ}(-\kk)$, hence we can write the free part of the BdG Hamiltonian as
\begin{equation}\label{eq:freeBdG0}
\calH^{(\eta,0)}(\kk) = H^{(\eta)}(\kk) \cdot \xi_z
\end{equation}
where $\xi_z$ is the third Pauli matrix in the particle-hole space.

Then we consider generic BdG Hamiltonian with nonzero pairing term.
Besides the valley-U(1) symmetry, we further require $\hat{H}$ to be invariant under spin-SU(2) and all the discrete symmetries defined in the above subsection.
According to \cref{eq:g-action,eq:TRS-D}, the time-reversal symmetry acts on the BdG basis as
\begin{align}
\label{eq:symm-T}
& T \psi_{\kk,\QQ,\eta,\alpha,s} T^{-1} =
\begin{cases}
c_{-\kk,-\QQ,-\eta,\alpha,\uparrow},\qquad & \text{if}\ s=\uparrow \\
c_{\kk,\QQ,\eta,\alpha,\downarrow}^\dagger,\qquad & \text{if}\ s=\downarrow
\end{cases} \nono\\
=&  \psi_{-\kk,-\QQ,-\eta,\alpha,s}
\end{align}
Thus $\mathcal{H}^{(+)}$ and $\mathcal{H}^{(-)}$, the BdG Hamiltonians with pairing, also are related by the {spinless} time-reversal symmetry as
\begin{equation} \label{eq:BdG-constraint-T}
\mathcal{H}^{(\eta)}_{\QQ \alpha s\, ;\, \QQ'\beta s'}(\kk) = \mathcal{H}^{(-\eta)*}_{-\QQ,\alpha,s\, ;\, -\QQ',\beta s'} (-\kk)\ .
\end{equation}
We find that the spin-SU(2) rotation $i\hat{s}_y$ {on $\psi$} also interchanges the two valleys {in $\psi$} as
\begin{align}
& (i \hat{s}_y )\psi_{\kk,\QQ,\eta,\alpha,s} (i\hat{s}_y)^{-1} =
\begin{cases}
c_{\kk,\QQ,\eta,\alpha,\downarrow},\qquad & \text{if}\ s=\uparrow \\
- c_{-\kk,-\QQ,-\eta,\alpha,\uparrow}^\dagger,\qquad & \text{if}\ s=\downarrow
\end{cases} \nono\\
=&  \sum_{s'} [i\xi_y]_{s's} \psi_{-\kk,-\QQ,-\eta,\alpha,s'}^\dagger ,
\end{align}
where $\xi_y$ is the second Pauli matrix in the particle-hole space.
Implementing this symmetry to the BdG Hamiltonian, we have
\begin{equation} \label{eq:BdG-constraint-sy}
[\xi_y\calH^{(\eta)*}(\kk)\xi_y]_{\QQ,\alpha,s;\, \QQ',\beta,s'} = -
\calH^{(-\eta)}_{-\QQ,\alpha,s;\, -\QQ',\beta,s'}(-\kk) \ .
\end{equation}
Combining \cref{eq:BdG-constraint-T,eq:BdG-constraint-sy} we have the constraint imposed by $i\hat{s}_y T$ (spinful time-reveral symmetry)
\begin{equation}\label{eq:BdG-constraint-syT}
\{ \xi_y, \mathcal{H}^{(\eta)} \} = 0.
\end{equation}
Therefore, the $i\hat{s}_y T$ symmetry, a spinful time-reversal symmetry on the original $c$ fermions, is equivalent to a chiral symmetry $i\xi_y$ in the single-particle first quantized BdG formalism in Eq.~\ref{eq:BdG-basis}. We gauge away the $i$ in $i\xi_y$ to define the chiral symmetry $S = \xi_y$, as chiral symmetry usually is chosen to square to $+1$.  While this does not affect the physics of the system, choosing $S = \xi_y$ instead of $S = i\xi_y$ alters commutation relations with anti-unitary symmetries: for example, $[S, \PH] = [\xi_y, i\xi_z \mu_y \sigma_x K] = 0$, while $\{i\xi_y, i\xi_z \mu_y \sigma_x K\} = 0$.

We turn to charge-conjugation symmetry $\PH_c$.
Its action on the BdG basis is
\begin{align} \label{eq:Pc-on-BdG-basis}
& \PH_c \psi_{\kk,\QQ,\eta,\alpha,s} \PH_c^{-1} \nono\\
= & \begin{cases}
\sum_{\QQ'\eta'\beta} D_{\QQ'\eta'\beta,\QQ\eta\alpha}^* (\PH) c^\dagger_{-\kk,\QQ',\eta,\beta,\uparrow},\qquad & \text{if}\ s=\uparrow \\
\sum_{\QQ'\eta'\beta} D_{-\QQ'\eta'\beta,-\QQ\eta\alpha} (\PH) c_{\kk,\QQ',\eta,\beta,\downarrow},\qquad & \text{if}\ s=\downarrow
\end{cases} \nono\\
=&  \sum_{\QQ' \eta \beta s'} \calD^*_{\QQ' \eta' \beta s'\,;\, \QQ \eta \alpha s}(\PH) \psi_{-\kk,-\QQ',\eta', s'}^\dagger
\end{align}
where
\begin{equation}
\calD_{\QQ',\QQ }(\PH) = \delta_{\QQ',-\QQ} \zeta_\QQ \cdot \tau_0 \cdot \sigma_x \cdot \xi_z
\end{equation}
according to \cref{eq:PH-D}.
The term $\delta_{\QQ',-\QQ} \zeta_\QQ$ switches layer and is represented by the Pauli matrix $i\mu_y$.
The $\xi_z$ matrix in $\calD_{\QQ',\QQ }(\PH)$ comes from the fact that the actions of $\PH$ in the particle space and the hole space are opposite: $D_{\QQ',\QQ}(\PH) = - D_{-\QQ',-\QQ}(\PH)$.
The relative negative sign that emerges can easily be seen from the action of $P$:
$c_{\kk,m} \rightarrow c^\dagger_{-\kk,m'}[i\mu_y\sigma_x]_{mm'}$, with $m$ short for the indices $\eta,\alpha,s$.
Applying the dagger reverses the sign because of the $i$:  $P: c^\dagger_{\kk,m} \rightarrow c_{-\kk,m'}[-i\mu_y\sigma_x]_{mm'}$.
Similar to the analysis of the $i\hat{s}_y T$ symmetry, applying $\PH_c$ to the BdG Hamiltonian, we obtain
\begin{equation}
\label{eq:symm-PH-D}
\calD (\PH) \calH^{{(\eta)}*}(\kk) \calD^T(\PH) = -\calH^{{(\eta)}}(-\kk).
\end{equation}
Thus the single-particle operator $\PH=\calD(\PH) K$, where $K$ is the complex conjugation, is a particle-hole symmetry for the BdG Hamiltonian.
Since $\PH$ commutes with the chiral symmetry $S$ and leaves each valley invariant, we can define an emergent ``time-reversal symmetry'' $\tilde{T} = S\PH$ that leaves each valley unchanged.
$\tilde{T}$ squares to $-1$ because $\tilde{T}^2 = S\PH S\PH = S^2 \PH^2=-1$.
Therefore, we identify the effective Altland-Zirnbauer symmetry class as CII.

The actions of the crystalline symmetries preserving the valley index --- $C_{3z}$, $C_{2x}$, $C_{2z}T$ --- on the BdG basis can be similarly derived.
Since the particle part of the BdG basis are annihilation operators in valley $\eta$ and the hole part are creation operators in the valley $-\eta$, we have $\calD^{(\eta)}(g) = D^{(\eta)}(g) \oplus D^{(-\eta)*}(g)$ for these three symmetries.
On the other hand, due to the time-reversal symmetry, there must be $D^{(-\eta)*}(g)=D^{(\eta)}(g)$ (one can check this from Eqs. (\ref{eq:C3z-D}) to (\ref{eq:TRS-D})) and hence we have $\calD^{(\eta)}(g) = D^{(\eta)} (g) \cdot \xi_0$.

To summarize, the two BdG Hamiltonian matrices in the two valleys are related by time-reversal $T$, and each valley has the symmetries $S$, $\PH$, $C_{3z}$, $C_{2x}$, $C_{2z}T$.
Since $[S,\PH]=0$ and $\PH^2=-1$, the effective symmetry class of each valley is CII.
The commutation and anti-commutation relations involving $S$ and $\PH$ are
\begin{equation}\small
[S,g]=0,\quad \text{for}\ g=\PH,C_{3z}, C_{2x},\quad
\{S,C_{2z}T\} = 0,
\end{equation}
\begin{equation}
\{\PH, C_{2x}\}=0,\quad [\PH,C_{3z}]=0,\quad [\PH,C_{2z}T]=0 .
\end{equation}

\subsection{Spin-singlet pairing} \label{app:uniform-s}

In this subsection we show that a uniform spin-singlet pairing satisfies all the BdG symmetries defined in \cref{app:BdG-symmetry}.
According to the Dirac theory in the main text, as a consequence of the anomaly of TBG, such a simple spin-singlet pairing will result in the higher order topological superconductor phase.
We consider spin-singlet pairing at at every carbon atom in TBG:
\begin{equation}
\Delta_0 \sum_{\alpha,l}  \sum_{\RR\in l} c_{\RR,l,\alpha,\uparrow}^\dagger c_{\RR,l,\alpha,\downarrow}^\dagger + h.c.
\end{equation}
Due to the time-reversal symmetry, $\Delta_0$ must be a real number.
Applying Eq. (\ref{eq:cR-ck}), we can write the pairing in terms of $c_{\kk,\QQ,\eta,\alpha,s}$ as
\begin{widetext}
\begin{align} \label{eq:pairing-fourier-transformation-s}
 & \Delta_0 \frac{1}{N_M N_0} \sum_{\alpha,l} \sum_{\RR \in l} \sum_{\eta \eta'} \sum_{\kk,\kk'\in{\rm MBZ}} \sum_{\QQ\in \mathcal{Q}_{\eta l}} \sum_{\QQ'\in \mathcal{Q}_{\eta' l}}
    e^{-i(\eta \mathbf{K}_l + \kk - \QQ + \eta' \mathbf{K}_l + \kk' - \QQ' )\cdot (\RR + \tt_\alpha)}
    c_{\kk,\QQ,\eta,\alpha,\uparrow}^\dagger c_{\kk',\QQ',\eta',\alpha,\downarrow}^\dagger + h.c. \nono\\
= & \Delta_0 \sum_{\alpha,l} \sum_{\eta} \sum_{\kk\in{\rm MBZ}} \sum_{\QQ\in \mathcal{Q}_{\eta l}}
    c_{\kk,\QQ,\eta,\alpha,\uparrow}^\dagger c_{-\kk,-\QQ,-\eta,\alpha,\downarrow}^\dagger + h.c.
= \Delta_0 \sum_{\eta,\alpha} \sum_{\kk\in{\rm MBZ}} \sum_{\QQ}
c_{\kk,\QQ,\eta,\alpha,\uparrow}^\dagger c_{-\kk,-\QQ,-\eta,\alpha,\downarrow}^\dagger  + h.c.
\end{align}
\end{widetext}
At the first equal sign, the summation over $\RR$ leads to the momentum conservation $\eta=-\eta'$, $\kk=-\kk'$, $\QQ=-\QQ'$.
First $\eta = -\eta'$, as no combination of the Moir\'e vectors will span the graphene vector $\mathbf{K}_l$.
The summation over $\RR$ then forces $\kk - \QQ + \kk' - \QQ' = 0$.
Because $\eta = -\eta'$, $\QQ$ and $\QQ'$ belong to opposite sublattices ${\cal Q}_+, {\cal Q}_-$, meaning $\QQ + \QQ' = n\bb_{M1} + m\bb_{M2}, ~n,m \in Z$, which is commensurate with the Moir\'e BZ lattice vectors.
Thus $\kk + \kk' = 0$ and $\QQ + \QQ' = 0$.
As this pairing operator adds a spin $\uparrow$ in one valley and spin $\downarrow$ in the other, it breaks spin and charge conservation in each valley.  However, \emph{total} charge and spin is still conserved.
In the BdG basis \cref{eq:BdG-basis}, this pairing term can be written as
\begin{equation}
\hat{H}_1 = \sum_{\kk\QQ\QQ'} \sum_{\eta \alpha\beta ss'}  \calH^{(\eta,1)}_{\QQ \alpha s, \QQ' \beta s'}(\kk) \psi^\dagger_{\kk,\QQ,\alpha,s} \psi_{\kk,\QQ',\beta,s'}
\end{equation}
where
\begin{equation}\label{eq:pairBdG1}
\calH^{(\eta,1)}_{\QQ, \QQ' }(\kk) = \Delta_0 \delta_{\QQ,\QQ'} \cdot  \sigma_0 \cdot \xi_x .
\end{equation}
One can verify that this pairing term satisfies the $T$, $S$, $\PH$, $C_{3z}$, $C_{2x}$, and $C_{2z}T$ symmetries.
The Bogoliubov bands with $\Delta_0=20$meV at $\theta=1.05^\circ, 1.2^\circ, 1.4^\circ$ are shown in \cref{fig:MB_WL}(c-e), respectively.

In realistic systems, the Cooper pairing might not be exactly on-site.
We consider the pairing in form of
\begin{align}
&\sum_{\alpha,\beta,l} \sum_{\RR\in l} \sum_{\substack{\RR'\in l\\ |\RR+\tt_\alpha-\RR'-\tt_\beta|\le a_0}} \Delta(|\RR+\tt_\alpha-\RR'-\tt_\beta|)  \nono\\
& \times c_{\RR,l,\alpha,\uparrow}^\dagger c_{\RR',l,\beta,\downarrow}^\dagger + h.c.,
\end{align}
where $a_0$ is the lattice constant of single-layer graphene.
In general $\Delta(|\mathbf{d}|)$ decays with $|\mathbf{d}|$, hence here we only keep the pairing to next nearest neighbor.

We denote the on-site pairing, the nearest neighbor pairing, and the next nearest neighbor pairing as $\bar{\Delta}_0=\Delta(0)$, $\bar{\Delta}_1=\Delta(\frac1{\sqrt3}a_0)$, and $\bar{\Delta}_2=\Delta(a_0)$, respectively.
For a simple nearly uniform spin-singlet pairing, $\bar{\Delta}_{0,1,2}$ have the same sign.
Following the same calculation as in \cref{eq:pairing-fourier-transformation-s}, we obtain the pairing term in momentum space as
{\small
\begin{align}
& \sum_{\eta,\alpha}  \sum_{\kk\in{\rm MBZ}} \sum_{\QQ} \Big(
(\Delta_0 + \widetilde{\Delta}^{(\eta)}_2(\kk-\QQ))
c_{\kk,\QQ,\eta,\alpha,\uparrow}^\dagger c_{-\kk,-\QQ,-\eta,\alpha,\downarrow}^\dagger \nono\\
& + \widetilde{\Delta}_1^{(\eta,\alpha)}(\kk-\QQ) c_{\kk,\QQ,\eta,\alpha,\uparrow}^\dagger c_{-\kk,-\QQ,-\eta,-\alpha,\downarrow}^\dagger \Big) + h.c.,
\end{align}}
where $-\alpha=B,A$ for $\alpha=A,B$ and
\begin{equation}
\widetilde{\Delta}^{(\eta)}_2(\kk-\QQ) = \bar{\Delta}_2
\sum_{\mathbf{d}}'' e^{i(\eta \mathbf{K}+\kk-\QQ)\cdot\mathbf{d}}
\end{equation}
\begin{equation}
\widetilde{\Delta}^{(\eta,\alpha)}_1(\kk-\QQ) = \bar{\Delta}_1
\sum_{\mathbf{d}_\alpha}' e^{i(\eta \mathbf{K}+\kk-\QQ)\cdot\mathbf{d}_\alpha}
\end{equation}
with $\sum_{\mathbf{d}_\alpha}'$ and $\sum_{\mathbf{d}}''$ summing over the three nearest neighbor vectors and the six next nearest neighbor vectors around $\tt_\alpha$, respectively.
Since $|\kk-\QQ|\ll 1/a_0$ in the low energy theory, the $\kk$ dependence of pairing can be neglected in the analysis of the topology of the superconductor phase.
However, in \cref{app:shiba}, we find a weak $\kk$-dependence is important for the appearance of Shiba states.
Thus for later reference, we expand $\widetilde{\Delta}^{(\eta)}_2(\kk-\QQ)$ and $\widetilde{\Delta}^{(\eta,\alpha)}_1(\kk-\QQ)$ to second order $a_0(\kk-\QQ)$:
\begin{equation}
\widetilde{\Delta}^{(\eta)}_2(\mathbf{p}) \approx - 3\bar{\Delta}_2 + \frac34 \bar{\Delta}_2 a_0^2 \mathbf{p}^2,
\end{equation}
\begin{align}
&\widetilde{\Delta}^{(\eta,A)}_1(\mathbf{p}) \approx \frac{\sqrt3}2 \bar{\Delta}_1 a_0 (-\eta p_x +i  p_y ) \nono\\
&\quad + \frac18 \bar{\Delta}_1 a_0^2 (p_x^2-p_y^2+2\eta i p_x p_y)
\end{align}
\begin{equation}
\widetilde{\Delta}^{(\eta,B)}_1(\mathbf{p})=\widetilde{\Delta}^{(\eta,A)*}_1(\mathbf{p}),
\end{equation}
where $\mathbf{p}=\kk-\QQ$.
Defining $\Delta_0=\bar{\Delta}_0-3\bar{\Delta}_2$, $m=-\frac34 \bar{\Delta}_2 a_0^2$, $v'=-\frac{\sqrt3}2\bar{\Delta}_1 a_0 $, $m'=\frac18\bar{\Delta}_1 a_0^2$, we can write the BdG Hamiltonian as
\begin{align} \label{eq:k-dependent-pairing}
& \calH^{(\eta,1)}_{\QQ, \QQ' }(\kk) = \delta_{\QQ,\QQ'} \xi_x \Big( (\Delta_0-m(\kk-\QQ)^2)  \sigma_0  \nono\\
& + v' \eta (k_x-Q_x) \sigma_x + v' (k_y-Q_y) \sigma_y \nono\\
& + m'((k_x-Q_x)^2-(k_y-Q_y)^2) \sigma_x + \nono\\
& - 2 m' \eta (k_x-Q_x)(k_y-Q_y) \sigma_y \Big) .
\end{align}
We will assume $m=v'=m'=0$ in the rest of this work unless  specified otherwise.
As mentioned above, since $|\kk-\QQ|\ll 1/a_0$, the quadratic term $(\kk-\QQ)^2$ in the mass term $\Delta_0-m (\kk-\QQ)^2$ is always much smaller than $\Delta_0$ within the cutoff of the single valley model.
Thus this quadratic term, as well as other $\kk$-dependent terms in the pairing, can be safely neglected in the analysis of the band topology.

\subsection{Wilson loop of the topological superconductor}
\label{app:WilsonLoop}

The topological nature of TBG-TSC can be easily verified via the Wilson loop construction.
As discussed in Ref.~\cite{Song:2020b}, the two symmetries $C_{2z}T$ and ${\cal P}$ protect a stable topology in the Wilson loop spectrum.
When considering the Wannier Hamiltonian
\begin{align}
W(k_x) &= e^{iH_W(k_x)} \\
&= \lim_{N \rightarrow \infty} \prod_{j=0}^{N-1} U^\dagger (k_x, \frac{2\pi j}{N}) U (k_x, \frac{2\pi (j+1)}{N}),
\end{align}
with $U$ defined to be the matrix of coefficients of the $M$ occupied BdG bands:
\begin{align}
    U(\kk) = (\ket{u_1(\kk)}, \ket{u_2(\kk)}, \cdots \ket{u_M(\kk)}),
\end{align}
defining $\ket{u_m(k_x,k_y)}$ as the eigenket for the $m$th band.
The product of unitaries becomes
\begin{align}
    [U^\dagger (\kk') U (\kk)]_{mn} = \braket{u_{m}(\kk')|u_{n}(\kk)}.
\end{align}  One finds that $S\cal P$ guarantees degeneracies in the Wilson loop spectrum at $k_x = 0, \pi$, while $C_{2z}T$ becomes a particle-hole symmetry and forces the bands to come in $E, -E$ pairs.
To see this, we examine the action of $S\PH$ on the Wilson loop.  As in Ref.~\onlinecite{Song:2020b}, under $S\PH$, defining the sewing matrix $B$ to act as
\begin{align}
    S\PH \ket{u_n(\kk)} &= \sum_m \ket{u_m(-\kk)} B^{S\PH}_{mn} (\kk), \\
    (S\PH)^2 &= -1 \implies B^{S\PH *} (\kk) B^{S\PH} (-\kk) = -1.
\end{align} we find
\begin{align}
    & \braket{u_m(\kk)|u_n(\kk')} = \braket{S\PH u_n(\kk')|S\PH u_m(\kk)} \\
    &= \sum_{n'm'} B^{S\PH *}_{n'n} (\kk') \braket{u_{n'}(-\kk') | u_{m'}(-\kk)} B^{S\PH}_{m'm} (\kk)
\end{align}  This implies that
\begin{align}
    W(k_x) &= B^{S\PH \dagger}(k_x,0) W^T(-k_x) B^{S\PH}(k_x,0).
\end{align}

Defining a time reversal symmetry $T'(k_x) = B^{S\PH \dagger}(k_x,0) K$, when $k_x = 0,\pi$:
\begin{align}
   & T'(k_x) H_W(-k_x) T'^{-1}(k_x) =  H_W(k_x), \\
& T'^2 = B^{S\PH}(k_x,0) B^{S\PH*}(k_x,0) = -1.
\end{align} Thus a Kramers' degeneracy is preserved at $k_x = 0, \pi$.

\begin{figure}
\includegraphics[width=\columnwidth]{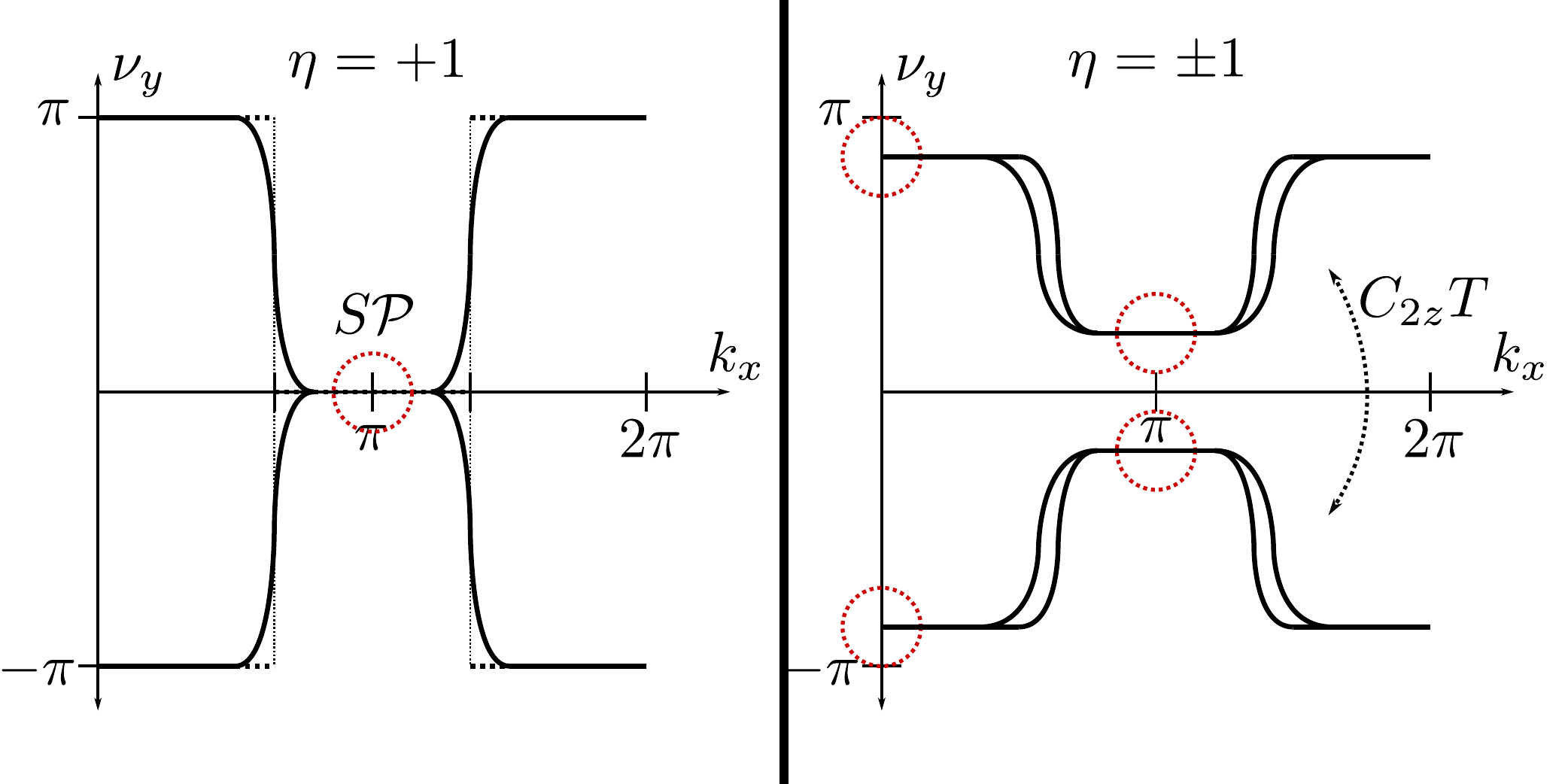}
\caption{\emph{Left panel:} Wannier band spectrum of TBG-TSC in a single valley $\tau = +1$.  Before pairing, the energy spectrum has gapless points when the momentum crosses the Dirac points.  Because there is one Dirac cone from $k_x \in [0,\pi)$, the Wannier bands' value will jump by $\pi$.  Adding pairing will smooth out the jump but cannot erase the nontrivial flow from left to right due to Kramers degeneracies at $k_x = 0, \pi$.
\emph{Right panel:} Both valleys are allowed to hybridize with each other, with strong valley-U(1) breaking.  This allows the nontrivial bands to pair up and avoid while still remaining Kramers' partners at the TRI momenta.  As twisted bilayer graphene is only anomalous within a single valley, this behavior is expected.
}
\label{WilsonFig}
\end{figure}

\emph{Without} pairing, the singularity (the Dirac crossing) forces a jump in the Wannier spectrum by $\pi$ as $k_x$ crosses the Dirac point location, as the Berry phase enclosed in a loop around the Dirac point is $\pi$.
The two Dirac points at $K_M, K_M'$ (we have restricted ourselves to the $\eta=+1$ graphene valley) thus yield two jumps of $\pi$.
After pairing is added, the sharp discontinuities are smoothed out (as the spectrum is now gapped), but the two Wannier bands still criss-cross, yielding a nontrivial Dirac flow for two bands.
Regardless of how many pairs of trivial bands one adds to the spectrum, the Wannier flow is preserved and hence the superconductor remains topological.
If one breaks valley-U(1) symmetry, the superconductor will turn trivial.
See Fig.~\ref{WilsonFig} for an illustration.

We verify numerically that the spin-singlet paired continuous TBG model is topological, using the Wilson loop.
Fixing $w_1=110$meV, $w_0=0.8w_1$, $v_F=5.944{\rm eV\cdot \mathring{A}}$, and $|\Delta_0|=20$meV, we calculated the Moir\'e bands and the Wilson loops of the -1 and -2 bands at three different twist angles, $\theta=1.05^\circ,1.20^\circ,1.40^\circ$, respectively, as is shown in Fig.\ref{fig:MB_WL}(c)-(h).
As expected, the spin-singlet superconducting phase at these angles are all topological.

\section{Dirac theory of the topological superconductor}
\label{app:Dirac}

We use a simplified Dirac theory to demonstrate the zero modes on the phase boundary and in the vortex of the superconductor order parameters.
With no pairing, BdG Hamiltonian in each valley has two Dirac points at $K_M$ and $K_M'$.
Hence we can write the low energy model of the BdG Hamiltonian as
\begin{equation} \label{eq:BdGDiracHam}
\calH = k_x \xi_z \tau_z \mu_0 \sigma_x + k_y \xi_z \tau_0 \mu_0 \sigma_y + \Delta \xi_x \tau_0 \mu_0 \sigma_0.
\end{equation}
Here $\xi_{0,x,y,z}$, $\tau_{0,x,y,z}$, $\mu_{0,x,y,z}$, $\sigma_{0,x,y,z}$ are the Pauli matrices (identity) in the space of particle-hole, valley, Moir\'e valley, and sublattice, respectively.
$\mu_z=1$ and $-1$ correspond to the Dirac points at $K_M$ and $K_M'$, respectively.
Similar to \cref{eq:BdG-basis}, the basis of the BdG Hamiltonian is defined as
\begin{equation} \label{eq:BdG-basis-dirac}
\psi_{\kk,\eta,l,\alpha,s} =
\begin{cases}
c_{\kk,\eta,l,\alpha,\uparrow},\qquad & \text{if}\; s=\uparrow \\
c_{-\kk,-\eta,-l,\alpha,\downarrow}^\dagger,\qquad & \text{if}\; s=\downarrow
\end{cases}\ ,
\end{equation}
where $\eta$, $l$, $\alpha$, $s$ stands for the {graphene valley,} Moir\'e valley, sublattice, and spin (particle-hole) indices, respectively.
The symmetry actions on the BdG basis are same as in \cref{app:sym-continuous-model,app:BdG-symmetry}.
(One can think that the $\mu_z=\pm1$ Moir\'e valleys are contributed by states on the lattices $\mathcal{Q}_{\pm}$, respectively. Then all the symmetry operators defined in \cref{app:sym-continuous-model,app:BdG-symmetry} in Eqs.~\ref{eq:C3z-D}-\ref{eq:PH-D}, \ref{eq:symm-T}-\ref{eq:symm-PH-D} also apply to the Dirac theory except that one needs to replace the matrices involving $\QQ$ vectors in \cref{eq:C3z-D}  to (\ref{eq:PH-D}) as $\delta_{\QQ',C_{3z}\QQ} \to \mu_0$,
$\delta_{\QQ',C_{2x}\QQ} \to \mu_x$,
$\delta_{\QQ',\QQ} \to \mu_0$,
$\delta_{\QQ',-\QQ} \to \mu_x$,
$\delta_{\QQ',-\QQ}\zeta_{\QQ} \to i\mu_y$.)
The effective symmetry operators for the BdG Hamiltonian can be similarly obtained as
\begin{equation} \label{eq:sym-BdG-dirac}
\begin{aligned}
& S=\xi_y,\quad \PH = i\xi_z \mu_y \sigma_x K,\quad T= \tau_x \mu_x K\\
& C_{2z}T=\sigma_x K,\quad C_{2x} = \mu_x \sigma_x,\quad C_{3z} = e^{i\frac{2\pi}3\tau_z \sigma_z}\ .
\end{aligned}
\end{equation}

\subsection{The pairing domain wall edge state in the \texorpdfstring{$x$}{x}-direction}
\label{app:edge-state-Dirac}

The BdG Hamiltonians in the two valleys are related by the time-reversal symmetry $T$ (or the $s_y$ spin rotation $i \hat{s}_y$).
The states in the valley $\eta=-1$ are simply the time-reversal partners of the states in the valley $\eta=+1$.
Thus in the following we will mainly focus on the valley $\eta=+1$.
Notice that for a single graphene and Moir\'e valley Eq.~(\ref{eq:BdGDiracHam}) resembles the Hamiltonian of a quantum spin Hall edge, with time-reversal $T' = i\sigma_y K$.
As the edge of a quantum spin Hall system carries a helical edge mode, TBG-TSC carries four, two for graphene valley and two for Moir\'e valley.
To derive the edge Hamiltonian, we model the domain wall with a jump in the pairing term $\Delta(x) = -\Delta_0 \text{sgn}(x)$ (take this to be the edge perpendicular to the $x$ axis, so the bulk is to the left of the edge), and take the ansatz
\begin{align}
\phi_{k_y}(x) \propto e^{ik_y y -\lambda |x|} u(k_y),\qquad (\lambda>0),
\end{align}
for some eight component vector $u$.
Since the translation symmetry along $y$ is respected, $k_y$ is still a good quantum number.
We first solve the zero modes at $k_y = 0$ and then apply the perturbation theory for finite $k_y$ to obtain the effective Hamiltonian on the edge.
For $k_y=0$, with Eq.~\ref{eq:BdGDiracHam} with $\eta = +1$, taking $k_x \rightarrow -i\partial_x$, we obtain the eigenequation
\begin{align}
(\lambda-\Delta_0 \xi_y \mu_0 \sigma_x) u = 0.
\end{align}
There are four degenerate bound states when $\lambda = \Delta_0$, and the wavefunctions are
{\small
\begin{align}
\label{eq:edge-basis-x}
u_1 &=
\frac12
\begin{bmatrix}
1 \\ i
\end{bmatrix}\otimes
\begin{bmatrix}
1 \\ 0
\end{bmatrix}\otimes
\begin{bmatrix}
1 \\ 1
\end{bmatrix} ,\;
u_2= \frac12
\begin{bmatrix}
1 \\ i
\end{bmatrix}\otimes
\begin{bmatrix}
0 \\ 1
\end{bmatrix}\otimes
\begin{bmatrix}
1 \\ 1
\end{bmatrix} , \nono\\
u_3 &= \frac12
\begin{bmatrix}
-1 \\ i
\end{bmatrix}\otimes
\begin{bmatrix}
1 \\ 0
\end{bmatrix}\otimes
\begin{bmatrix}
1 \\ -1
\end{bmatrix}, \;
u_4 = \frac12
\begin{bmatrix}
-1 \\ i
\end{bmatrix}\otimes
\begin{bmatrix}
0 \\ 1
\end{bmatrix}\otimes
\begin{bmatrix}
1 \\ -1
\end{bmatrix} .
\end{align} }
The effective Hamiltonian is obtained by projecting $k_y \xi_z \mu_0 \sigma_y$ onto the zero mode basis:
\begin{equation} \label{eq:edge-Dirac-x}
\calH_{\rm eff} = k_y \xi_y' \mu_0 \ .
\end{equation}
Here $\xi_{0,x,y,z}'$ and  $\mu_{0,x,y,z}$ are Pauli (identity) matrices defined in the projected four-dimensional space.
While $\mu$ is the same as the original Pauli matrix used in the Dirac Hamiltonian, $\xi_z'$ is a combination of $\xi, \sigma$ operators.
The projected symmetry operators for $S$, $\PH$ and $C_{2x}$ are
\begin{equation} \label{eq:edge-sym-x}
S=\xi_z',\quad \PH = i\xi_z' {\mu_y} K,\quad C_{2x}=\xi_z' {\mu_x}\ .
\end{equation}

The gapless edge state is protected by $S$, $\PH$ and $C_{2x}$.
There are eight terms that anti-commute with $\calH_{\rm eff}$ and hence would open full gaps: $\xi_{x}'\mu_{0,x,y,z}$, $\xi_{z}'\mu_{0,x,y,z}$.
Only the first four terms are allowed by $S=\xi_z'$.
Among these four terms, only $\xi_x' \mu_0$ is allowed by $\PH=i\xi_z'\mu_y K$.
However, this term breaks the $C_{2x}=\xi_z'\mu_x$ symmetry.
Therefore, no gap term is allowed by the symmetries.
Terms like $\xi_y' \mu_y$ that commute with the Hamiltonian will not open up gaps, but shift the bands up and down relative to one another and move the crossing locations.

We can diagonalize the edge Hamiltonian.
Here we choose the wavefunctions of the two right movers ($\xi_y'=1$) as $\frac{1+i}{2} (u_1+iu_3)$, $\frac{1+i}{2} (u_2+iu_4)$, and the wavefunctions of the two left movers ($\xi_y'=-1$) as $\frac{1+i}{2} (u_2-iu_4)$, $\frac{1+i}{2} (u_1-iu_3)$.
The annihilation operators of the two right movers are given by
\begin{widetext}
\begin{equation}
\psi_{1R\uparrow} (y) = \frac{\sqrt{\lambda}}{2} \int dx e^{-\lambda |x|}
    ( c_{+,+,A,\uparrow}(x,y) -i c_{+,+,B,\uparrow}(x,y) -
      c^\dagger_{-,-,A,\downarrow}(x,y) - i c^\dagger_{-,-,B,\downarrow}(x,y) )
      \label{eq:fermionMoversFirst}
\end{equation}
\begin{equation}
\psi_{2R\uparrow} (y) = \frac{\sqrt{\lambda}}{2} \int dx e^{-\lambda |x|}
    ( c_{+,-,A,\uparrow}(x,y) -i c_{+,-,B,\uparrow}(x,y) -
      c^\dagger_{-,+,A,\downarrow}(x,y) - i c^\dagger_{-,+,B,\downarrow}(x,y) )
\end{equation}
\begin{equation}
\psi_{1L\uparrow} (y) = \frac{\sqrt{\lambda}}{2} \int dx e^{-\lambda |x|}
    ( -i c_{+,-,A,\uparrow}(x,y) + c_{+,-,B,\uparrow}(x,y)
     -i c^\dagger_{-,+,A,\downarrow}(x,y) - c^\dagger_{-,+,B,\downarrow}(x,y) )
\end{equation}
\begin{equation}
\psi_{2L\uparrow} (y) = \frac{\sqrt{\lambda}}{2} \int dx e^{-\lambda |x|}
    ( -i c_{+,+,A,\uparrow}(x,y) + c_{+,+,B,\uparrow}(x,y)
      -i c^\dagger_{-,-,A,\downarrow}(x,y) -  c^\dagger_{-,-,B,\downarrow}(x,y) )
\label{eq:fermionMovers}
\end{equation}
\end{widetext}
Here the subscripts $\eta,l,\alpha,s$ in the operator $c_{\eta,l,\alpha,s}$ represent the valley, Moir\'e valley, sublattice, and spin, respectively.
The spin of the operator $\psi$ is chosen to be in the same direction as the annihilation operators $c$ and opposite to the creation operators $c^\dagger$ that compose $\psi$.
The four edge states in the other valley sector can be obtained by acting spin rotation (or time-reversal) on the above four states:
\begin{equation}
    \psi_{\alpha, n, \downarrow} = -i\hat{s}_y \psi_{\alpha, n, \uparrow} (-i\hat{s}_y)^\dagger, \quad (\alpha=1,2,\ n=R,L)
\end{equation}
with $-i\hat{s}_y$ being the $\pi$-rotation of spin along the $y$-direction.
Therefore, we have in total four right movers and four left movers on the edge perpendicular to $x$-direction.
Each of the eight modes is a complex fermion, not a Majorana as we are working in the non-redundant basis.

\subsection{Edge states in domain walls in other directions}

We consider a $\Delta$ domain wall rotated (anticlockwise) by an angle $\theta$ from the edge perpendicular to the $x$ direction.
We define $k_\perp = k_x \cos\theta + k_y \sin\theta$, $k_\parallel = -k_x \sin\theta +  k_y \cos\theta$, $x_\perp = x \cos\theta + y \sin\theta$, $r_\parallel = -x \sin\theta +  y \cos\theta$.
Then the {2D} Dirac Hamiltonian can be rewritten as
\begin{equation}
\calH^{(+)} = k_\perp \xi_z \mu_0 \tilde{\sigma}_x^{(\theta)} +
k_\parallel \xi_z \mu_0 \tilde{\sigma}_y^{(\theta)} + \Delta ( r_\perp ) \xi_x \mu_0 \sigma_0,
\end{equation}
where $\Delta (r_\perp) = -\Delta_0 \mathrm{sgn}(r_\perp)$, $\tilde{\sigma}_{x,y}^{(\theta)} = e^{-i\frac{\theta}2\sigma_z} \sigma_{x,y} e^{i\frac{\theta}2\sigma_z}$.
The basis of the low-energy states along this rotated edge is $u_n^{(\theta)} = e^{-i\frac{\theta}2\sigma_z} u_{n}$ ($n=1,2,3,4$) with $u_n$ defined in \cref{eq:edge-basis-x} such that the effective edge Hamiltonian and the $S$, $\PH$ symmetries take the same form as \cref{eq:edge-Dirac-x,eq:edge-sym-x}.
The $S$ and $\PH$ symmetric gap term $\xi_x'\mu_0$ is forbidden by $C_{2x}$ for $\theta=0$, as under $C_{2x} = \xi_z'\mu_x, M_1(y) \xi_x'\mu_0 \rightarrow -M_1(-y) \xi_x'\mu_0$, forcing $M_1(y) = -M_1(y)$.  With translation symmetry this forces $M_1 = 0$.
This term is also forbidden for $\theta=n \frac{2\pi}6$ ($n=1\cdots 5$) due to the $C_{3z}$ symmetry.
However, for generic $\theta$, no crystalline symmetry can forbid this gap term.
For $\theta=\frac{\pi}2$ ($n\frac{2\pi}6+\frac{\pi}6$), the edge has the $C_{2y}T=C_{2x}\cdot C_{2z}T$ symmetry ($C_{3z}^n C_{2y}T C_{3z}^{-n}$).
{The symmetries obey}
\begin{align}
    [S, \PH] = 0, ~~\{\PH, C_{2y}T\} = 0,~~\{S,C_{2y}T\} = 0.
\end{align}
However, the $C_{2y}T$ symmetry cannot protect the edge states.
To be specific, for $\theta=\frac{\pi}2$, we find the projected $C_{2y}T$ operator as $\xi_x' \mu_x K$.
(This is derived from the Pauli matrix form of $C_{2y}T = C_{2x} \cdot C_{2z}T = \mu_x K$, which projecting into the low-energy basis gives $\xi_z'\mu_x K$.)
The gap term $\xi_x'$ is allowed by the $S$, $\PH$, $C_{2y}T$ symmetries.
Therefore, the edge state for generic $\theta \neq n \frac{2\pi}{6}$  is unstable.

As shown in \cref{app:num_TBG}, for the particular Hamiltonian of TBG and  pairings that obey $\Delta (r_\perp) = -\Delta(-r_\perp)$, the system has an accidental inversion-like unitary particle-hole symmetry if the domain wall passes through a $C_{2z}$ center of the Moir\'e unit cell.
This accidental symmetry, when present, will protect the gapless edge state in any direction.
As detailed in \cref{app:num_TBG}, additional terms are added in the numerical calculation to remove this accidental symmetry.

\subsection{Corner states and the higher-order topology}
\label{app:corner-dirac}

To conveniently see the corner states, let us break the translation symmetry along $y$ but preserve $C_{2x}$ symmetry on the edge perpendicular to $x$.
In practice, the translation breaking can be realized by a corner geometry (\cref{fig:hex}) or a circular geometry (\cref{HexagonFig}b and \cref{fig:circ}).
As discussed in \cref{app:edge-state-Dirac}, for the edge Hamiltonian \cref{eq:edge-Dirac-x}, the only $S$ and $\PH$ allowed gap term $M_1(y) \xi_x' \mu_0$ is forbidden by $C_{2x}$ if the translation symmetry is assumed, which enforces $M_1(y)$ to be a $y$-independent constant.
Since now the translation symmetry is broken, a $y$-dependent $M_1(y)$ is allowed and $M_1(y)$ must be odd in $y$ due to $\{C_{2x},\xi_x'\mu_0\}=0$.
The edge Hamiltonian now reads
\begin{align}
    {\cal H} = (-i\partial_y) \xi_y' \mu_0 + M_1(y) \xi_x' \mu_0.
\end{align}
Now we show that the oddness of $M_1(y)$ gives rise two zero modes per valley, or four total.
This is expected from general arguments: a helical one-dimensional Dirac mode with a domain wall will bind a fermionic zero mode;  four copies will bind four.
Without loss of generality, we assume $M_1(\infty)=-M_1(-\infty)>0$, taking ansatz
\begin{align}
\phi(y) \propto e^{-\lambda \int_0^{y} dy' M_1(y')} v ,
\end{align}
with $\lambda>0$ and $v$ being a four-component vector, we have the eigenequation
\begin{align}
(\lambda - \xi_z' \mu_0 )v=0.
\end{align}
There are two solutions for $\lambda = 1$, and the corresponding wavefunctions are
\begin{equation}
v_1 = (1,0,0,0)^T,\qquad  v_2 = (0,1,0,0)^T \ . \label{eq:corner-basis-x}
\end{equation}
The projected symmetry operators in this 2D space are
\begin{equation}
S=1,\qquad \PH = i\mu_y'' K,\qquad C_{2x} = \mu_x''\ ,
\end{equation}
with $\mu_{x,y}''$ being the Pauli matrices in the 2D space.
One can see that the two zero modes have the same chiral eigenvalue.
Due to the fact $\PH$ is local in real space, anti-unitary, and squares to $-1$, Kramers theorem guarantees that the two zero modes must locate at the same position.
We call the two zero modes a zero mode doublet.
Thus the two zero modes per valley are stable against any local perturbation that respects the chiral symmetry.
According to the wavefunctions \cref{eq:corner-basis-x,eq:edge-basis-x}, we can write the annihilation operators of the two zero modes as
\begin{widetext}
\begin{equation}
\label{eq:zero_mode_second_quantized}
\Phi_{1\uparrow} = \frac{1}{\sqrt{\mathcal{N}}} \int dxdy\ e^{-\Delta_0|x| - \int^y_0 dy' M_1(y') } ( c_{+,+,A,\uparrow}(x,y) + c_{+,+,B,\uparrow}(x,y)
    - i c_{-,-,A,\downarrow}^\dagger(x,y) -i c_{-,-,B,\downarrow}^\dagger(x,y) ),
\end{equation}
\begin{equation}
\Phi_{2\uparrow} = \frac{1}{\sqrt{\mathcal{N}}} \int dxdy\ e^{-\Delta_0|x| - \int^y_0 dy' M_1(y') } ( c_{+,-,A,\uparrow}(x,y) + c_{+,-,B,\uparrow}(x,y)
    - i c_{-,+,A,\downarrow}^\dagger(x,y) -i c_{-,+,B,\downarrow}^\dagger(x,y) ),
\end{equation}
\end{widetext}
where $\eta,l,\alpha,s$ in the operator $c_{\eta,l,\alpha,s}$ represent the valley, Moir\'e valley, sublattice, and spin, respectively, and $\mathcal{N}$ is a normalized constant.
The two corner states in the other valley sector can be obtained by acting spin rotation (or time-reversal) on the above two states:
\begin{equation}
    \Phi_{\alpha, \downarrow} = -i\hat{s}_y \Phi_{\alpha,\uparrow} (-i\hat{s}_y)^\dagger, \quad (\alpha=1,2),
\end{equation}
with $-i\hat{s}_y$ being the $\pi$-rotation of spin along the $y$-direction.
Therefore, we have in total four complex fermion zero modes (or eight Majorana zero modes) at this corner.

Now we consider a circular geometry of the phase domain wall shown in \cref{HexagonFig}b and \cref{fig:circ}.
The above calculation applies to the corner  at $y = 0$.
Due to the $C_{3z}$ and $C_{2z}T$ symmetry, the system must have zero mode doublets at other five positions, which are rotated (anticlockwisely) from the corner in the $x$-direction by the angles $\theta=n \frac{2\pi}6$ ($n=1\cdots 5$).
Since $[C_{3z},S]=0$ and $\{C_{2z}T,S\}=0$, the zero mode doublets at $\theta=0,\frac{2\pi}{3}, -\frac{2\pi}3$ have the chiral eigenvalue $+1$; whereas zero mode doublets at $\theta=\frac{\pi}3, \pi, -\frac{\pi}3$ have the chiral eigenvalue $-1$.

We emphasize that the higher-order topology is protected by $C_{2z}T$, $\PH$, $S$.
The presence of $C_{2x}$ and $C_{3z}$ only pins the corner states at the specific positions ($\theta=n \frac{2\pi}6$, $n=0\cdots 5$).
We can consider breaking $C_{3z}$ to create (annihilate) a pair of zero mode doublets (in each valley) with opposite chiral eigenvalues at the angle $\theta$.
Then due to the $C_{2z}T$ symmetry, we must create (annihilate) another pair of zero mode doublets (in each valley) at the angle $\theta+\pi$.
Hence we can change the number of zero mode doublets by 4.
Starting with 6 zero mode doublets, such $C_{2z}T$-symmetric process may change the number of zero mode doublets to $4n+2$ ($n\in \mathbb{N}$).
Therefore, one cannot remove all the zero modes in our system without breaking $C_{2z}T$.
This is another proof that our system is always higher-order topological.

\subsection{Robustness of the corner states against \texorpdfstring{$\PH$-breaking}{PH-breaking}}
\label{app:corner-split}

Recall that the $\PH$ symmetry is an approximate, although very good, symmetry of the TBG Hamiltonian.
In practice, it is weakly broken by the small $\theta$ dependence of the two Dirac Hamiltonians from the two graphene layers \cite{Song:2020b}.
In the BdG formalism, the $\PH$ symmetry can also be broken by finite chemical potential.
In this section we study the robustness of the zero modes when $\PH$ is perturbatively broken.
As shown in \cref{app:corner-dirac}, the two zero modes (in each valley) at a corner have the same chiral eigenvalue.
Hence they must be stable against any perturbation that respects the chiral symmetry.
To see this, assume the states $\psi_{1,2}$ are at nonzero energy $H\psi_1 = E \psi_1, H\psi_2 = -E \psi_2$.
Since $\{S, H\} = 0$, we have $S\psi_1 = \psi_2$ and $S\psi_2 = \psi_1$ up to a phase factor that can be gauged away.
This means that the combinations $\psi_1 \pm \psi_2$ are eigenvectors of $S$ with eigenvalue $\pm 1$, which do \emph{not} have the same chiral eigenvalue.
Thus if two states have the same chiral eigenvalue, they cannot tunnel into one another.

We also find that the $\PH$-breaking term will not destroy the two zero modes, but only split them in real space.
To be specific, we consider a corner geometry described by the edge Hamiltonian \cref{eq:edge-Dirac-x} plus a translation breaking term $M_1(y) \xi_x' \mu_0$.
There are two terms, \ie $\xi_x'\mu_y$, $\xi_x'\mu_z$, that break $\PH$ and preserve $S$ and $C_{2x}$.
(The symmetry operators are given in \cref{eq:edge-sym-x}.)
Since the two terms are related by a gauge transformation that maps $\mu_y$ to $\mu_z$ and leaves the edge Hamiltonian \cref{eq:edge-Dirac-x} and $M_1(y) \xi_x' \mu_0$ unchanged, we only need to calculate one of them.
Thus we assume the Hamiltonian as
\begin{equation}
\calH = -i \partial_y \xi_y' \mu_0 + M_1(y) \xi_x' \mu_0 + M_2(y) \xi_x' \mu_z,
\end{equation}
where $M_1(y)$ is odd in $y$ (and yields the $2$ corner modes per valley) and $M_2(y)$ is even in $y$.
Without loss of generality, we assume $M_1(y) = m_1 \tanh (y) $ and $M_2(y) = m_2$.
For $m_2<m_1$, the gap term $ M_1(y) \xi_x' \mu_0 + M_2(y) \xi_x' \mu_z$ has zero eigenvalues at $y_\pm = \pm \tanh^{-1} \frac{m_2}{m_1}$.
Therefore, the two zero modes per valley are split in space to the positions $y_\pm$.
We can see that, when $m_2\to m_1 - 0^+$, which means strong $\PH$ breaking, the two zero modes are sent to $\pm \infty$.
As discussed in the main text, for a circular domain wall, under a strong $\PH$ breaking, two zero modes with opposite chiral eigenvalues from two nearby corners will meet and annihilate each other.

\subsection{Zero modes in the Abrikosov vortex}\label{app:vortex}

The two zero modes per valley also appear in the center of a vortex of the superconductor order parameter.
We introduce the polar coordinate system $x=r\cos\theta$, $y=r\sin\theta$ and assume the pairing term as $\Delta(x,y) = e^{-i\theta} \Delta(r)$.
Substituting this pairing term into the Dirac Hamiltonian (\cref{eq:BdGDiracHam}), we can write the BdG Hamiltonian (in valley $\eta=+$) as
\begin{align}
\calH ^{(\eta=+)}&= \xi_z \mu_0 (k_x  \sigma_x + k_y \sigma_y) \nonumber \\
&+ \Delta(r) (\cos{\theta} \xi_x + \sin{\theta} \xi_y)\mu_0\sigma_0.
\end{align}
The new term $\Delta(r) \sin{\theta} \xi_y$, even choosing $\rr$ at a $C_{2z}$ center, violates the symmetries $S,\PH$ defined in Eq.~\eqref{eq:sym-BdG-dirac}, while $\Delta(r) \cos{\theta} \xi_x$ violates $C_{2z}T$ (which rotates $\theta \rightarrow \theta + \pi$).

It will be convenient for us to define the operator $C_{2z}'T = \xi_z C_{2z}T$; this operator fails to commute with $\Delta(r) \sin{\theta} \xi_y$ instead of the $\xi_x$ term.  Since now all three of these symmetries $C_{2z}'T, S, \PH$ all fail to commute with the $\Delta(r) \sin{\theta} \xi_y$ operator, the combined symmetries, $C_{2z}'TS,S\PH$ are preserved.

To solve the Hamiltonian, we switch to polar coordinates
\begin{align}
    \partial_r &= \cos{\theta} \partial_x + \sin{\theta} \partial_y \\
    \frac{1}{r}\partial_\theta &= -\sin{\theta} \partial_x + \cos{\theta} \partial_y \\
    \partial_x + i\partial_y &= e^{i\theta} (\partial_r + \frac{i}{r}\partial_\theta):
\end{align}
\begin{widetext}
\begin{equation}
\calH ^{(\eta=+)}=
\begin{pmatrix}
0 & -i e^{-i\theta} (\partial_r - \frac{i}{r} \partial_\theta) \mu_0  & e^{-i\theta} \Delta(r) \mu_0 & 0  \\
-i e^{i\theta} (\partial_r + \frac{i}{r} \partial_\theta) \mu_0 & 0 & 0 & e^{-i\theta } \Delta(r) \mu_0 \\
e^{i\theta} \Delta(r) \mu_0 & 0 & 0 & i e^{-i\theta} (\partial_r - \frac{i}{r} \partial_\theta) \mu_0 \\
0 & e^{i\theta} \Delta(r) \mu_0 & i e^{i\theta} (\partial_r + \frac{i}{r} \partial_\theta) \mu_0 & 0
\end{pmatrix},
\end{equation}
\end{widetext}
where $\mu_0$ is the two-by-two identity in the Moir\'e valley index.
Here we have ordered the basis as $\psi_{\eta=+,l=+,\alpha=A,\uparrow}$, $\psi_{+,-,A,\uparrow}$, $\psi_{+,+,B,\uparrow}$, $\psi_{+,-,B,\uparrow}$,
$\psi_{+,+,A,\downarrow}$, $\psi_{+,-,A,\downarrow}$, $\psi_{+,+,B,\downarrow}$, $\psi_{+,-,B,\downarrow}$.
We take the ansatz of the zero modes as
\begin{equation}
\phi(x,y) = \frac{1}{\sqrt{\mathcal{N}}} [ g_1(r), g_2(r), g_3(r), g_4(r) ]^T \otimes v
\end{equation}
where $v$ is a two-component vector, $g_{1,2,3,4}(r)$ are functions of $r$ to be determined and $\mathcal{N}$ is a normalization factor.
For either $v=(1,0)^T$ or $v=(0,1)^T$, we find that the zero mode satisfies $g_1=g_4=0$ and
\begin{equation}
-i \partial_r g_2 + \Delta(r) g_3 = 0, \qquad  \Delta(r) g_2 + i \partial_r g_3= 0.
\end{equation}
Combining the two equations, we obtain the equations
\begin{equation}
\Delta(r) g_2 - \partial_r \big( \frac{1}{\Delta(r)} \partial_r g_2 \big) = 0,\quad g_3 = \frac{i}{\Delta(r)} \partial_r g_2(r)
\end{equation}
and the solution
\begin{equation}
g_2(r) = e^{- \int_0^r dr' \Delta(r')},\quad g_3(r) = -i g_2(r).
\end{equation}
A vortex of opposite vorticity $\Delta(\rr) = e^{i\theta} \Delta(r)$ will instead have $g_2,g_3 = 0$ and $g_1,g_4$ nonzero.

The combined symmetries, $C_{2z}'TS = \xi_z \sigma_x K$ and $ S\PH = -\xi_x \mu_y \sigma_x K$ are sufficient to protect the two (fermionic) Abrikosov modes per valley and pin them to $0$ energy.  To see this, $S\PH$ acts as a time-reversal symmetry and squares to $-1$, guaranteeing Kramers' degeneracy, while $C_{2z}'TS$ acts as particle-hole and squares to $+1$, forcing zero modes remain at $0$ energy, so long as the vortex is located at a $C_{2z}$-symmetric point.

\section{Numerical verification of the boundary states of TBG-TSC}\label{app:num_TBG}
To study the edge states or corner states located around the domain walls in the TBG, we bring Eq. (\ref{eq:freeBdG0}) and Eq. (\ref{eq:pairBdG1}) into the real-space form under the basis $(c_{\mathbf{r},l,+,\alpha,\uparrow}, c^\dagger_{\mathbf{r},l',-,\beta,\downarrow})^T$,
\begin{eqnarray}\label{eq:Hr}
\label{eq:realH}
\mathcal{H}^{(+,0)} (\mathbf{r}) &=& v_F k_D \xi_z \left(
\begin{array}{cc}
-i\partial_{\mathbf{r}} \cdot \boldsymbol{\sigma} & 0  \\
0& -i\partial_{\mathbf{r}} \cdot \boldsymbol{\sigma}  \\
\end{array}
\right)  \\ \nonumber
& & + \xi_z \left(
\begin{array}{cc}
 0 & T(\mathbf{r})  \\
T^\dagger(\mathbf{r})& 0\\
\end{array}
\right)  \\
\mathcal{H}^{(+,1)}(\mathbf{r})&=& \Delta(\mathbf{r}) \xi_x,
\end{eqnarray}
where $T(\mathbf{r})=\sum_{j=1}^3 e^{-i\mathbf{q}_j\cdot \mathbf{r}}T^j$, and $\sigma$, $\xi$, and the explicitly written $2\times2$ matrices denote sublattice, particle-hole, and the layer degrees of freedom, respectively. The continuous position $\mathbf{r}$ is the coarse-grained graphene lattice.
In the following subsections, we will consider different geometries of dividing the material into pairing domains with positive and negative $\Delta(\mathbf{r})$, and discuss whether the edge states or corner states would appear.

\subsection{Edge states}

\begin{figure*}[htbp]
\centering
\includegraphics[width=1.0\linewidth]{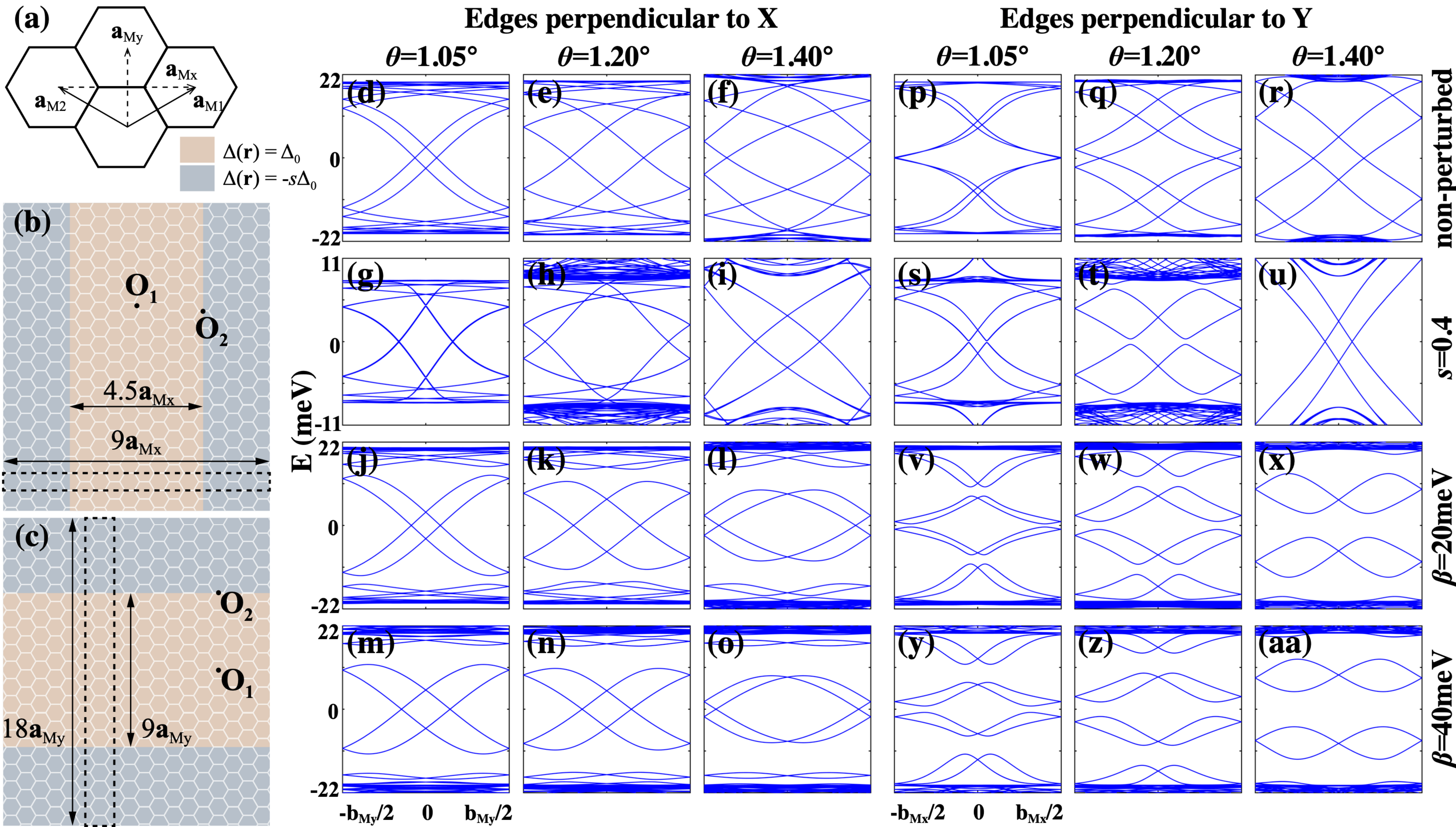}
\caption{\label{fig:edgeXY} (a) Definitions of $\mathbf{a}_{\rm Mx}$ and $\mathbf{a}_{\rm My}$, the Moir\'e lattice vectors. (b)(c) Pairing domain geometries used to solve for edge states in the $x$ and $y$ directions, respectively. The dashed squares denotes supercells in these geometries, and the specific sizes used in the numerics are indicated.
The orange (blue) regions denote domains with $\Delta(\mathbf{r})=\Delta_0>0$ ($\Delta(\mathbf{r})=-s\Delta_0<0$).
Domain walls that bind zero modes are formed at the boundaries between positive and negative pairings.
$s>0$ controls the relative magnitude of {the} pairing potential between the positive and negative regions, and is designed to break accidental symmetries Eq.~\eqref{eq:accidental_symm_1},~\eqref{eq:accidental_symm_2}.
$O_1$ denotes an ``inversion'' center of the $P$ symmetry, and $O_2$ denotes an ``inversion'' center of the $P_2$ symmetry. (d)-(o) Edge bands in the $x$ direction. (p)-(aa) Edge bands in the $y$ direction. Three different twisted angles are investigated: $\theta=1.05^\circ,1.20^\circ$, and $1.40^\circ$. Four different sets of parameters are chosen: $s=1$, $\beta=0$ (1st row); $s=0.4$, $\beta=0$ (2nd row); $s=1$, $\beta=20$meV (3rd row); $s=1$, $\beta=40$meV (4th row).
In the calculation we have chosen the parameters $w_1=110$meV, $w_0=0.8w_1$, $v_F=5.944{\rm eV\cdot \mathring{A}}$, and $|\Delta_0|=20$meV. }
\end{figure*}

To solve edge states in the direction perpendicular to $x$ ($y$), we take the periodic boundary condition in the $x$ ($y$) direction with $x$ ($y$) ranging from $-L/2$ to $L/2$.
Then we assign positive and negative $\Delta(x)$ for $|x|<L/4$ ($|y|<L/4$) and $|x|>L/4$ ($|y|>L/4$), respectively, as shown in Fig. \ref{fig:edgeXY}(b)(c).
Two domain walls that bind zero modes are formed at $x=\pm L/4$ ($y=\pm L/4$).
Using periodic boundary condition greatly simplifies the calculation as the basis states are plane waves.
The matrix elements of the real-space Hamiltonian $H(\rr)$ are calculated with respect to plane waves $\ket{\kk, \mathbf{G}, i}$, where $\kk$ indexes the Moir\'e Brillouin zone momentum, $\mathbf{G}$ the element of the $\QQ$ lattice, and $i$ the remaining indices of the TBG Hamiltonian, which are sublattice, valley, Moir\'e valley, and particle-hole.  The Hamiltonian matrix
\begin{align}
    \braket{\kk, \mathbf{G}, i|H(\rr)|\kk, \mathbf{G}', i'}
\end{align} is then numerically diagonalized.

\par
Using plane waves as the basis of Eq. (\ref{eq:Hr}), we obtain the edge bands in the $x$ direction (i.e. when the pairing domain walls are set perpendicular to the $x$ direction) at three twist angles $\theta=1.05^\circ$, $1.2^\circ$, $1.4^\circ$, as shown in Fig. \ref{fig:edgeXY}(d)-(f), as well as the edge bands in the $y$ direction, as shown in Fig. \ref{fig:edgeXY}(p)-(r).
It is seen that, gapless edge states not only appear in the $x$ direction, as analyzed in the main text, but also appear accidentally in the $y$ direction, where no known symmetry could protect such gapless modes.

The origin of the accidental edge states in the $y$ direction may be ascribed to an additional emergent symmetry on the domain wall, which we would denote by $P_2\equiv \xi_z P$, where $P=\delta_{\mathbf{Q,-Q'}}\zeta_{\mathbf{Q}}\sigma_0\xi_z$ is the unitary symmetry defined in Appendix \ref{app:BdG-TBG} that anti-commutes with the bulk Hamiltonian,
\begin{eqnarray}
P\mathcal{H}^{(+,0)}(\mathbf{r})P^{-1} = -\mathcal{H}^{(+,0)}(-\mathbf{r}) \label{eq:accidental_symm_1} \\
P\mathcal{H}^{(+,1)}(\mathbf{r})P^{-1} = -\mathcal{H}^{(+,1)}(-\mathbf{r}) \label{eq:accidental_symm_2}
\end{eqnarray}
where $\mathcal{H}^{(\eta,1)}(\mathbf{r})=\xi_z {H}^{(+)}(\mathbf{r})$ is the free part, and $\mathcal{H}^{(\eta,1)}(\mathbf{r})=\Delta_0\xi_x$ is a uniform spin-singlet pairing potential. However, at the domain wall, $\Delta(\mathbf{r})$ no longer stays constant, but has opposite signs at the two sides, \ie $\Delta(\mathbf{r})=-\Delta(\mathbf{-r})$.
(Here $\mathbf{r}=0$ corresponds to a point at the domain wall.)
Therefore, $P$ is broken by $\mathcal{H}^{(+,1)}(\mathbf{r})$ by the domain wall.
However, both $\mathcal{H}^{(+,0)}(\mathbf{r})$ and $\mathcal{H}^{(+,1)}(\mathbf{r})$ anti-commute with $P_2$ instead.
In real space, $P_2$ is $P$, which inverts position $\rr \rightarrow -\rr$ and switches layer, combined with $\xi_z$, which introduces a relative sign of the transformation depending on whether the operator being transformed is a particle or a hole.

Within the Dirac theory of the edge states, we can show that this emergent symmetry $P_2$ can indeed protect edge states in any arbitrary direction. As has been presented in Appendix \ref{app:edge-state-Dirac}, one can show that $P_2=i \xi_0' \mu_y$ on the edge by definition, and that $P_2\mathcal{H}_{\text{eff}}(k_y)P_2=-\mathcal{H}_{{\rm eff}}(-k_y)$. Therefore, the only gap term allowed by both $S$ and $\mathcal{P}$, $\xi_x'\mu_0$, would not obey $P_2$, in the sense that $\{P_2, \xi_x'\mu_0\} \not= 0$, thus not permitted. Therefore, although $C_{2x}$ is absent in an arbitrary direction, the emergent $P_2$ would still rule out other possible gap terms.

Enlightened by this, we tried several methods to break the $P_2$ symmetry in order to gap the edge states in the $y$ direction. One natural way is to tune the relative magnitude between the negative and the positive domains, so that
\begin{eqnarray}
\Delta(\mathbf{r})&=&\Delta_0, ~~~~\text{for the positive domains} \\
\Delta(\mathbf{r})&=&-{s}\Delta_0, ~~~\text{for the negative domains}
\end{eqnarray}
The numerical results are shown in Fig. \ref{fig:edgeXY}(g)-(i)(s)-(u).
Only at $\theta=1.2^\circ$, the edge state in the $y$ direction has a non-negligible gap.
However, the edge states in the $y$ direction at $\theta=1.05^\circ, 1.4^\circ$ are still almost gapless, meaning our perturbation does not do a very good job of breaking the accidental symmetry.
Another term that  breaks $P_2$ is given by
\begin{equation}
\mathcal{H}^{(+,2)} = f(\mathbf{r}) \xi_z \left(
\begin{array}{cc}
 0 & \sigma_0  \\
\sigma_0 & 0\\
\end{array}
\right)
\end{equation}
where $f(\mathbf{r})$ is odd on the domain wall,
\begin{eqnarray}
f(\mathbf{r})=+\beta, ~\text{if}~ \Delta(\mathbf{r})>0 \\
f(\mathbf{r})=-\beta, ~\text{if}~ \Delta(\mathbf{r})<0
\end{eqnarray}
With different magitude of $\beta$, we plot the edge bands in Fig. \ref{fig:edgeXY}(j)-(o)(v)-(aa). A significant gap in the $y$ direction is opened and the gap increases with $\beta$, while the zero modes in the $x$ direction, as expected, are stable against this perturbation.  Note that the low-lying states are always gapped from the bulk; this is expected as the zero modes are pointlike: instead of giving a continuous set of energies like an edge, the energies from the zero modes are discrete.

\subsection{Corner states}\label{app:num-corner}
\begin{figure}[t]
\centering
\includegraphics[width=1.0\linewidth]{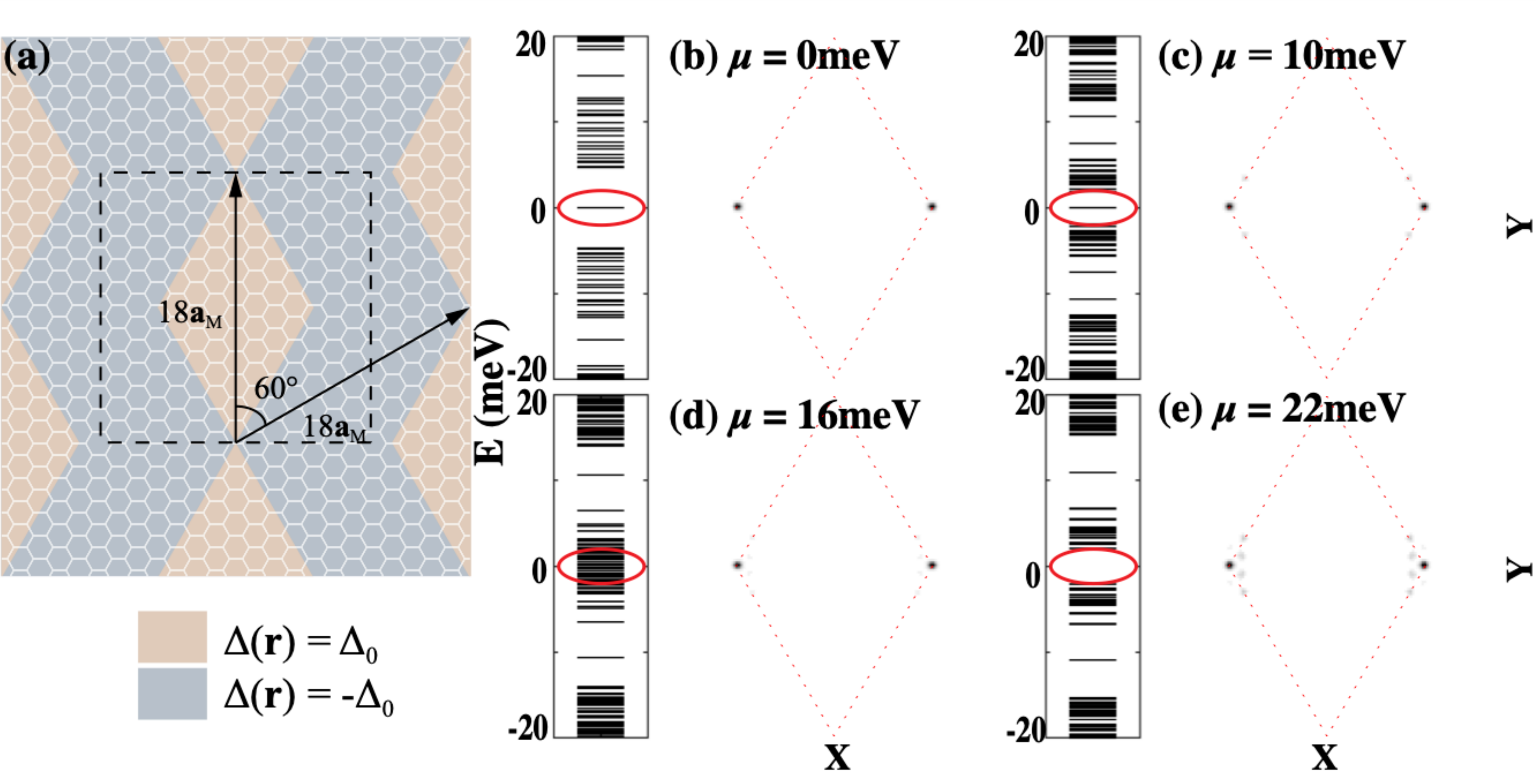}
\caption{\label{fig:hex}Rhombic pairing domains.
The periodic boundary condition is realized by a supercell spanned by $18\aa_{M1}$ and $18\aa_{M2}$.
(a) The configuration of the domains. The dashed square indicates the spatial region displayed in the following spatial density diagrams. (b)-(e) energy spectra near the zero-energy and the spatial densities of the 4 lowest states. The red dashed rhomboids indicate the  the domain walls. Results for $\mu=0,10,16,22$meV are displayed. Through all the numerics, we choose $w_1=110$meV, $w_0=0.8w_1$, $v_F=5.944{\rm eV\cdot \mathring{A}}$, $\theta=1.40^\circ$, $\beta=40$meV, and $|\Delta_0|=20$meV.}
\end{figure}

\begin{figure}[t]
\centering
\includegraphics[width=1.0\linewidth]{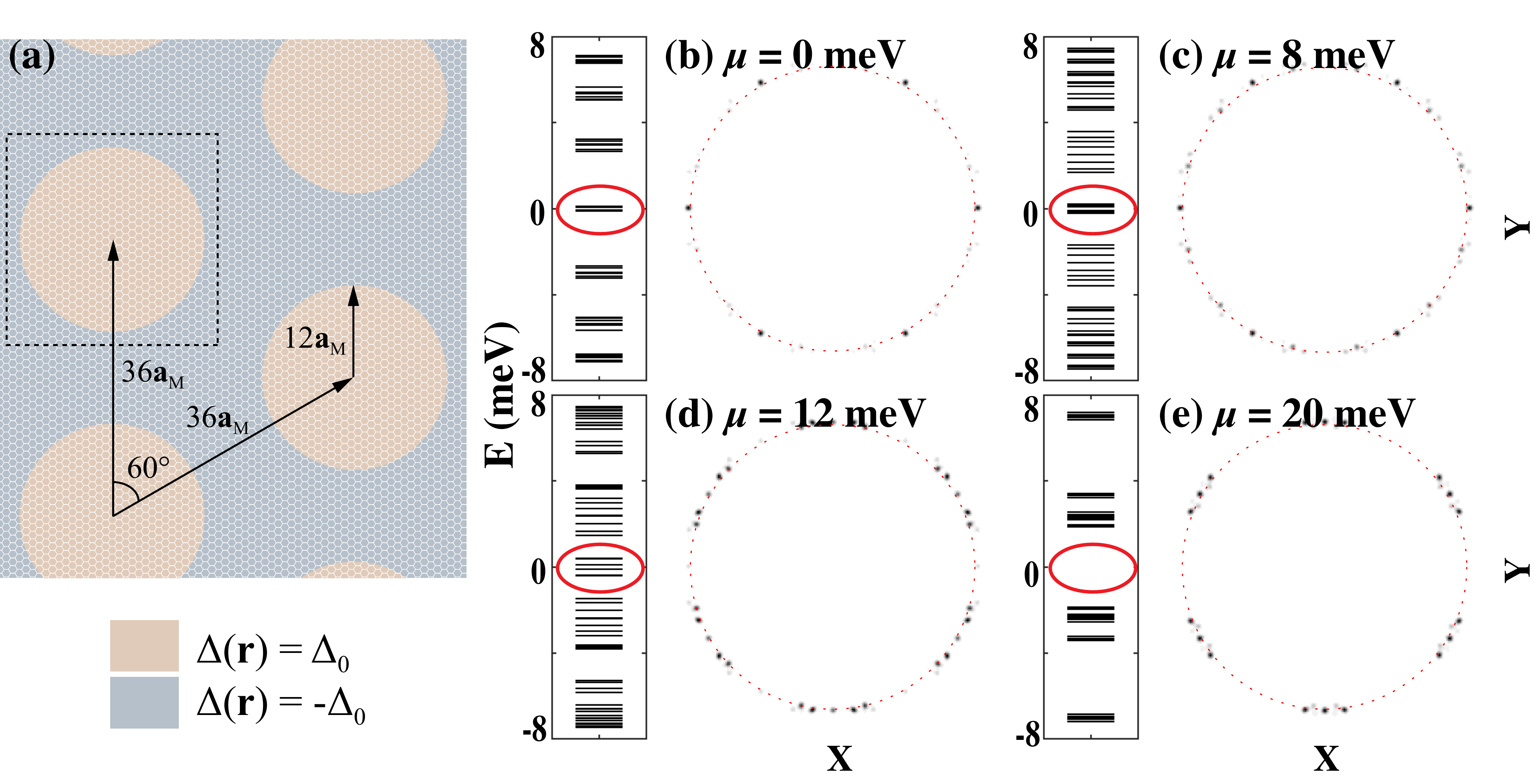}
\caption{\label{fig:circ}Circular pairing domains. The periodic boundary condition is realized by a supercell spanned by $36\aa_{M1}$ and $36\aa_{M2}$. (a) The configuration of the domains. The dashed square indicates the spatial region displayed in the following spatial density diagrams. (b)-(e) energy spectra near the zero-energy and the spatial densities of the 12 lowest states. The red dashed circles indicate the the domain walls. Results for $\mu=0,8,12,20$meV are displayed. Through all the numerics, we choose $w_1=110$meV, $w_0=0.8w_1$, $v_F=5.944{\rm eV\cdot \mathring{A}}$, $\theta=1.40^\circ$, $\beta=40$meV, and $|\Delta_0|=20$meV.}
\end{figure}

Now that the accidental edge states are gapped out, we can numerically investigate the evolution of the symmetry-protected corner modes when the chemical potential is tuned to finite values.
The largest gap in the $y$-direction appears at $\theta=1.40^\circ$ with $\beta=40$meV (\cref{fig:edgeXY}).
We hence choose $\theta=1.4^\circ$, $\beta=40$meV in the following numerical calculations to demonstrate the existence of the corner states.
Two different pairing domain geometries are devised. One is the rhombic domains as shown in Fig. \ref{fig:hex}(a), chosen to break the $C_{3z}$ symmetry, and the other is the circular domains as shown in Fig. \ref{fig:circ}(a). \par
In the first geometry, when $\mu=0$, as illustrated in Fig. \ref{fig:hex}(b), there are 4 zero modes that are well separated from other states in the spectrum, and their total spatial densities are localized at the two $C_{2x}$-invariant corners, which shows perfect accordance with the theoretical prediction in the main text. As $\mu$ is tuned up, as illustrated in Fig. \ref{fig:hex}(c), the zero modes slightly ``diffuse'' along the edges, but are still well separated from other states in energy. At some critical value (Fig. \ref{fig:hex}(d)), the bulk states collapse onto the zero modes in the spectrum, and for chemical potential beyond the critical value, zero modes disappear, and the spatial densities of the 4 lowest states start to spread into the bulk (Fig. \ref{fig:hex}(e)). The reason the zero modes spread, and not move, is due to the straight edges of the rhombic geometry.

To see how zero modes from different $C_{3z}$ corners evolve with chemical potential and annihilate each other, a second geometry is investigated, as shown in Fig. \ref{fig:circ}(a). When $\mu=0$ (Fig. \ref{fig:circ}(b)), there are  12 zero modes per valley well separated from other states in the spectrum, which corresponds to 3 copies of the states in the rhombic geometry, and their spatial distributions are localized at the two $C_{2x}$-invariant corners and four other equivalent corners related by $C_{3z}$.
As $\mu$ is tuned up (Fig. \ref{fig:circ}(c)), while their energies are still pinned close to zero and separated from other states, they start moving towards their adjacent corners. At some critical value, states from adjacent corners meet (Fig. \ref{fig:circ}(d)), and correspondingly, the gap between the zero modes and other states vanish (<1meV). For chemical potential beyond this critical value (Fig. \ref{fig:circ}(e)), zero modes from adjacent corners gap each other out.

The numerical results are highly consistent with the theoretical predictions presented in the main text.

\section{Other pairing symmetries}
\label{app:other-pairing}

In \cref{app:BdG-TBG,app:Dirac,app:num_TBG} we have analytically and numerically shown that the TBG Hamiltonian with a uniform spin-singlet pairing is a topological superconductor guaranteed by the $C_{2z}T$ and $\PH$ symmetries.
Now we show that the superconducting phase with the $B_1$ triplet pairing (defined in the next paragraph) is also topological and has the similar edge and corner states as the uniform spin-singlet pairing.

Only four intervalley terms open full gaps in the Bogoliubov bands: $A_1$ (spin-singlet), $A_2$ (spin-singlet), $B_1$ (spin-triplet), $B_2$ (spin-triplet) \cite{christos2020superconductivity}.
$A_{1}, A_2$ and $B_{1}, B_2$ are even and odd under the $C_{2z}$ operation, respectively.
And, $A_1,B_1$ and $A_2,B_2$ are even and odd under the $C_{2x}$ operation, respectively.
The four terms are invariant under the other crystalline symmetry operations.
We find that the $A_1$ and $B_1$ pairings are topological whereas the $A_2$ and $B_2$ pairings are trivial.
In fact, the $B_1$ pairing is very similar to the uniform spin-singlet ($A_1$) pairing: their BdG Hamiltonians in a single valley are the same, but the sign of the pairing switches for the opposite valley.
Explicitly, the pairing forms read $\xi_x\tau_0\mu_0\sigma_0, \xi_x\tau_z\mu_z\sigma_0, \xi_x\tau_z\mu_0\sigma_0,\xi_x\tau_0\mu_z\sigma_0$, for the $A_1, A_2, B_1, B_2$ pairings, respectively.

\subsection{The \texorpdfstring{$A_2$}{A2} singlet pairing}
\label{app:A2-pairing}

In this subsection we show the triviality of the $A_2$ pairing.
We assume the valley-U(1) symmetry and use the same BdG basis (\cref{eq:BdG-basis-dirac}) as in the Dirac theory of the $A_1$ pairing.
As discussed in the main text and in \cref{app:BdG-symmetry}, due to the time-reversal and spin-SU(2) symmetries, the Dirac theory of any spin-singlet pairing takes the form
\begin{equation}
\calH = k_x \xi_z \tau_z \mu_0 \sigma_x + k_y \xi_z \tau_0 \mu_0 \sigma_y + \xi_x \otimes \Delta,
\end{equation}
with $\Delta$ being an eight by eight Hermitian matrix, built of $\tau, \mu, \sigma$ operators.
Here $\tau$, $\mu$, $\sigma$ are Pauli (identity) matrices for the valley, Moir\'e valley, and sublattice degrees of freedoms, respectively.
For $A_1$, the pairing matrix $\Delta = \tau_0 \mu_0 \sigma_0 = I$.
Anti-commutation with $\xi_y$ forces the pairing term to $\xi_x \otimes \Delta.$

Due to the valley-U(1) symmetry, $\Delta$ is diagonal in the valley index, as Eq.~\ref{eq:BdG-basis} combines creation operators of one valley with annihilation operators of the other.
The symmetry operators in the BdG formalism are  given in \cref{eq:sym-BdG-dirac}.
Among these symmetries, $S$ and $\PH$ are constraints respectively given by the spinful time-reversal symmetry and the charge conjugation $\PH_c$ (\cref{eq:Pc-on-BdG-basis}), and $C_{3z}$, $C_{2x}$, $T$, $C_{2z}T$ are crystalline symmetries.
For simplicity, we only consider the pairings that are diagonal in the Moir\'e valley index $\mu_0$, \ie pairings preserving the Moir\'e translation symmetry.
The $A_2$ pairing is invariant under $C_{3z}$, $C_{2z}$, $C_{2z}T$ and is odd under $C_{2x}$.
For the pairing to be invariant (with eigenvalue $1$) under $C_{3z}$ ($e^{i\frac{2\pi}3\tau_z{\mu_0}\sigma_z}$), $\Delta$ must be proportional to $\sigma_0$ or $\sigma_z$; For it to be odd under $C_{2x}$ (${\tau_0}\mu_x\sigma_x$), $\Delta$ can be chosen as $\tau_{0,z}\mu_z\sigma_0$ or $\tau_{0,z}\mu_0\sigma_z$; For it to be invariant under $C_{2z}$ ($\tau_x\mu_x\sigma_x$), only $\tau_{z}\mu_z\sigma_0$ remains.
The term $\tau_{z}\mu_z\sigma_0$ also preserves the $C_{2z}T$ (${\tau_0\mu_0}\sigma_x K$) symmetry.
This form of the $A_2$ pairing is also given in Table II of Ref. \cite{christos2020superconductivity}.
Therefore, the BdG Hamiltonian on the basis \cref{eq:BdG-basis-dirac} reads
\begin{equation} \label{eq:Dirac-A2}
\calH = k_x \xi_z \tau_z \mu_0 \sigma_x + k_y \xi_z \tau_0 \mu_0 \sigma_y + \Delta_0 \xi_x \tau_z \mu_z \sigma_0.
\end{equation}
The $C_{2x}$ and $\PH$ (given by the charge conjugation $\PH_c$ \cref{eq:Pc-on-BdG-basis}) are broken by the $A_2$ pairing.
Nevertheless, there exists a projective $C_{2x}$ and a projective $\PH$
\begin{equation}
C_{2x}' = \xi_z \tau_0 \mu_x \sigma_x,\qquad \PH' = i\xi_0 \tau_0 \mu_y \sigma_x K\ ,
\end{equation}
which correspond to $C_{2x}$ followed by a charge-U(1) rotation and $\PH_c$ followed by a charge-U(1) rotation, respectively \cite{Fang:2017}.
We will show these projective symmetries do not protect topology.

Under a charge rotation $c \rightarrow e^{i\gamma}c, c^\dagger \rightarrow e^{-i\gamma} c^\dagger$, $\psi \rightarrow e^{i\gamma \xi_z} \psi$, as $\psi$ consists of both creation and annihilation operators.  The charge rotation implemented is $\gamma = \pi/2$, with additional factors of $i$ placed to preserve the algebra of the operators $C_{2x}'$ and $\PH'$.
Following the calculation in \cref{app:edge-state-Dirac}, we  solve for the effective edge theory (in valley $\eta=+$) in the $x$ direction.
First take ansatz $\phi_{k_y}(x) \propto e^{ik_yy - \lambda |x|} u(k_y)$ and solve the Hamiltonian with domain wall in $\Delta = -\Delta_0 \text{sgn}(x)$ for $k_y = 0$, $\Delta_0 > 0$:
\begin{align}
    & {\cal H} = -i \partial_x \xi_z \tau_z \mu_0 \sigma_x + \Delta_0 \text{sgn}(x) \xi_x \tau_z \mu_z \sigma_0\\
    & i\lambda \text{sgn}(x) \xi_z \tau_z \mu_0 \sigma_x u -\Delta_0 \text{sgn}(x) \xi_x \tau_z \mu_z \sigma_0 u = 0 \\
    & \lambda u = \Delta_0 \xi_y \mu_z \sigma_x u.
\end{align}
These equations have solutions when $\lambda = \Delta_0$, $\xi_y \mu_z \sigma_x = 1$.
Notice that the valley operator $\tau$ drops out of the solution, allowing us to focus on a single sector $\tau_z = +1$:
{\small
\begin{align}
\label{eq:A2edge-basis-x}
u_1 &=
\frac12
\begin{bmatrix}
1 \\ i
\end{bmatrix}\otimes
\begin{bmatrix}
1 \\ 0
\end{bmatrix}\otimes
\begin{bmatrix}
1 \\ 1
\end{bmatrix} ,\;
u_2= \frac12
\begin{bmatrix}
1 \\ i
\end{bmatrix}\otimes
\begin{bmatrix}
0 \\ 1
\end{bmatrix}\otimes
\begin{bmatrix}
1 \\ -1
\end{bmatrix} , \nono\\
u_3 &= \frac12
\begin{bmatrix}
-1 \\ i
\end{bmatrix}\otimes
\begin{bmatrix}
1 \\ 0
\end{bmatrix}\otimes
\begin{bmatrix}
1 \\ -1
\end{bmatrix}, \;
u_4 = \frac12
\begin{bmatrix}
-1 \\ i
\end{bmatrix}\otimes
\begin{bmatrix}
0 \\ 1
\end{bmatrix}\otimes
\begin{bmatrix}
1 \\ 1
\end{bmatrix} .
\end{align} }
We find
\begin{equation}
\calH = k_y \xi_y' \mu_z\ ,
\end{equation}
with the projected symmetries on edge
\begin{align}
S = \xi_z', \quad \PH'=-i\xi_y'\mu_x K,\quad C_{2x}' = \xi_y'\mu_y.
\end{align}
The gap term $\xi_y'\mu_y$ is allowed by the symmetries and hence there is no protected edge state.
On the edge in the $y$ direction, we can obtain the same effective Hamiltonian with the projected symmetries
\begin{equation}
S = \xi_z', \quad \PH'=i\xi_y'\mu_x K,\quad C_{2x}'C_{2z}T = \xi_0' \mu_x K.
\end{equation}
The action of the symmetries on the edge in the $y$ direction is found by rotating the wavefunctions $u_i' = e^{-i\frac{\pi}{4}\sigma_z} u_i$.  The prefactor $e^{-i\frac{\pi}{4}\sigma_z}$ commutes with both $S, \PH'$, so no change occurs to those symmetries, but $C_{2y}T$ is modified.
The gap term $\xi_x' \mu_0$ is allowed by the symmetries and hence there is no protected edge state.
For edges in generic directions the only symmetries on the edge are $S$ and $\PH'$, which, as shown in the $x$ and $y$ directions, cannot forbid gap terms.
Therefore, there is no gapless edge in any direction.

There is also no corner state.
Unlike the $A_1$ pairing, where $[S,\PH]=0$, for the $A_2$ pairing we have $\{S,\PH'\}=0$.
Thus, even there existed zero energy doublet at a corner, the two states in the doublet must have opposite chiral eigenvalues and hence could be gapped out.
To be specific, we can choose the $S$ and $\PH'$ operators in the doublet space as $\sigma_z$ and $i\sigma_yK$, respectively.
Then both $\sigma_x$ and $\sigma_y$ gap terms are allowed to appear.

From the absence of gapless boundary mode, we conclude the triviality of the $A_2$ pairing phase.

\subsection{The \texorpdfstring{$B_1$ and $B_2$}{B1 and B2} spin triplet pairings}
\label{app:B1B2-pairing}

We assume the valley-U(1) symmetry and the spinful time-reversal symmetry in the triplet pairing.
The triplet pairing must break the spin-SU(2) symmetry, as the $S_z = \pm 1, 0$ sectors will rotate into each other.
For example, rotating spins by $S_z$ will introduce phases into the $S_z = \pm 1$ sectors.
Here we assume a spin-U(1) remains.
Without loss of generality, we assume the rotation axis of the spin-U(1) in the $z$ direction.
Due to the valley-U(1) and spin-U(1), the operator $c^{\dagger}_{\kk,\eta,l,\alpha,\uparrow}$ can only couple to $c^{\dagger}_{-\kk,-\eta,l,\alpha,\downarrow}$ in the pairing (we are focusing on the $S_z = 0$ sector of the triplet pairing).
Hence we can still use the BdG basis \cref{eq:BdG-basis-dirac} for the triplet pairings.
Due to the spinful time-reversal symmetry, the BdG Hamiltonian still has the chiral symmetry $\{\xi_y, \calH^{(\eta)}\}=0$ (\cref{eq:BdG-constraint-syT}).
We can write the Dirac BdG Hamiltonian as
\begin{equation}
\calH = k_x \xi_z \tau_z \mu_0 \sigma_x + k_y \xi_z \tau_0 \mu_0 \sigma_y + \xi_x \otimes \Delta,
\end{equation}
with the symmetry operators given in \cref{eq:sym-BdG-dirac}.
Due to the valley-U(1), $\Delta$ must be proportional to $\tau_{0,z}$.
For simplicity, we only consider $\Delta$'s that are diagonal in the Moir\'e valley index, \ie pairings that preserve the Moir\'e translation symmetry.
Since the triplet pairing is even under the spinful time-reversal ($i\hat{s}_yT$) and odd under the $i\hat{s}_y$ spin-rotation, the triplet pairing must be {\it odd} under the spinless time-reversal symmetry $T=\tau_x\mu_xK$.
Thus the options for $\Delta$ include: $\tau_z \mu_{0} \sigma_{0,x,z}$, $\tau_z \mu_{z} \sigma_{y}$, $\tau_0 \mu_{z} \sigma_{0,x,z}$, $\tau_0 \mu_{0} \sigma_{y}$.

We now determine the forms of the $B_1$ and $B_2$ pairings.
For the pairing to be invariant under $C_{3z}=e^{i\frac{2\pi}3\tau_z\sigma_z}$, $\Delta$ must be proportional to $\sigma_0$ or $\sigma_z$, hence only $\tau_z \mu_{0} \sigma_{0,z}$ and $\tau_0 \mu_{z} \sigma_{0,z}$ are valid.
Because the $B_1$ and $B_2$ by definition are odd under $C_{2z}$ and they are also odd under the spinless time-reversal ($T$) as discussed in the last paragraph, they must be even under $C_{2z}T$.
For it to be invariant under $C_{2z}T=\sigma_x K$, only  $\tau_z \mu_{0} \sigma_{0}$ and $\tau_0 \mu_{z} \sigma_{0}$ are valid.
$\tau_z \mu_{0} \sigma_{0}$ and $\tau_0 \mu_{z} \sigma_{0}$ are even and odd under $C_{2x}=\mu_x\sigma_x$, thus they form the $B_1$ and $B_2$ representations, respectively.
These forms of $B_1$ and $B_2$ pairings are also given in Table VI of Ref.  \cite{christos2020superconductivity}.
The BdG Hamiltonian of the $B_1$ pairing and the symmetries are given by
\begin{equation}
\calH = k_x \xi_z \tau_z \mu_0 \sigma_x + k_y \xi_z \tau_0 \mu_0 \sigma_y + \Delta_0 \xi_x \tau_z \mu_0\sigma_0,
\end{equation}
and \cref{eq:sym-BdG-dirac}, respectively.
Since the Hamiltonian and symmetries in the $\eta=+1$ block are the same as those of the $A_1$ pairing (\cref{eq:BdGDiracHam,eq:sym-BdG-dirac}), the $B_1$ pairing must have the same topology and the same boundary states as the $A_1$ pairing.
An identical argument holds in the opposite $\eta = -1$ valley.

The BdG Hamiltonian of the $B_2$ pairing is
\begin{equation}
\calH = k_x\xi_z \tau_z\mu_0\sigma_x + k_y\xi_z\tau_0\mu_0\sigma_y + \Delta_0 \xi_x  \tau_0 \mu_z\sigma_0.
\end{equation}
It breaks the single valley symmetries $C_{2x}=\mu_x\sigma_x$  and $\PH=i\xi_z\mu_y\sigma_x K$.
Nevertheless, there exists a projective $C_{2x}$ and a projective $\PH$
\begin{equation}
C_{2x}' = \xi_z \tau_0 \mu_x \sigma_x,\qquad \PH' = i\xi_0 \tau_0 \mu_y \sigma_x K\ ,
\end{equation}
which correspond to $C_{2x}$ and $\PH_c$ (\cref{eq:Pc-on-BdG-basis}) followed by a charge-U(1) rotation {$c \rightarrow ic$}, respectively.
The Hamiltonian and symmetries in the $\eta=+1$ block are the same those of the $A_2$ pairing (\cref{app:A2-pairing}), thus the $B_2$ pairing must have the same topology as the $A_2$ pairing, which is trivial.

In the above we have shown that the Dirac BdG Hamiltonians of the $B_{1}$ and $B_2$ pairings are the same as the Dirac BdG Hamiltonians of $A_1$ and $A_2$ pairings respectively, up to a sign depending on the valley (which will not affect the physics of TBG because the valleys are independent).
In fact, this correspondence between singlet pairing and triplet pairing is in general true if the TBG Hamiltonian has two independent spin-SU(2) symmetries in the two valleys, which is usually assumed in theoretical works on TBG.
Suppose we have a singlet ($\uparrow\downarrow-\downarrow\uparrow$) pairing
\begin{equation}
\sum_{\kk} \sum_{\eta} \sum_{ll'\alpha\beta} \Delta_{l,\alpha;l',\beta}^{(\eta)}(\kk) c^\dagger_{\kk,\eta,l,\alpha,\uparrow} c^\dagger_{-\kk,-\eta,-l',\beta,\downarrow},
\end{equation}
where $\Delta$ is Hermitian due to spinful time-reversal symmetry ($i\tau_x\mu_x s_y K$) and satisfies $\Delta^{(\eta)}_{l,\alpha;l',\beta}(\kk) = \Delta^{(-\eta)}_{-l',\beta;-l,\alpha}(-\kk)$ due to the $i\hat{s}_y$ spin-rotation.
The pairing must be degenerate under any charge and spin rotations in the two valleys separately.
We apply charge-spin rotation $e^{i\frac{\pi}4 (s_0-s_z)}$ in the valley $\eta=+$ and $e^{i\frac{\pi}2(s_0+s_z)}$ in the valley $\eta=-$ and obtain
\begin{equation}
\sum_{\kk} \sum_{\eta} \sum_{ll'\alpha\beta} \eta \Delta_{l,\alpha;l',\beta}^{(\eta)}(\kk) c^\dagger_{\kk,\eta,l,\alpha,\uparrow} c^\dagger_{-\kk,-\eta,-l',\beta,\downarrow}.
\end{equation}
The rotated pairing becomes triplet {$(\uparrow \downarrow + \downarrow \uparrow)$} since it is odd under $i\hat{s}_y$.
Therefore, the triplet pairings and singlet pairings are related by a unitary transformation.
The BdG Hamiltonian in the valley $\eta=+$ is invariant under this transformation.
Thus the single-valley symmetries and topologies of the singlet pairing and the corresponding rotated triplet pairing must be the same.

\section{Valley-U(1) breaking on a hard edge}
\label{app:U1-breaking}

In this section we discuss the valley-U(1) symmetry, and how in a system of TBG with no pairing either the valley symmetry or the anti-unitary particle-hole $\PH$ \emph{must} be strongly broken on the edge.

Consider a circular sample of TBG from radius $0<r<a$.
The Hamiltonian per spin sector without pairing reads
\begin{align}
    H = k_x \tau_z \mu_0\sigma_x + k_y \tau_0\mu_0\sigma_y,
\end{align} with the symmetries
\begin{align}
    C_{2z}T = \sigma_x K, ~~ {\cal P} = i\mu_y\sigma_x K.
    \label{eq:TBGBreakingSymm}
\end{align}
Outside the sample is a vacuum and is fully gapped.
The vacuum must respect $C_{2z}T$ and we {\it assume} that the vacuum also respects $\PH$.
Then there must be terms consistent with the symmetries in \cref{eq:TBGBreakingSymm} to gap out the Dirac terms \emph{outside} the sample to properly model the symmetry-preserving vacuum.

Working in graphene valley $\tau_z = +1$, we wish to gap the spectrum by adding terms that anti-commute with the Hamiltonian: $\mu_i \sigma_z$.
Terms that commute with $H$ and affect Moir\'e valley only, like $\mu_{x,z}$, do not open up gaps; they simply shift the Dirac crossing points away from each other.
Forcing consistency with $\cal{P}$ further restricts the gap terms to the form
\begin{align}
    M_1(\rr) \mu_0 \sigma_z,
\end{align}  which to be symmetric under $C_{2z}T$ forces $M_1(\rr) = -M_1(-\rr)$, with origin at $\rr = 0$.
Adding this term will gap out the Dirac cones, except along lines where $M_1(\rr)$ vanishes.
By $M_1(\rr) = -M_1(-\rr)$ and the intermediate value theorem, there must be at least two lines extending from the sample to infinity that are gapless, leading to a contradiction.
For a simple way to see the gapless lines, take $\rr = re^{i\theta}$, and consider a circle of constant radius $r_0$ so that $\rr$ lies outside the sample of TBG.
$C_{2z}T$ symmetry forces $M_1(r_0,\theta) = -M_1(r_0,\theta+\pi)$.
If $M_1(r_0,\theta_0) > 0$, then $M_1(r_0,\theta_0+\pi) < 0$, and there must be at least two diametric points along the circle of constant radius $r_0$ where $M_1$ vanishes.
This argument holds for all $r_0$ outside the sample of TBG.

What this analysis implies is that outside the TBG sample, in order to reproduce the fully gapped vacuum that respects $\PH$ we must involve terms that mix the valleys: e.g. $M_2(\rr) \tau_x \mu_{x,z} \sigma_x $.
Alternatively,  breaking $\PH$ allows the term $M_3(\rr) \mu_y\sigma_z$.
Both $M_2(\rr)$ and $M_3(\rr)$ will gap out the Dirac cone everywhere, because they are symmetric under $C_{2z}T$ and is not required to vanish anywhere in the vacuum.
This means that along the edge of TBG, at those points where $M_1(\rr)$ vanishes we must gap out the system instead with $M_2$ or $M_3$, and will strongly break either valley-$U(1)$ or $\PH$.

\section{Shiba States}
\label{app:shiba}
We study the response of TBG-TSC to impurities.
Similar techniques involving Shiba states \cite{Shiba:1968} have been proposed to detect topological superconductivity in 1D and topological insulators in higher dimensions \cite{Sau:2013,Slager:2015}, and even build topological phases \cite{Pientka:2013}.
These works study the Green's function of the Hamiltonian and how it reacts to delta-function impurities.
However, the Dirac version of our Hamiltonian has a problem: it does not appropriately distinguish between the trivial and topological phases on its own.
The Dirac equation possess a symmetry $\Delta \rightarrow -\Delta$, with $\Delta$ the mass, while the two phases are meant to be distinct.
The standard technique is to modify the Dirac Hamiltonian mass (i.e. superconducting gap) to $\Delta - mk^2$ for some mass $m$ (\cref{eq:k-dependent-pairing}) \cite{Shen:2011}.
This term is employed to properly calculate, for example, the Chern number for the Dirac theory of topological insulators and superconductors: with $m < 0$ for $\Delta > 0$, the system is trivial, and with $m > 0$, the system is topological \cite{Shen:2011}.
Notice that the topological phase has a band inversion while the trivial phase does not.
In this Appendix, we calculate the behavior of Shiba states for the Dirac Hamiltonian with quadratic correction, and demonstrate a qualitative difference between the $m < 0$ and $m > 0$ phases.

For TBG-TSC, however, we argue that $m < 0$ and $m > 0$ \emph{do not} correspond to the topological and trivial phases.
The distinction between topological and trivial phases works for non-anomalous systems, like a Chern insulator in two dimensions.  In that situation $m<0$ and $m>0$ indeed do correspond to topological and trivial phases.  In TBG-TSC however, the single-valley Hamiltonian is anomalous, analogous to the surface of a 3DTI.  The two distinct phases correspond to $\Delta > 0$ and $\Delta < 0$ (not $m$), and even then are not topological/trivial themselves but only have topological distinction when compared to one another.
Properly capturing the topological-trivial distinction requires $|m\Lambda^2|\gg \Delta$, with $\Lambda$ being the UV cutoff for $k$.
In this limit the band inversion is present and the Hamiltonian is well-defined even as $k \rightarrow \infty$.

The opposite limit, $|m \Lambda^2| \ll| \Delta|$, does not have a well-defined large momentum limit.
Only the difference between $\Delta < 0$ and $\Delta > 0$ has a well-defined topological invariant.
In this work we have assumed the second limit because the nearly uniform spin-singlet pairing (\cref{app:uniform-s}) has no $\kk$-dependence or very weak $\kk$-dependence (\cref{eq:k-dependent-pairing}): $\Delta - m\kk^2$ is close to being uniform, so $m\kk^2 \ll |\Delta|$ for all $\kk$ in the BZ.
This means that while Shiba states do distinguish between $m < 0$ and $m > 0$ phases (the former corresponding to singlet-pairing TBG-TSC), it does not offer a distinguishing signature of TBG-TSC.  Nevertheless, we examine the behavior of impurities in the TBG-TSC Hamiltonian for different values of $m$.

\subsection{Dirac Hamiltonian with quadratic correction}
Given a superconducting Hamiltonian ${\cal H}_0$ and an impurity potential $V \delta(r) X$, for some matrix $X$ with eigenvalues $\pm 1$, zero modes can be found by solving the equation
\begin{align}
    ({\cal H}_0({\bf r}) &+ V\delta({\bf r}) X) \psi = E \psi, \\
    \text{Det}[&V^{-1} - G(E, {\bf r} = 0) X] = 0.
    \label{eq:ShibaSolns}
\end{align} where the Green's function $G$ has been integrated over momentum to yield $G(E, {\bf r}=0)$.
A solution to Eq.~(\ref{eq:ShibaSolns}) is indicative of Shiba states.
If there is a solution for $E$ smaller than the gap, then there are in-gap states.
Take Hamiltonian
\begin{align}
    {\cal H}(\kk) &= \xi_z (k_x \tau_z \mu_0 \sigma_x  + k_y \tau_0 \mu_0  \sigma_y) \nonumber \\
    &+ (\Delta - mk^2)  \xi_x \tau_0 \mu_0 \sigma_0,
    \label{eq:DiracRegulatedTI}
\end{align} which describes TBG-TSC.

For convenience we set $\tau_z=\mu_z = 1$.
The Green's function reads
\begin{align}
    & G(E,\kk) = (E - {\cal H}(\kk))^{-1} \\
    &= \dfrac{E + \xi_z (k_x \sigma_x + k_y \sigma_y) + (\Delta - mk^2)  \xi_x}{E^2 - k_x^2 - k_y^2 - (\Delta-mk^2)^2},
\end{align} and the terms odd in $k$ will vanish over the momentum integral required to obtain $G(E, \rr = 0) = \int_k G(E,\kk)$.
To prove that there are Shiba states, we follow the procedure performed in Ref.~\onlinecite{Slager:2015}.

\begin{figure}
    \centering
    \includegraphics[width=\columnwidth]{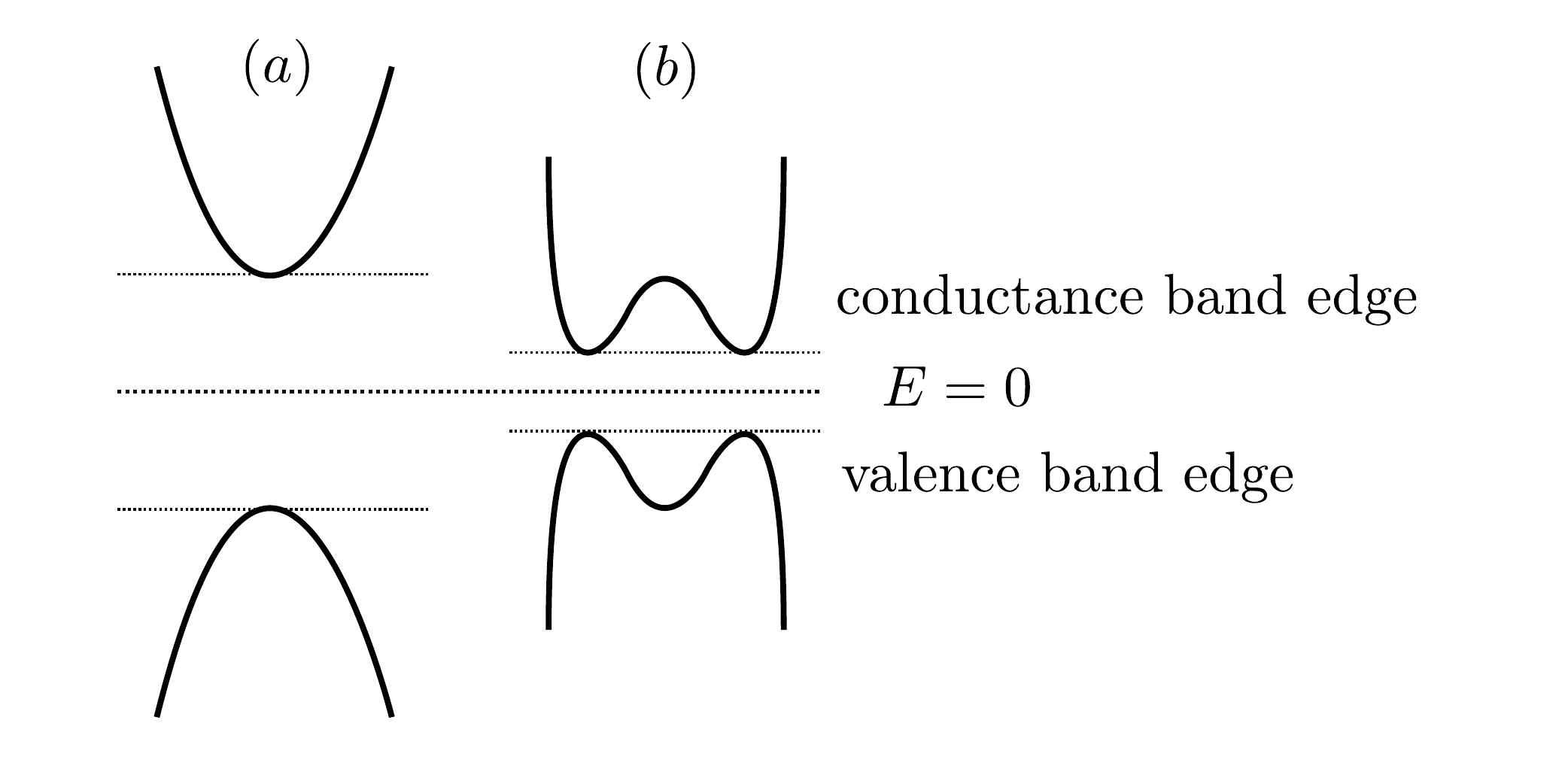}
    \caption{Locations where we evaluate the Green's function.  In (a), for $m<\frac{1}{2\Delta}$, the dispersion is parabolic.  In (b), for $m < \frac{1}{2\Delta}$, the dispersion is a Mexican hat.}
    \label{fig:GreenFcn}
\end{figure}

In Ref.~\onlinecite{Slager:2015}, the authors examine the Green's function at the edges of the positive and negative energy bands and show they may diverge to $\pm \infty$.
In certain cases one may prove the existence of Shiba states via the intermediate value theorem: if the Green's function diverges to $\infty$ at the positive energy band edge and $-\infty$ at the negative energy band edge, then for \emph{any} nonzero impurity strength $V$ Eq.~(\ref{eq:ShibaSolns}) has at least one solution in the gap.
We apply their method to the Dirac Hamiltonian with quadratic correction.
At the band edges, we expect van Hove singularities, though the exact location and nature depends upon the relative strength of $\Delta, m$.
As $k^2 + (\Delta - mk^2)^2 = \Delta^2 + (1-2m\Delta) k^2 + m^2 k^4$, the nature of the gap depends on the sign of $1-2m\Delta$, where we have set $v_F = 0$.
If $m < \frac{1}{2\Delta}$, then the dispersion about $k = 0$ is still parabolic.
If $m > \frac{1}{2\Delta}$, the dispersion at the band edge is actually about $k = k_c$ and resembles the Mexican hat potential, see Fig.~\ref{fig:GreenFcn} \footnote{We have set Fermi velocity $v_F = 1$.  Restoring $v_F$ gives the two limits for the mass as $m> v_F^2/(2\Delta)$ and $m<v_F^2/(2\Delta)$}.
In this situation, we actually expand around
\begin{align}
    \Delta^2 + (1-2m\Delta) k^2 &+ m^2 k^4 = m^2 (k^2 - k_c^2)^2 + \Delta'^2, \\
    k_c^2 &= \frac{1}{2m^2} (2m\Delta - 1) \\
    \Delta' &= \frac{1}{2m}\sqrt{4m\Delta - 1}.
\end{align}

Notice that the only Pauli matrix that enters in final equation for $G$ is $\xi_x$.
This means while the impurity potential $X$ may be composed of Pauli matrices $X = \xi_i \tau_j \mu_k \sigma_l$, the term $\tau_j \mu_k \sigma_l$ will not affect the integrals and can be chosen to be $\pm 1$.
We thus only need to consider the case with $X = \xi_0$ and $\xi_{x,y,z}$.
The Green's function has eigenvalues
\begin{align}
    &\int_k G(E, k) = g_0(E) + g_1(E) \xi_x \\
    \text{Eig}&[\int_k G(E, k)] = g_0(E) \pm g_1(E), \\
    g_0 &= \int \dfrac{k dk}{2\pi} \dfrac{E}{E^2 - k^2 - (\Delta - mk^2)^2} \label{eq:g0}\\
    g_1 &= \int \dfrac{k dk}{2\pi} \dfrac{\Delta - mk^2}{E^2 - k^2 - (\Delta - mk^2)^2}  \label{eq:g1}.
\end{align}

\subsection{$m < \frac{1}{2\Delta}$, Parabolic Dispersion}
As in Ref.~\onlinecite{Slager:2015}, we calculate $g_0, g_1$ at the band edges.
If $m < \frac{1}{2\Delta}$ (we have chosen $\Delta > 0$), the bands are parabolic and the band edges occur at $k = 0$.
We evaluate the Green's function at energy $E = \Delta - \delta \epsilon$ for the positive energy band and $E = -\Delta + \delta \epsilon$ for the negative energy band, $\delta \epsilon$ infinitesimal and positive.
The integrals are
\begin{align}
    g_0 &= \int_0^{\Lambda} \dfrac{k dk}{2\pi} \dfrac{\pm \Delta \mp \delta \epsilon}{-2\Delta \delta \epsilon - (1-2m\Delta)k^2 - m^2 k^4} \\
    g_1 &= \int_0^{\Lambda} \dfrac{k dk}{2\pi} \dfrac{\Delta - mk^2}{-2\Delta \delta \epsilon - (1-2m\Delta)k^2 - m^2 k^4}.
\end{align}
The upper bound $\Lambda$ is the UV cutoff, chosen to be much smaller than $|\mathbf{K}|$ with $\mathbf{K}$ the $C_{3z}$-invariant K point in the graphene Brillouin zone.
We have thrown away $\delta \epsilon ^2$ terms, as they are small.

Separating the integrals into $R_1, R_2$, and capturing leading divergences:
\begin{align}
    &R_1 = \int_0^{\Lambda} \dfrac{k dk}{2\pi} \dfrac{1}{-2\Delta \delta \epsilon - (1-2m\Delta)k^2 - m^2 k^4} \nonumber \\
    &\sim -\dfrac{1}{2\pi (1-2m\Delta)}\int_0^{0+} \dfrac{dx x}{2\Delta \delta \epsilon + x^2} \nonumber \\
    &\sim \dfrac{1}{4\pi (1-2m\Delta)} \log{\delta\epsilon} \qquad {\rightarrow -\infty}.
\end{align}
\begin{align}
    &R_2 = \int_0^{\Lambda} \dfrac{k dk}{2\pi} \dfrac{k^2}{-2\Delta \delta \epsilon - (1-2m\Delta)k^2 - m^2 k^4} \nonumber \\
    &= \text{finite}, <0
\end{align}
\begin{equation}
g_0 = \pm \Delta R_1 \mp \delta \epsilon R_1,~~g_1 = \Delta R_1 - m R_2.
\end{equation}
$R_2$ is negative as $\Delta > 0, (1-2m\Delta) > 0, m^2 > 0$.  We finally calculate the value of the eigenvalues of the Green's function at the band edges:
\begin{align}
    &{\text{conductance band edge:}} \\
    &\qquad {g_0 + g_1 \sim 2\Delta R_1 \rightarrow -\infty} \nono \\
    &\qquad {g_0 - g_1 = m R_2} \nono \\
    &{\text{valence band edge:}} \\
    &\qquad {g_0 + g_1 = -m R_2} \nono \\
    &\qquad {g_0 - g_1 \sim -2\Delta R_1 \nono \rightarrow \infty}
\end{align}
If $m > 0$, the intermediate value theorem guarantees solutions to the Shiba state equation Eq.~(\ref{eq:ShibaSolns}) for arbitrary impurity potential $V$.
If $ m < 0$, there is no guarantee, consistent with topological/trivial phases of the Dirac equation.
Again, in order for $m > 0$ and $m < 0$ to correspond to topological and trivial phases, we require that $|m| \Lambda^2 \gg |\Delta|$.
Without this condition, the analysis of the Shiba states still holds, but the two cases $m>0, m<0$ do not correspond to topological and trivial phases.

\begin{table*}
    \centering
    \begin{tabular}{c|c|c|c}
        Quantity & Positive energy band edge & Negative energy band edge & Band center \\
        \hline
        Parabolic dispersion, $m < 0$ & & &  \\
        $g_0 + g_1$ & $-\infty$ & $-mR_2 < 0$ & --- \\
        $g_0 - g_1$ & $mR_2 > 0$ & $\infty$ & --- \\
        $g_0^2 - g_1^2$ & $-\infty$ & $-\infty$ & $-\infty$ \\
        \hline
        Parabolic dispersion, $0 < m < \frac{1}{2\Delta}$ & & &  \\
        $g_0 + g_1$ & $-\infty$ & $-mR_2 > 0$ & --- \\
        $g_0 - g_1$ & $mR_2 < 0$ & $\infty$ & --- \\
        $g_0^2 - g_1^2$ & $\infty$ & $\infty$ & $-\infty$ \\
        \hline
        Mexican hat, $m > \frac{1}{2\Delta}$ & & &  \\
        $g_0 + g_1$ & $-\infty$ & $\infty$ & --- \\
        $g_0 - g_1$ & $-\infty$ & $\infty$ & --- \\
        $g_0^2 - g_1^2$ & $\infty$ & $\infty$ & $-\infty$
    \end{tabular}
    \caption{Eigenvalues/determinant of the Green's function evaluated at the band edges and band center $E = 0$.  Horizontal lines indicate that the value of said quantity is unknown, but unnecessary for general arguments.}
    \label{tab:shibaEigenvals}
\end{table*}

\subsection{$m > \dfrac{1}{2\Delta}$, Mexican Hat Dispersion}
We now consider the case $m > \frac{1}{2\Delta}$.
In this situation the gap resembles the Mexican hat potential illustrated in Fig.~\ref{fig:GreenFcn}.
We again expand about the conductance and valence band edges, using the formula for $g_0, g_1$ in Eq.~(\ref{eq:g0}), (\ref{eq:g1}), now located at energy $\pm \Delta'$ and momentum $k_c$:
\begin{align}
    g_0 &= \int_0^{\Lambda} \dfrac{k dk}{2\pi} \dfrac{\pm \Delta' \mp \delta \epsilon}{-2\Delta' \delta \epsilon - m^2 (k^2-k_c^2)^2} \\
    g_1 &= \int_0^{\Lambda} \dfrac{k dk}{2\pi} \dfrac{\Delta - mk^2}{-2\Delta' \delta \epsilon - m^2 (k^2-k_c^2)^2}.
\end{align}
with $mk_c^2 =  \Delta - \frac{1}{2m}, \Delta' = \sqrt{\frac{\Delta}{m} - \frac{1}{4m^2}}$.
Define
\begin{align}
    &R_3 = \int_0^{\Lambda} \dfrac{k dk}{2\pi} \dfrac{1}{-2\Delta' \delta \epsilon - m^2 (k^2-k_c^2)^2} \\
    &\sim \int_{k_c - \epsilon}^{k_c + \epsilon} \dfrac{k dk}{2\pi} \dfrac{1}{-2\Delta' \delta \epsilon - m^2 (k^2-k_c^2)^2} \\
    \sim &-\dfrac{1}{2\pi} \int_0^{0^+} du \dfrac{1}{2\Delta' \delta \epsilon + m^2 u^2} \\
    \propto & -\frac{1}{2\pi m} (2\Delta' \delta \epsilon)^{-1/2} \qquad \rightarrow -\infty.
\end{align}

The integrals become
\begin{align}
    g_0 &= (\pm \Delta' \mp \delta \epsilon) R_3,~~\\
    &\sim \pm \frac{1}{2m}\sqrt{4m\Delta - 1} R_3. \\
    g_1 &\sim \int_{k_c - \epsilon}^{k_c + \epsilon} \dfrac{k dk}{2\pi} \dfrac{\Delta - mk^2}{-2\Delta' \delta \epsilon - m^2 (k^2-k_c^2)^2} \nonumber \\
    &\approx \int_{k_c - \epsilon}^{k_c + \epsilon} \dfrac{k dk}{2\pi} \dfrac{\Delta - mk_c^2}{-2\Delta' \delta \epsilon - m^2 (k^2-k_c^2)^2} \\
    &= \frac{1}{2m}R_3.
\end{align}
As $\Delta - mk_c^2 = \frac{1}{2m}$, $\Delta' = \frac{1}{2m}\sqrt{4m\Delta - 1}$.
We thus have
\begin{align}
    &{\text{conductance band edge:}} \\
    &{g_0 + g_1 \sim (\frac{1}{2m}[\sqrt{4m\Delta - 1} + 1]) R_3 \rightarrow -\infty} \\
    &{g_0 - g_1 \sim (\frac{1}{2m}[\sqrt{4m\Delta - 1} - 1]) R_3 \rightarrow -\infty} \\
    &{\text{valence band edge:}} \\
    &{g_0 + g_1 \sim (\frac{1}{2m}[-\sqrt{4m\Delta - 1} + 1]) R_3 \rightarrow \infty} \\
    &{g_0 - g_1 \sim (\frac{1}{2m}[-\sqrt{4m\Delta - 1} - 1]) R_3 \rightarrow \infty},
\end{align}
as $m > \frac{1}{2\Delta}$, $\sqrt{4m\Delta - 1} > 1$.

\subsection{Shiba States}
We are now in a position to argue the existence/nonexistence of zero modes.
For impurity potential $V X = V \xi_0 \tau_i \mu_j \sigma_k$, for arbitrary $i, j, k$, the impurity violates chiral symmetry $S$ and corresponds to a time-reversal breaking perturbation.
In both scenarios of $m\Delta > 0$ (either in the parabolic or Mexican hat dispersion) the eigenvalues of the Green's function both go through $0$ somewhere inside the gap.
Contrast this to $m\Delta < 0$, which never reaches $0$ inside the gap.
Eq.~(\ref{eq:ShibaSolns}) may still have solutions.  However, similar to Ref.~\onlinecite{Slager:2015},  increasing $V$ arbitarily high will drive the Shiba states into the bulk modes.
For the case with $m\Delta > 0$, however, regardless of how strongly we increase $V$ since $G$ has a zero there is always a solution.  This is a qualitative difference between the two cases.

For impurities  $X = \xi_z \tau_i \mu_j \sigma_k$, however, the situation greatly changes.
We are required to solve the equation (taking $\tau_i \mu_j \sigma_k= +1)$
\begin{align}
    \text{Det}[V^{-1} - G_0(E, {\bf r}=0)\xi_z] = 0,
    \label{eq:ShibaModified}
\end{align} and $G_0(E, {\bf r}=0)\xi_z$ is not Hermitian!  Its eigenvalues are no longer real, and we are not guaranteed solutions even if $G$ crosses $0$.

Solving Eq.~(\ref{eq:ShibaModified}) yields
\begin{align}
    V^{-2} = g_0^2 - g_1^2,
\end{align}
with $g_0, g_1$ defined as before in Eq.~(\ref{eq:g0}), (\ref{eq:g1}).
To find zero modes, observe that at $E= 0$ $g_0 = 0$ (trivially by Eq.~(\ref{eq:g0})), so the quantity $g_0^2 - g_1^2$ is negative {at the band center $E = 0$}.
In the case $m\Delta < 0$, $g_0^2 - g_1^2 = (g_0 + g_1)(g_0 - g_1)$ diverges to $-\infty$ in both conductance and valence edges.
In fact it is negative throughout the gap, and thus there are no solutions.
In the case $m\Delta > 0$, $g_0^2 - g_1^2 = (g_0 + g_1)(g_0 - g_1)$ diverges to $\infty$ in both conductance and valence edges.
Thus two in-gap modes are found, at $\pm E'$ for some $E'$, again regardless of impurity strength.
The quantity $g_0^2 - g_1^2$ is easily calculated by multiplying together the eigenvalues $ (g_0 + g_1),(g_0 - g_1)$, calculated for the parabolic and Mexican hat dispersions.  We have collected these results in Table~\ref{tab:shibaEigenvals}.

\subsection{Multiple Valleys}
We have demonstrated that zero modes exist for arbitrary impurity strengths for $m\Delta > 0$, while for $m\Delta < 0$ Shiba states will either not exist or can be pushed into the bulk with decreasing impurity strength.
As calculated in App.~\ref{app:uniform-s}, in TBG-TSC with nearly uniform pairing we have $m\Delta < 0$, where the on-site pairing strength is much greater than the NNN pairing strength, $\bar{\Delta}_0 \gg \bar{\Delta}_2$.
However, in TBG-TSC the observation of Shiba states is complicated by the presence of multiple valleys and zero modes.
A delta-function impurity itself will break valley-U(1) symmetry, allowing for the zero modes of different valleys to hybridize.

Take the real-space perturbation
\begin{align}
    & {~~~~~~m \sum_{\alpha,l,s} \sum_{\RR \in l} \delta_{\RR,0} c^\dagger_{\RR,l,\alpha,s} c_{\RR,l,\alpha,s}} \\
    &= {\frac{m}{N_M N_0} \sum_{\alpha,l,s,\eta, \eta', \kk, \kk', \QQ, \QQ'} e^{-i(\eta \mathbf{K}_l + \kk - \QQ)\cdot \mathbf{t}_\alpha}} \nonumber \\
    &{\times e^{i(\eta' \mathbf{K}_l + \kk' - \QQ')\cdot \mathbf{t}_\alpha}  c^\dagger_{\kk,\QQ,\eta,\alpha,s} c_{\kk',\QQ',\eta',\alpha,s}}.
\end{align}
We see that this perturbation mixes the valleys, and thus require that the perturbation not be a delta-function impurity, but ramp up gradually over the length scale of the graphene lattice (but quickly on the scale of the Moir\'e lattice).

\section{Bosonization}
\label{app:bosonization}
We examine the fate of the gapless edge modes along the $x$-direction (Section~\ref{app:edge-state-Dirac} in the presence of strong interactions.
We begin with an edge perpendicular to the x-axis.
As the dispersion is linear in free fermions and the edge is one-dimensional, interactions can be captured via bosonization.
Under the symmetries of TBG-TSC, the fermionic movers listed in Eqs.~(\ref{eq:fermionMoversFirst})-(\ref{eq:fermionMovers}) transform as
\begin{align}
  U(1)_\text{valley} &: \psi_{\alpha n s} \rightarrow e^{i\gamma} \psi_{\alpha n s} \label{eq:boson_symm_1} \\
  U(1)_\text{spin} &: \psi_{\alpha n s} \rightarrow e^{is\gamma} \psi_{\alpha n s} \\
  T &: \psi_{\alpha n s} \rightarrow i s \psi_{-\alpha -n -s}^\dagger \label{eq:T_dirac_symm} \\
  i\hat{s}_y &: \psi_{\alpha n s} \rightarrow s \psi_{\alpha n -s}  \\
  C_{2x} &: \psi_{\alpha n s} \rightarrow \psi_{\alpha -n s} \\
  {\cal P}_c &: \psi_{\alpha n s} \rightarrow i \alpha n \psi_{-\alpha n s}^\dagger \label{eq:boson_symm_2},
\end{align}
with $\alpha = 1,2, n = R,L, s = \uparrow,\downarrow$, where again $\PH_c$ has been defined in Eq.~\eqref{eq:Pc-on-BdG-basis}.
There are $8$ total modes: two pairs of helical modes per valley.
Note that using the BdG basis has turned $T$ into a BdG symmetry (Eq.~\eqref{eq:T_dirac_symm}).
For example:
\begin{widetext}
\begin{align}
\psi_{1R\uparrow} (y) &= \frac{\sqrt{\lambda}}{2} \int dx e^{-\lambda |x|}
    ( c_{+,+,A,\uparrow}(x,y) -i c_{+,+,B,\uparrow}(x,y) -
      c^\dagger_{-,-,A,\downarrow}(x,y) - i c^\dagger_{-,-,B,\downarrow}(x,y) ) \\
T \psi_{1R\uparrow} (y) T^{-1} &= \frac{\sqrt{\lambda}}{2} \int dx e^{-\lambda |x|}
    ( c_{-,-,A,\uparrow}(x,y) +i c_{-,-,B,\uparrow}(x,y) -
      c^\dagger_{+,+,A,\downarrow}(x,y) + i c^\dagger_{+,+,B,\downarrow}(x,y) ) \nono\\
      &= i \psi^\dagger_{2L\downarrow} (y).
\end{align}
\end{widetext}

The kinetic energy has a linear dispersion, and is simply
\begin{align}
    H_\text{kinetic} = \sum_{\alpha ns} \int_y -i n v_F \psi_{\alpha ns}^\dagger \partial_y \psi_{\alpha ns}.
\end{align}
$C_{2x}$ is preserved as it switches $y \rightarrow -y$.
We first consider the possible bilinears allowed under the above symmetries, and determine under what conditions the edge can be fully gapped.
The kinetic term has spin rotation invariance, but when we add interactions we will not enforce full spin-$SU(2)$ invariance, and work with the simpler $i\hat{s}_y$ rotation for ease in bosonization.
By spin and valley symmetry, we can add
\begin{align}
    H_\text{gap} = \sum_{\alpha \beta mn} \int_y a_{\alpha \beta mn} ( \psi^\dagger_{\alpha m \uparrow} \psi_{\beta n \uparrow} &+ \psi^\dagger_{\alpha m \downarrow} \psi_{\beta n \downarrow}) \nonumber \\
    &+ H.c.,
\end{align}
and the remaining symmetries further constrain the form of $a$.
We further restrict ourselves to bilinears that scatter right-movers to left-movers, as terms mixing two right movers and two left movers do not open up gaps in the spectrum.
The effect of such scattering terms that hop between modes of the same chirality, e.g. $\psi_{1L\uparrow}^\dagger \psi_{2L\uparrow}$, will only shift the linearized spectrum to the left or right in momentum space.  They will not create gaps, so we ignore them.
\begin{align}
    H_\text{gap} &= \sum_{\alpha \beta} \int_y a_{\alpha \beta} ( \psi^\dagger_{\alpha R \uparrow} \psi_{\beta L \uparrow} + \psi^\dagger_{\alpha R \downarrow} \psi_{\beta L \downarrow}) \nonumber \\
    &+ a_{\alpha \beta}^* (\psi^\dagger_{\beta L \uparrow} \psi_{\alpha R \uparrow} + \psi^\dagger_{\beta L \downarrow} \psi_{\alpha R \downarrow}),
\end{align} and under $T$ and ${\cal P}_c$:
\begin{align}
    T H_\text{gap} T^{-1} &= \sum_{\alpha \beta} \int_y -a_{\alpha \beta}^* ( \psi^\dagger_{-\beta R \downarrow} \psi_{-\alpha L \downarrow} \nonumber \\
    &+ \psi^\dagger_{-\beta R \uparrow} \psi_{-\alpha L \uparrow}) + H.c.\\
    {\cal P}_c H_\text{gap} {\cal P}_c^{-1} &= \sum_{\alpha \beta} \int_y \alpha \beta a_{\alpha \beta} ( \psi^\dagger_{-\beta L \uparrow} \psi_{-\alpha R \uparrow} \nonumber \\
    &+ \psi^\dagger_{-\beta L \downarrow} \psi_{-\alpha R \downarrow}) + H.c.,
\end{align} which constrains
\begin{align}
    a_{\alpha \beta} = -a_{-\beta-\alpha}^* = \alpha\beta a_{-\alpha-\beta}^*.
\end{align}
The diagonal entries with $\alpha\beta = 1$ vanish, and the off-diagonal entries $a_{1,-1} = -a_{1,-1}^* = -a_{-1,1}^*$.
Thus
\begin{align}
    H_\text{gap} = \int_y ia(\psi^\dagger_{1R\uparrow} \psi_{2L\uparrow} + \psi^\dagger_{2R\uparrow} \psi_{1L\uparrow}) + H.c. + (\uparrow \leftrightarrow \downarrow).
\end{align}
This term is odd under $C_{2x}$.
Thus, if the edge preserves $C_{2x}$ symmetry, it is gapless on the level of free fermions, as we know from our symmetry analysis in App.~\ref{app:edge-state-Dirac}.

To study the gapless edge in the presence of interactions, we bosonize, following Ref.~\onlinecite{Fisher:1996}.
Take
\begin{align}
\psi_{\alpha n s} \sim e^{i(\varphi_{\alpha s} + n\theta_{\alpha s})},
\end{align} with bosonized fields $\varphi, \theta$ obeying
\begin{align}
    [\varphi_{\alpha s}(y), \theta_{\beta s'}(y')] = \frac{i\pi}{2} \delta_{\alpha \beta} \delta_{ss'}\text{sgn} (y-y'),
\end{align} to ensure anti-commutation between fermions at different positions.
We do not enforce translation symmetry, meaning that we will not concern ourselves with momentum conservation when studying gap terms that may appear in the interacting Hamiltonian.

This yields
\begin{align}
  &U(1)_\text{valley} : \varphi_{\alpha s} \rightarrow \varphi_{\alpha s} + \gamma,  ~\theta_{\alpha s} \rightarrow \theta_{\alpha s} \label{eq:bosonizationRules1}\\
  &U(1)_\text{spin} :  \varphi_{\alpha s} \rightarrow \varphi_{\alpha s} + s\gamma,~\theta_{\alpha s} \rightarrow \theta_{\alpha s}\\
&T : \varphi_{\alpha s} \rightarrow \varphi_{-\alpha -s} - \frac{\pi}{2}s,  ~\theta_{\alpha s} \rightarrow -\theta_{-\alpha -s} \\
  &i\hat{s}_y : \varphi_{\alpha s} \rightarrow \varphi_{\alpha -s} - \frac{\pi}{2} (1-s),  ~\theta_{\alpha s} \rightarrow \theta_{\alpha -s} \\
  &C_{2x} : \varphi_{\alpha s} (y)\rightarrow \varphi_{\alpha s}(-y), ~\theta_{\alpha s} (y) \rightarrow -\theta_{\alpha s} (-y) \\
  &{\cal P}_c : \varphi_{\alpha s} \rightarrow -\varphi_{-\alpha s},  ~\theta_{\alpha s} \rightarrow -\theta_{-\alpha s} + \frac{\pi}{2}\alpha \text{~(incorrect)} \label{eq:bosonizationRulesN}.
\end{align}

However, there is a subtlety regarding the transformation law for $\PH_c$.  It is not consistent with the bosonization scheme: consider a bilinear which bosonizes as

\begin{align}
    e^{i\gamma} & \psi^\dagger_{\alpha m s} \psi_{\beta n s} + H.c. \nonumber \\
    \sim&  \sin(-\varphi_{\alpha s} -m\theta_{\alpha s} + \varphi_{\beta s} + n\theta_{\beta s} + \gamma).
\end{align}
The bilinears transform under $\PH_c$ as:
\begin{widetext}
\begin{align}
    {\cal P}_c: -e^{i\gamma} & \alpha\beta mn \psi^\dagger_{-\beta n s} \psi_{-\alpha m s} + H.c. \sim -\sin(-\varphi_{-\beta s} -n\theta_{-\beta s} + \varphi_{-\alpha s} + m\theta_{-\alpha s} + \gamma + \frac{\pi}{2}(\alpha m - \beta n)) \label{eq:ConsistencyFermP}.
\end{align}
However, directly transforming the bosonized cosines gives
\begin{align}
    \PH_c:& \sin(\varphi_{-\alpha s} + m\theta_{-\alpha s} - \varphi_{-\beta s} - n\theta_{-\beta s} + \gamma - \frac{\pi}{2}(m\alpha - n\beta)) \label{eq:ConsistencyBosP}.
\end{align}
\end{widetext}
These two results differ by an overall a minus sign. Because we concern ourselves only with four-fermion cosines, there will always be two pairs of fermions, so the overall sign of the quartic form is restored.  However, it is interesting to see how to resolve the apparent inconsistency in the $\PH_c$ symmetry.  We resolve this issue below, by introducing additional transformation properties that resemble Klein factors, as performed in Ref.~\cite{Mross_2017}.  Define an ordering on the 8 flavors of fermion operators $\psi_{\alpha n s}$ by
\begin{align}
    f(\alpha, n, s) = j, ~ j \in \{0,1,\cdots 7\}. \\
    f(1, R, \uparrow) = 0, f(1, L, \uparrow) = 1, \\
    f(2, R, \uparrow) = 2, f(2, L, \uparrow) = 3, \\
    f(1, R, \downarrow) = 4, f(1, L, \downarrow) = 5, \\
    f(2, R, \downarrow) = 6, f(2, L, \downarrow) = 7.
\end{align}
Define operator $\xi_j$ as
\begin{align}
    \xi_j &= 2\pi \sum_{j' < j} N_{j'}, \\
    [\xi_j, \varphi_{\alpha s} + n\theta_{\alpha s}] &= 2\pi i \Theta(j - f(\alpha, n, s)- 0.5) \\
    N_{j'} =\int_y \psi^\dagger_{\alpha n s} \psi_{\alpha n s} &, ~j' = f(\alpha, n, s).
\end{align} Here $\xi_j$ is a Jordan-Wigner type string that plays a role similar to a Klein factor, giving anti-commutation of different flavors of fermions.  This $\xi_j$ will introduce an additional negative sign that fixes the transformation rules for $\PH_c$.
\begin{align}
    \PH_c: \varphi_{\alpha s} + n\theta_{\alpha s} \rightarrow & -(\varphi_{-\alpha s} + n\theta_{-\alpha s}) \nonumber \\
    &+ \frac{\pi}{2}\alpha n + \xi_{f(-\alpha,n,s)}.
\end{align}
Under this new transformation, the $\PH_c$ transformed cosine becomes
\begin{widetext}
\begin{align}
    \PH_c:& \sin(\varphi_{-\alpha s} + m\theta_{-\alpha s} - \varphi_{-\beta s} - n\theta_{-\beta s} + \gamma - \frac{\pi}{2}(m\alpha - n\beta) - \xi_{f(-\alpha,m,s)} + \xi_{f(-\beta,n,s)}),
\end{align} which using the BCH formula,
\begin{align}
    &= \frac{1}{2i} [e^{i(\varphi_{-\alpha s} + m\theta_{-\alpha s} - \varphi_{-\beta s} - n\theta_{-\beta s} + \gamma - \frac{\pi}{2}(m\alpha - n\beta) - \xi_{f(-\alpha,m,s)} + \xi_{f(-\beta,n,s)})} - H.c.], \\
    &= \frac{1}{2i} [e^{i(\varphi_{-\alpha s} + m\theta_{-\alpha s} - \varphi_{-\beta s} - n\theta_{-\beta s} + \gamma - \frac{\pi}{2}(m\alpha - n\beta))}e^{i(-\xi_{f(-\alpha,m,s)} + \xi_{f(-\beta,n,s)})} \nonumber \\
    & e^{\frac{1}{2}[(\varphi_{-\alpha s} + m\theta_{-\alpha s} - \varphi_{-\beta s} - n\theta_{-\beta s} + \gamma - \frac{\pi}{2}(m\alpha - n\beta)),(-\xi_{f(-\alpha,m,s)} + \xi_{f(-\beta,n,s)})] } - H.c.] \\
    &= \frac{1}{2i} [e^{i(\varphi_{-\alpha s} + m\theta_{-\alpha s} - \varphi_{-\beta s} - n\theta_{-\beta s} + \gamma - \frac{\pi}{2}(m\alpha - n\beta))} \nonumber \\
    & e^{\frac{1}{2}([\varphi_{-\alpha s} + m\theta_{-\alpha s}, \xi_{f(-\beta,n,s)})] + [\varphi_{-\beta s} + n\theta_{-\beta s}, \xi_{f(-\alpha,m,s)}]) } - H.c.] \\
    &= -\sin(\varphi_{-\alpha s} + m\theta_{-\alpha s} - \varphi_{-\beta s} - n\theta_{-\beta s} + \gamma - \frac{\pi}{2}(m\alpha - n\beta))
\end{align}
\end{widetext} where in the third line we have used $e^{i\xi_j} = 1$ (as $N_j$ are integer-valued), and in the last line we have used that $[\xi_{f(\alpha,n,s)}, \varphi_{\alpha' s'} + n' \theta_{\alpha' s'}] + [\xi_{f(\alpha',n',s')}, \varphi_{\alpha s} + n \theta_{\alpha s}] = 2\pi i( \Theta (f(\alpha, n, s) - f(\alpha', n', s') - 0.5) + \Theta (f(\alpha', n', s') - f(\alpha, n, s) - 0.5)) = 2\pi i$ (when $f(\alpha,n,s) \neq f(\alpha',n',s')$), so the commutator yields the required sign $-1$.
Adding the $\xi$ factor fixes the sign in the transformation law for bilinears.  We will, however, only consider 4-fermion interactions (there are no cosine bilinears in our problem), so the problematic sign fixes itself (as two negative signs cancel).

Now we add interactions.
According to the rules of bosonization,
\begin{align}
    n_{\alpha m s} &= \psi_{\alpha m s}^\dagger \psi_{\alpha m s} - \langle \psi_{\alpha m s}^\dagger \psi_{\alpha m s} \rangle \\
    &= \dfrac{\partial_y (m\varphi_{\alpha s} + \theta_{\alpha s})}{2\pi}.
\end{align}
Beginning with the bosonized Hamiltonian for a free wire:
\begin{align}
H_\text{kin} =\sum_{\alpha s} \frac{v_F}{2\pi} \int_{y} ( \partial_y \varphi_{\alpha s})^2 + (\partial_y \theta_{\alpha s})^2,
\label{LuttingerHam}
\end{align} we add symmetric density-density interactions (obeying Eq.~\ref{eq:boson_symm_1}-\ref{eq:boson_symm_2}) of the form
\begin{align}
& u_1 [n_{1L\uparrow} n_{1R\uparrow} + n_{1L\downarrow} n_{1R\downarrow} + n_{2L\uparrow} n_{2R\uparrow} + n_{2L\downarrow} n_{2R\downarrow}], \nonumber \\
& u_2 [n_{1L\uparrow} n_{1R\downarrow} + n_{1L\downarrow} n_{1R\uparrow} + n_{2L\uparrow} n_{2R\downarrow} + n_{2L\downarrow} n_{2R\uparrow}], \nonumber \\
& u_3 [n_{1L\uparrow} n_{1L\downarrow} + n_{1R\uparrow} n_{1R\downarrow} + n_{2L\uparrow} n_{2L\downarrow} + n_{2R\uparrow} n_{2R\downarrow}], \nonumber \\
& u_4 [n_{1L\uparrow} n_{2L\uparrow} + n_{1R\uparrow} n_{2R\uparrow} + n_{1L\downarrow} n_{2L\downarrow} + n_{1R\downarrow} n_{2R\downarrow}], \nonumber \\
& u_5 [n_{1L\uparrow} n_{2R\uparrow} + n_{1L\downarrow} n_{2R\downarrow} + n_{1R\uparrow} n_{2L\uparrow} + n_{1R\downarrow} n_{2L\downarrow}], \nonumber \\
& u_6 [n_{1L\uparrow} n_{2L\downarrow} + n_{1L\downarrow} n_{2L\uparrow} + n_{1R\uparrow} n_{2R\downarrow} + n_{1R\downarrow} n_{2R\uparrow}], \nonumber \\
& u_7 [n_{1L\uparrow} n_{2R\downarrow} + n_{1L\downarrow} n_{2R\uparrow} + n_{1R\uparrow} n_{2L\downarrow} + n_{1R\downarrow} n_{2L\uparrow}].
\end{align}
These interactions renormalize the quadratic terms in bosonized language.
For example,
\begin{align}
    & u_1 [n_{1L\uparrow} n_{1R\uparrow} + n_{1L\downarrow} n_{1R\downarrow} + n_{2L\uparrow} n_{2R\uparrow} + n_{2L\downarrow} n_{2R\downarrow}] \nonumber \\
    &= \frac{u_1}{4\pi^2} [(\partial_y(\varphi_{1\uparrow} + \theta_{1\uparrow}))(\partial_y (-\varphi_{1\uparrow} + \theta_{1\uparrow})) \nonumber \\
    &+ (1 \leftrightarrow 2) + (\uparrow \leftrightarrow \downarrow)] \nonumber \\
    &= \frac{u_1}{4\pi^2}  [-(\partial_y \varphi_{1\uparrow})^2 + (\partial_y \theta_{1\uparrow})^2 + (1 \leftrightarrow 2) + (\uparrow \leftrightarrow \downarrow)].
\end{align}
In a similar calculation to Ref.~\onlinecite{Isobe:2015}, we gather all quadratic fields into matrices $M_{\varphi, \theta}$.
\begin{align}
H_\text{int} = \frac{1}{4\pi^2} \int_x
\partial_x {\vec \varphi}^T  M_\varphi \partial_x {\vec \varphi} + \partial_x {\vec \theta}^T  M_\theta \partial_x {\vec \theta},
\end{align} with
\begin{align}
\partial_x {\vec \varphi}^T = \begin{bmatrix}
\partial_x\varphi_{1\uparrow} & \partial_x\varphi_{2\uparrow} & \partial_x\varphi_{1\downarrow} & \partial_x\varphi_{2\downarrow}
\end{bmatrix}, \\
\partial_x {\vec \theta}^T = \begin{bmatrix}
\partial_x\theta_{1\uparrow} & \partial_x\theta_{2\uparrow} & \partial_x\theta_{1\downarrow} & \partial_x\theta_{2\downarrow}
\end{bmatrix}.
\end{align}
The matrices are given by
\begin{align}
M_{\varphi / \theta} = \begin{bmatrix}
\mp u_1 & u_4 \mp u_5 & \mp u_2 + u_3 & u_6 \mp u_7 \\
u_4 \mp u_5 & \mp u_1 & u_6 \mp u_7 & \mp u_2 + u_3 \\
\mp u_2 + u_3 & u_6 \mp u_7 & \mp u_1 & u_4 \mp u_5 \\
u_6 \mp u_7 & \mp u_2 + u_3 & u_4 \mp u_5 & \mp u_1
\end{bmatrix}.
\end{align}
It will be convenient for us to switch into the spin-charge basis by defining
\begin{align}
\varphi_c &= (\varphi_{1\uparrow} + \varphi_{2\uparrow} + \varphi_{1\downarrow} + \varphi_{2\downarrow})/2 \label{eq:newBasis1} \\
\varphi_s &= (\varphi_{1\uparrow} + \varphi_{2\uparrow} - \varphi_{1\downarrow} - \varphi_{2\downarrow})/2 \\
\varphi_a &= (\varphi_{1\uparrow} - \varphi_{2\uparrow} + \varphi_{1\downarrow} - \varphi_{2\downarrow})/2 \\
\varphi_b &= (\varphi_{1\uparrow} - \varphi_{2\uparrow} - \varphi_{1\downarrow} + \varphi_{2\downarrow})/2 \\
\theta_c &= (\theta_{1\uparrow} + \theta_{2\uparrow} + \theta_{1\downarrow} + \theta_{2\downarrow})/2 \\
\theta_s &= (\theta_{1\uparrow} + \theta_{2\uparrow} - \theta_{1\downarrow} - \theta_{2\downarrow})/2 \\
\theta_a &= (\theta_{1\uparrow} - \theta_{2\uparrow} + \theta_{1\downarrow} - \theta_{2\downarrow})/2 \\
\theta_b &= (\theta_{1\uparrow} - \theta_{2\uparrow} - \theta_{1\downarrow} + \theta_{2\downarrow})/2 \label{eq:newBasisN},
\end{align} which obeys the same commutation relations as the original $\varphi, \theta$ variables.
This basis was obtained by diagonalizing the $M_{\varphi, \theta}$ matrix.
In this new basis the Hamiltonian reads
\begin{align}
H_\text{kin}+H_\text{int} = \sum_i \frac{v_i}{2\pi} \int_y K_i (\partial_y \varphi_i)^2 + \frac{1}{K_i} (\partial_y \theta_i)^2,
\end{align}
with$v_i K_i$ and $v_i/K_i$ being the eigenvalues of
\begin{align}
    v_F + 2\pi M_\varphi, v_F + 2\pi M_\theta.
\end{align}
\begin{align}
v_c &= \frac{1}{2\pi}[(2\pi v_F - u_1 - u_2 + u_3 + u_4 - u_5 + u_6 - u_7) \nonumber \\
& (2\pi v_F + u_1 + u_2 + u_3 + u_4 + u_5 + u_6 + u_7)]^{1/2}, \nonumber \\
K_c &= \sqrt{\dfrac{(2\pi v_F - u_1 - u_2 + u_3 + u_4 - u_5 + u_6 - u_7)}{(2\pi v_F + u_1 + u_2 + u_3 + u_4 + u_5 + u_6 + u_7)}} \nonumber \\
v_s &= \frac{1}{2\pi}[(2\pi v_F - u_1 + u_2 - u_3 + u_4 - u_5 - u_6 + u_7) \nonumber \\
& (2\pi v_F + u_1 - u_2 - u_3 + u_4 + u_5 - u_6 - u_7)]^{1/2}, \nonumber \\
K_s &= \sqrt{\dfrac{(2\pi v_F - u_1 + u_2 - u_3 + u_4 - u_5 - u_6 + u_7)}{(2\pi v_F + u_1 - u_2 - u_3 + u_4 + u_5 - u_6 - u_7)}} \nonumber \\
v_a &= \frac{1}{2\pi}[(2\pi v_F - u_1 - u_2 + u_3 - u_4 + u_5 - u_6 + u_7) \nonumber \\
& (2\pi v_F + u_1 + u_2 + u_3 - u_4 - u_5 - u_6 - u_7)]^{1/2}, \nonumber \\
K_a &= \sqrt{\dfrac{(2\pi v_F - u_1 - u_2 + u_3 - u_4 + u_5 - u_6 + u_7)}{(2\pi v_F + u_1 + u_2 + u_3 - u_4 - u_5 - u_6 - u_7)}} \nonumber \\
v_b &= \frac{1}{2\pi}[(2\pi v_F - u_1 + u_2 - u_3 - u_4 + u_5 + u_6 - u_7) \nonumber \\
& (2\pi v_F + u_1 - u_2 - u_3 - u_4 - u_5 + u_6 + u_7)]^{1/2}, \nonumber \\
K_b &= \sqrt{\dfrac{(2\pi v_F - u_1 + u_2 - u_3 - u_4 + u_5 + u_6 - u_7)}{(2\pi v_F + u_1 - u_2 - u_3 - u_4 - u_5 + u_6 + u_7)}}.
\end{align}

The previous symmetry-preserving density-density interactions simply renormalized the Luttinger parameters.
To fully gap the Luttinger liquid, one must add symmetric commuting cosines involving $\varphi_i, \theta_i$.
Notice that the effect of valley-U(1) symmetry is to shift $\varphi_c \rightarrow \varphi_c + 2\gamma$, with $\gamma$ the phase factor.
No other bosonized variables in Eqs.~(\ref{eq:newBasis1})-(\ref{eq:newBasisN}) are altered!
Hence $\varphi_c$ cannot appear in a cosine -- including $\varphi_c$ would violate valley charge conservation.
The same holds true for $\varphi_s$ and spin-U(1) conservation.
Because $\varphi_c, \varphi_s$ cannot appear in any cosines,  in the limit of strong pairing their conjugate variables $\theta_c, \theta_s$ may be pinned, for example via the terms:
\begin{align}
&~ w_1 [\cos(2\theta_s + 2\theta_c) + \cos(2\theta_s - 2\theta_c)] \nonumber \\
\propto &~ w_1 [\psi_{1R\uparrow}^\dagger \psi_{2R\uparrow}^\dagger \psi_{2L\uparrow} \psi_{1L\uparrow} + (\uparrow\leftrightarrow\downarrow) + H.c.]
\end{align}
Note that lower harmonics of $\theta_c, \theta_s$, such as $\cos(2\theta_s)$, cannot be expressed in terms of fermions.
For example
\begin{align}
    \cos(2\theta_s) = \cos(\theta_{1\uparrow} + \theta_{2\uparrow} - \theta_{1\downarrow} - \theta_{2\downarrow}),
\end{align} while a generic hopping term is constructed out of $\psi$ operators that bosonize as $e^{i(\varphi_{\alpha s} \pm \theta_{\alpha s})}$.  This implies, for a gap term reading
\begin{align}
    \cos (\sum_{\alpha s} (m_{\alpha s} \varphi_{\alpha s} + n_{\alpha s} \theta_{\alpha s})),
\end{align}
that the parity of the $\varphi$ and $\theta$ coefficients must be equal, that is:
\begin{align}
    m_{\alpha s} = n_{\alpha s} ~\text{mod}~2.
\end{align}
$\cos(2\theta_s)$ clearly violates this constraint.
With the charge and spin sectors gapped, we gap out the $a, b$ modes.
There are many choices; for example, gapping with the $\theta$ terms
\begin{align}
&~w_2[\cos{(2\theta_a + 2\theta_b)} + \cos{(2\theta_a - 2\theta_b)}], \nonumber \\
\propto &~w_2 [\psi_{1R\uparrow}^\dagger \psi_{2L\uparrow}^\dagger \psi_{2R\uparrow} \psi_{1L\uparrow} + (\uparrow\leftrightarrow\downarrow) + H.c.],
\end{align} or the $\varphi$ terms,
\begin{align}
&~w_3[\cos{(2\varphi_a + 2\varphi_b)} + \cos{(2\varphi_a - 2\varphi_b)}], \nonumber \\
\propto &~w_3 [\psi_{1R\uparrow}^\dagger \psi_{1L\uparrow}^\dagger \psi_{2L\uparrow} \psi_{2R\uparrow} + (\uparrow\leftrightarrow\downarrow) + H.c.],
\end{align} or a mix of both,
\begin{align}
&~w_4[\cos{(2\theta_a + 2 \varphi_b)} + \cos{(2\theta_a - 2 \varphi_b)}], \nonumber \\
\propto &~w_4 [\psi_{1L\downarrow}^\dagger \psi_{2R\uparrow}^\dagger \psi_{2L\downarrow} \psi_{1R\uparrow} + (\uparrow\leftrightarrow\downarrow) + H.c.], \\
&~w_5[\cos{(2\theta_b + 2\varphi_a)} + \cos{(2\theta_b - 2\varphi_a)}], \nonumber \\
\propto &~w_5 [\psi_{2L\downarrow}^\dagger \psi_{2R\uparrow}^\dagger \psi_{1R\uparrow} \psi_{1L\downarrow} + (\uparrow\leftrightarrow\downarrow) + H.c.].
\end{align}
The charge and spin sectors can be gapped with the $w_1$ cosine, while we have a choice of different gap terms to choose to gap out the $a, b$ sectors.
All five terms respect the symmetries of the problem, and for any given choice of $w_{2,3,4,5}$, we can pin both the $a$ and $b$ sectors, whether by $\varphi$ or $\theta$.
The cosines commute with the $w_1$ cosine, and so in principle it is possible to fully gap out the $C_{2x}$-symmetric wire.

As the Luttinger Hamiltonian is quadratic in these new variables Eq.~\ref{eq:newBasisN}, the scaling of the cosine terms follows from standard analysis of the sine-Gordon model, see Ref.~\onlinecite{Giamarchi:2004}.

Our RG flow equations thus read
\begin{align}
\frac{dw_1}{dl} &= (2 - K_s - K_c) w_1 \\
\frac{dw_2}{dl} &= (2 - K_a - K_b)w_2,~~\frac{dw_3}{dl} = (2 - \frac{1}{K_a} - \frac{1}{K_b})w_3 \\
\frac{dw_4}{dl} &= (2 - K_a - \frac{1}{K_b})w_4,~~\frac{dw_5}{dl} = (2 - \frac{1}{K_a} - K_b)w_5.
\end{align}
Regardless of the Luttinger parameters $K_a, K_b$, there will always be a choice (except for the point where $K_a, K_b = 1$ that makes at least one of the terms $w_{2,3,4,5}$ relevant; thus these sectors will always be gapped.
The charge and spin sectors will be gapped so long as $K_c, K_s < 1$, or repulsive interactions.

\section{Interacting zero modes}
\label{app:corner-interactions}

\begin{table}[t]
    \centering
    \begin{tabular}{c|c|c|c|c|c|c}
        & $U(1)_v$ & $U(1)_s$ & $i\hat{s}_y$ & $C_{2x}$ & $T$ & ${\cal P}_c$ \\
        \hline
         $\Phi_{1\uparrow}$ & $e^{i\gamma}\Phi_{1\uparrow}$ & $e^{i\gamma}\Phi_{1\uparrow}$ & $\Phi_{1\downarrow}$ & $\Phi_{2\uparrow}$ & $i\Phi_{1\downarrow}^\dagger$ & $-\Phi_{2\uparrow}^\dagger$\\
         $\Phi_{2\uparrow}$ & $e^{i\gamma}\Phi_{2\uparrow}$ & $e^{i\gamma}\Phi_{2\uparrow}$ & $\Phi_{2\downarrow}$  & $\Phi_{1\uparrow}$ & $i\Phi_{2\downarrow}^\dagger$ & $\Phi_{1\uparrow}^\dagger$\\
         $\Phi_{1\downarrow}$ & $e^{i\gamma}\Phi_{1\downarrow}$ & $e^{-i\gamma}\Phi_{1\downarrow}$ & $-\Phi_{1\uparrow}$ & $\Phi_{2\downarrow}$ & $-i\Phi_{1\uparrow}^\dagger$ & $-\Phi_{2\downarrow}^\dagger$\\
         $\Phi_{2\downarrow}$ & $e^{i\gamma}\Phi_{2\downarrow}$ & $e^{-i\gamma}\Phi_{2\downarrow}$ & $-\Phi_{2\uparrow}$  & $\Phi_{1\downarrow}$ & $-i\Phi_{2\uparrow}^\dagger$ & $\Phi_{1\downarrow}^\dagger$
    \end{tabular}
    \caption{Transformation laws of the zero modes $\Phi$, to be employed in the interacting formalism.}
    \label{tab:zeroModeInteractions}
\end{table}

We now consider the corner states and whether interactions can gap them out completely in the presence of all symmetries.
This is to be expected, as the edge modes can be completely gapped, as shown in Appendix~\ref{app:bosonization}.
The zero modes $\Phi$ transform as Table~\ref{tab:zeroModeInteractions}.
\begin{widetext}
For example, under $T$,
\begin{align}
\Phi_{1\uparrow} &= \frac{1}{\sqrt{\mathcal{N}}} \int dxdy\ e^{-\Delta_0|x| - \int^y_0 dy' M_1(y') } ( c_{+,+,A,\uparrow}(x,y) + c_{+,+,B,\uparrow}(x,y)
    - i c_{-,-,A,\downarrow}^\dagger(x,y) -i c_{-,-,B,\downarrow}^\dagger(x,y) ) \\ &\rightarrow \frac{1}{\sqrt{\mathcal{N}}} \int dxdy\ e^{-\Delta_0|x| - \int^y_0 dy' M_1(y') } ( c_{-,-,A,\uparrow}(x,y) + c_{-,-,B,\uparrow}(x,y)
    + i c_{+,+,A,\downarrow}^\dagger(x,y) +i c_{+,+,B,\downarrow}^\dagger(x,y)) \\
    &= i \Phi_{1\downarrow}^\dagger
\end{align}
\end{widetext}
Under a ${\cal P}_c$ breaking perturbation the zero modes $\Phi_{1\uparrow}, \Phi_{1\downarrow}$ will shift away from $\Phi_{2\uparrow}, \Phi_{2\downarrow}$.
Let us first work in this limit, and ask if interactions can gap out the modes $\Phi_{1\uparrow}, \Phi_{1\downarrow}$.
The density terms $\Phi_{1\uparrow}^\dagger \Phi_{1\uparrow}$ and $\Phi_{1\downarrow}^\dagger \Phi_{1\downarrow}$ are forbidden by $i\hat{s}_yT$.
We can however add the interaction
\begin{align}
(\Phi_{1\uparrow}^\dagger \Phi_{1\uparrow} - \frac{1}{2})(\Phi_{1\downarrow}^\dagger \Phi_{1\downarrow} - \frac{1}{2})
\end{align} to partially gap out the system.
A twofold degeneracy still remains, and this cannot be lifted.
Note that this interaction is invariant under the full spin-SU(2) rotation symmetry.
The subtraction of $-1/2$ in the densities is required to be consistent with time-reversal, which exchanges creation and annihilation operators as listed in Table~\ref{tab:zeroModeInteractions}:
\begin{align}
    (\Phi_{1\uparrow}^\dagger \Phi_{1\uparrow} - \frac{1}{2}) \rightarrow (\Phi_{1\downarrow} \Phi_{1\downarrow}^\dagger - \frac{1}{2}) =
    -(\Phi_{1\downarrow}^\dagger \Phi_{1\downarrow} - \frac{1}{2}).
\end{align}

Restoring ${\cal P}_c$ symmetry brings all $4$ corner modes at a single point.
In class BDI the system with eight Majorana modes has been analyzed in the presence of interactions and has been shown to be fully gapped \cite{Fidkowski:2010}.
However, our system is in class CII, with different symmetries.
Density-density interactions partially gap the system:
\begin{align}
&-r_1(\Phi_{1\uparrow}^\dagger \Phi_{1\uparrow} - \frac{1}{2})(\Phi_{1\downarrow}^\dagger \Phi_{1\downarrow} - \frac{1}{2}) \nonumber \\
&-r_1(\Phi_{2\uparrow}^\dagger \Phi_{2\uparrow} - \frac{1}{2})(\Phi_{2\downarrow}^\dagger \Phi_{2\downarrow} - \frac{1}{2}),
\end{align} which for $r_1 > 0$ leave behind four degenerate ground states
\begin{align}
\ket{1, 0, 1, 0}, ~~ \ket{0,1,0,1},~~\ket{1, 1, 1, 1}, ~~ \ket{0,0,0,0},
\end{align} where $0, 1$ label the occupation of the $\Phi_{1\uparrow},\Phi_{2\uparrow},\Phi_{1\downarrow},\Phi_{2\downarrow}$ fermions respectively.
To lift this degeneracy completely we add
\begin{align}
r_2\Phi_{1\uparrow}^\dagger \Phi_{2\uparrow} \Phi_{1\downarrow}^\dagger \Phi_{2\downarrow} + H.c.,
\end{align} which, with $r_1, r_2 > 0$, leaves behind the unique ground state
\begin{align}
\ket{G.S.} = \frac{1}{\sqrt{2}} (\ket{1,0,1,0} - \ket{0,1,0,1}).
\end{align}
All perturbations are symmetric, even under spin-SU(2) rotation, and as both terms commute we have completely gapped out the zero modes.
In fact, we do not even need the density-density terms to find a unique ground state:  the $r_2$ term is sufficient.
Our results mirror the situation in class BDI: 8 Majoranas can be fully gapped but 4 cannot \cite{Fidkowski:2010}.

\section{Josephson Junction}
\label{app:josephson}

\subsection{Setup}

Consider a Josephson junction built out of two sheets of TBG-TSC; on the left the superconductor is held at phase $0$, while the right the superconductor is held at phase $\phi$.  The Hamiltonian reads
\begin{align}
    H[\phi] &= \xi_z \mu_0 (-i\partial_x \tau_z \sigma_x -i\partial_y \sigma_y) + \Delta \Theta(-x) \xi_x \mu_0 \tau_0 \sigma_0 \nonumber \\
    &+ \Delta \Theta(x) [\cos{\phi} \xi_x \mu_0 \tau_0 \sigma_0 + \sin{\phi} \xi_y \mu_0 \tau_0 \sigma_0],
\end{align} with $\Theta(x)$ the Heaviside step function.  This Hamiltonian describes a $(\phi_L, \phi_R) = (0, \phi)$ domain wall, where the first number of the pair is the phase on the left and the second the phase on the right.

For $\phi = 0$ the spectrum of the TBG-TSC Josephson junction is simple -- the system is one continuous sheet of TBG-TSC.  There is a gap everywhere on order of the superconducting pairing strength $\Delta$.  This argument assumes that the domain wall is perfectly sharp; eventually it will be advantageous to relax this condition, allowing the phase to vary smoothly across the transition.  As we will show in Section~\ref{app:soft_domain}, this allows for the formation of in-gap states even at $\phi = 0, 2\pi$.

For $\phi = \pi$ the Josephson junction reduces to the pairing domain wall; we understand there are four \emph{total} zero modes for a single corner, two per graphene valley. We wish to understand the system away from $\phi = \pi$; what is the fate of the zero modes?

\subsection{Symmetries of the Josephson junction}
We examine which symmetries of TBG-TSC persist away from $\phi = \pi$.  Chiral symmetry $S$, time-reversal $T$, and particle-hole symmetry $\PH$ are broken by the inclusion of the $\Delta \xi_y \sin{\phi}$ term.  However, they can be restored if we take the additional operation $\phi \rightarrow -\phi$: thus, these three symmetries conspire to relate the Hamiltonian $H[\phi]$ to the Hamiltonian at $H[-\phi]$.

$C_{2x}$, being a mirror symmetry that flips $y \rightarrow -y$, is unaffected by the Josephson junction (as the phase difference is along the $x$-direction).  Interestingly, $C_{2z}T$ is \emph{also} preserved, even when $\phi = \pi$, when the origin of rotation is located at the junction.  Even though $T$ does not commute with the additional $\xi_y$ term (effectively sending $\phi \rightarrow -\phi$), the inversion $C_{2z}$ (when centered at the domain wall) switches the left and right sections of the domain wall.  So beginning with a domain wall with phase $0$ on the left and $\phi$ on the right (a $(0, \phi)$ domain wall for short), $T$ will change this to a $(0, -\phi)$ domain wall, and $C_{2z}$ flips this to $(-\phi, 0)$.  This is identical to a $(0, \phi)$ domain wall via a gauge rotation $U(\phi) = e^{-i\phi (\xi_z - 1)/2}$.  The symmetries are listed in Table~\ref{SymmetryTableJJ}.

\begin{table}
\begin{tabular}{c|c|c|c}
Symmetry & Action on $H[\phi]$ & $\kk \rightarrow$ & $\phi \rightarrow$ \\
\hline
$C_{2x}$ & $\mu_x\sigma_x$ & $C_{2x} \kk$ & $\phi$  \\
$U(\phi) C_{2z}T$ & $e^{-i\phi(\xi_z-1)/2}\sigma_x K$ & $\kk$  & $\phi$ \\
$T$ & $\tau_x \mu_x K$ & $-\kk$  & $-\phi$ \\
${\cal P}$ & $i\xi_z\mu_y\sigma_x K$ & $-\kk$  & $-\phi$ \\
$S$ & $\xi_y$ & $\kk$  & $-\phi$
\end{tabular}
\caption{Table of symmetries in TBG-TSC Josephson junction, along with their action on $\phi$.  The latter $3$ symmetries $T, \PH, S$ are not symmetries of $H[\phi]$; they relate $H[\phi]$ to $H[-\phi]$.}
\label{SymmetryTableJJ}
\end{table}

Though $C_{2z}T$ is present at $\phi \neq \pi$, chiral symmetry $S$ is not, and both were required to enforce the four zero modes at $0$ energy.  Thus, the modes will move away from $0$.  The exact behavior of the four in-gap states, which we call $\Phi_{as}[\phi]$, can be deduced from general arguments:
\begin{itemize}
    \item 1.  Because of spin rotation $i\hat{s}_y$ symmetry, the four in-gap states can be thought of as two pairs $\Phi_{as}[\phi]$, with $\Phi_{1\uparrow}[\phi]$ and $\Phi_{1\downarrow}[\phi]$ having identical spectra, and the same for the $\Phi_{2\uparrow}[\phi], \Phi_{2\downarrow}[\phi]$ pair.  When $\PH_c$ is broken, these two pairs of in-gap states separate in position space, with one pair moving up and one moving down.

    \item 2.  Even if $\PH_c$ is broken, the two pairs of corner in-gap states are related by $C_{2x}$ and \emph{also} have identical spectra.  If $C_{2x}$ is broken then this is no longer true, but for weak $C_{2x}$-breaking the qualitative nature of the spectrum will not change dramatically.

    \item 3.  At $\phi = \pi$, the in-gap states are zero modes $\Phi_{as}[\phi = \pi] = \Phi_{as}$, and at $\phi = 0, 2\pi$, they are fully merged with the bulk states.  This will no longer be true when we consider the soft domain wall; in this situation at $\phi = 0, 2\pi$ there are in-gap corner states.

    \item 4.  Because of $S$ symmetry relating the Hamiltonians at $\phi$ and $-\phi$, the spectrum at $\phi, E[\phi]$, is related to the opposite spectrum by $E[\phi] = -E[-\phi]$.
\end{itemize}

We thus deduce the form of the in-gap state spectra to be as depicted in Fig.~\ref{fig:softDomainWall}: there are four in-gap modes corresponding to four zero modes at $\phi = \pi$: two $(\mu_z = \pm 1)$ per graphene valley.  When $\PH_c$ is broken, the two pairs of in-gap states will separate in space; one pair moves up and the other moves down in real space along the domain wall.  Nevertheless, the energy spectra of the in-gap states will not change: they are forced to all be identical to one another by either $C_{2x}$ or spin-U(1).  For the rest of this Appendix, we assume that the two pairs of in-gap states are separated and only consider a single pair.

\subsection{Soft Domain Wall}
\label{app:soft_domain}
\begin{figure}
    \centering
    \includegraphics[width=\columnwidth,trim=80 0 0 0, clip]{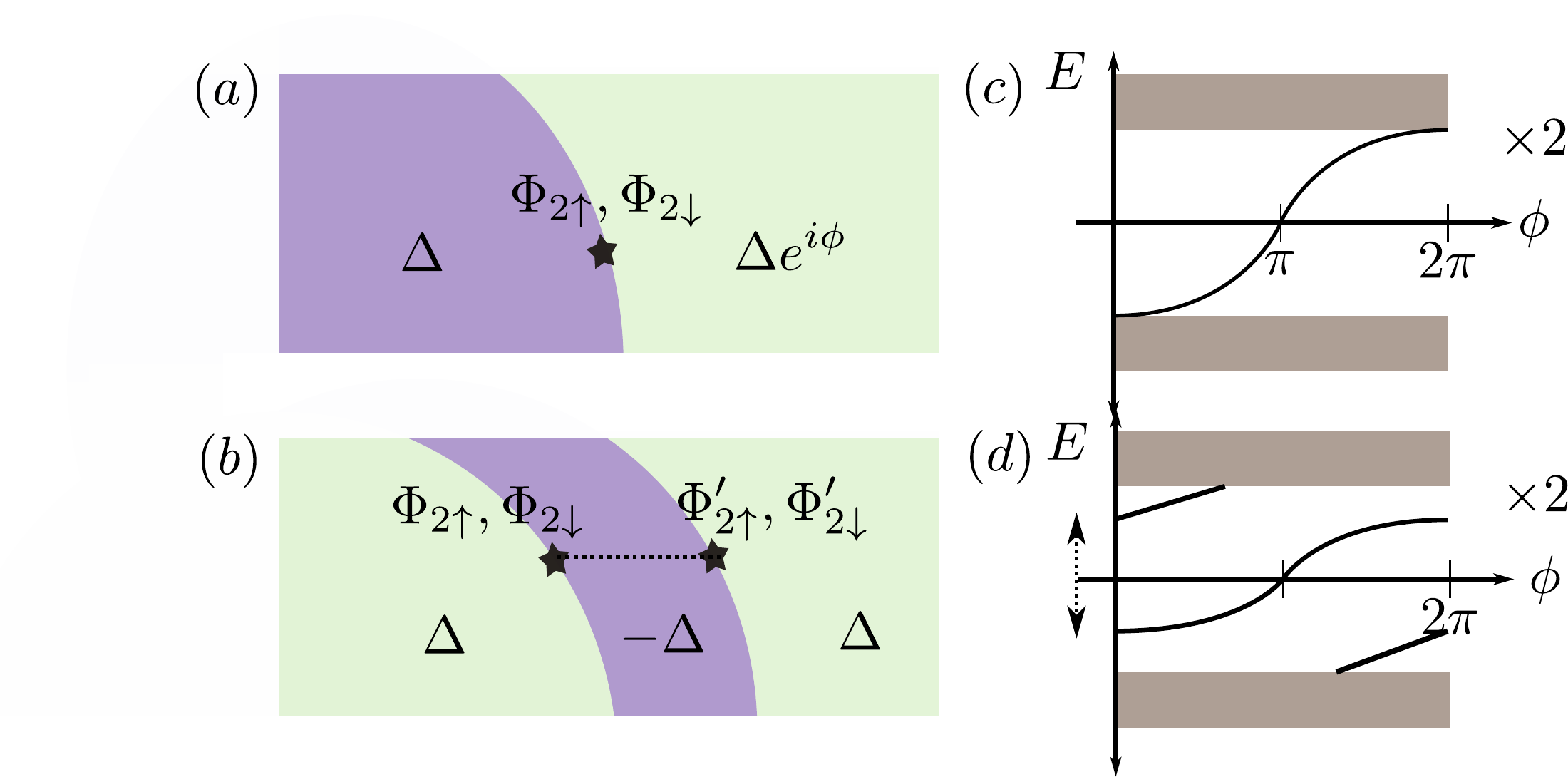}
    \caption{A soft potential at the domain wall creates in-gap states.  For a hard edge, when the superconducting phase difference is $0$, if the sheet of TBG-TSC is homogeneous (as depicted in (a)), the system is merely a continuous sheet of TBG-TSC and the in-gap states (depicted in (c)) flow into the bulk.  Spin-rotation symmetry keeps the in-gap states corresponding to $\uparrow, \downarrow$ degenerate.  We model a soft potential in (b), where the left and right superconductors are separated by a pairing region at superconducting phase $\pi$.  Corner modes are bound to both domain walls, and can hybridize as the boundary is shrunk, as depicted in (d).  The in-gap energy states are now well-separated from the bulk.  The gap arising between the dashed line stems from hybridization in the dashed line in (b), and can be in principle smaller than the gap $\Delta$.}
    \label{fig:softDomainWall}
\end{figure}

Our previous calculations have assumed that the transition between the two superconductors forming the Josephson junction is a sharp step, while in reality the transition will be smooth.  In fact, in order to preserve valley-U(1), we actually require that the phase jump be smooth on the order of the graphene lattice (but sharp on the order of the Moir\'e lattice).

When the domain wall separating the superconductors is spread out instead of a sharp step, additional in-gap states are pulled from the continuum.  An easy way to see this is to pick a judicious form of the domain wall: consider a $(0, \pi, 0)$ junction, where there are \emph{three} superconducting regions at phase $\phi = 0, \pi, 0$ respectively, for the left, middle, and right regions, as depicted in Fig.~\ref{fig:softDomainWall}(b).  The middle region is new and is short on the order of the Moir\'e length scale.  There are thus \emph{two} domain walls, separated by a small region.  Because the left and the right regions are both at phase $\phi = 0$, this is a Josephson junction with phase difference $0$; a $(0, 0)$ domain wall.  The purpose of the middle $\phi = \pi$ region is to mimic a soft domain wall, with the great advantage that we already know the low-lying states that exist at the $(0, \pi)$ and $(\pi, 0)$ domain walls: call them $\Phi_{as}, \Phi_{as}'$ respectively.  These are allowed to hybridize because the center region is narrow; if this hybridization is small then there are in-gap states even at $\phi = 0, 2\pi$.

To calculate the zero modes and their properties under symmetries at the $(\pi,0)$ domain wall, we perform the gauge transformation $U(\pi) = e^{-i\pi (\xi_z - 1)/2} = \xi_z$ that takes $c \rightarrow ic$.  This has no effect on the hopping terms $c^\dagger c$, but it reverses the sign of the pairing terms $\Delta cc \rightarrow -\Delta cc$.  In the Dirac Hamiltonian notation, $\xi_z$ commutes with the kinetic term but anti-commutes with the pairing.
The gauge transformation thus maps a $(0,\pi)$ junction to a $(\pi,0)$ junction.
The primed zero mode operators $\Phi'$ read
\begin{widetext}
\begin{equation}
\Phi_{1\uparrow}' = \frac{1}{\sqrt{\mathcal{N}}} \int dxdy\ e^{-\Delta_0|x| - \int^y_0 dy' M_1(y') } (ic_{+,+,A,\uparrow}(x,y) + ic_{+,+,B,\uparrow}(x,y)
    - c_{-,-,A,\downarrow}^\dagger(x,y) - c_{-,-,B,\downarrow}^\dagger(x,y) ),
\end{equation}
\begin{equation}
\Phi_{2\uparrow}' = \frac{1}{\sqrt{\mathcal{N}}} \int dxdy\ e^{-\Delta_0|x| - \int^y_0 dy' M_1(y') } (ic_{+,-,A,\uparrow}(x,y) + ic_{+,-,B,\uparrow}(x,y)
    - c_{-,+,A,\downarrow}^\dagger(x,y) - c_{-,+,B,\downarrow}^\dagger(x,y) ),
\end{equation}
\end{widetext}
and under $T$ transform as $\Phi_{1\uparrow}' \rightarrow -i \Phi_{1\downarrow}'^\dagger, \Phi_{2\uparrow}' \rightarrow -i \Phi_{2\downarrow}'^\dagger.$
We are allowed to hybridize corner states across the soft domain wall as
\begin{align}
    H_\text{domain} = v_0 \Phi_{2\uparrow}^\dagger \Phi_{2\uparrow}' + v_0 \Phi_{2\downarrow}^\dagger \Phi_{2\downarrow}' + H.c.,
\end{align} with four eigenstates, at energies $E = \pm |v_0|$.
Without loss of generality, we take $v_0$ real and the low-energy eigenstates as
\begin{align}
    \Psi_{a\uparrow} = \frac{1}{\sqrt{2}} (\Phi_{2\uparrow} + \Phi_{2\uparrow}'), \\
    \Psi_{b\uparrow} = \frac{1}{\sqrt{2}} (\Phi_{2\uparrow} - \Phi_{2\uparrow}').
\end{align}
Under $T$,
\begin{align}
    \Psi_{a\uparrow} \rightarrow -i\Psi_{b\downarrow}^\dagger \\
    \Psi_{b\uparrow} \rightarrow -i\Psi_{a\downarrow}^\dagger.
\end{align}

\begin{align}
    H_\text{domain} = v_0 (\Psi_{a\uparrow}^\dagger \Psi_{a\uparrow} - \Psi_{b\uparrow}^\dagger \Psi_{b\uparrow}) + (\uparrow \leftrightarrow \downarrow).
    \label{eq:josephsonDomainWall}
\end{align}

Thus, if the hybridization is small $v_0 < |\Delta|$, then there are in-gap states at energy $\pm v_0$.

\subsection{Interactions}

\begin{figure}
    \centering
    \includegraphics[width=\columnwidth]{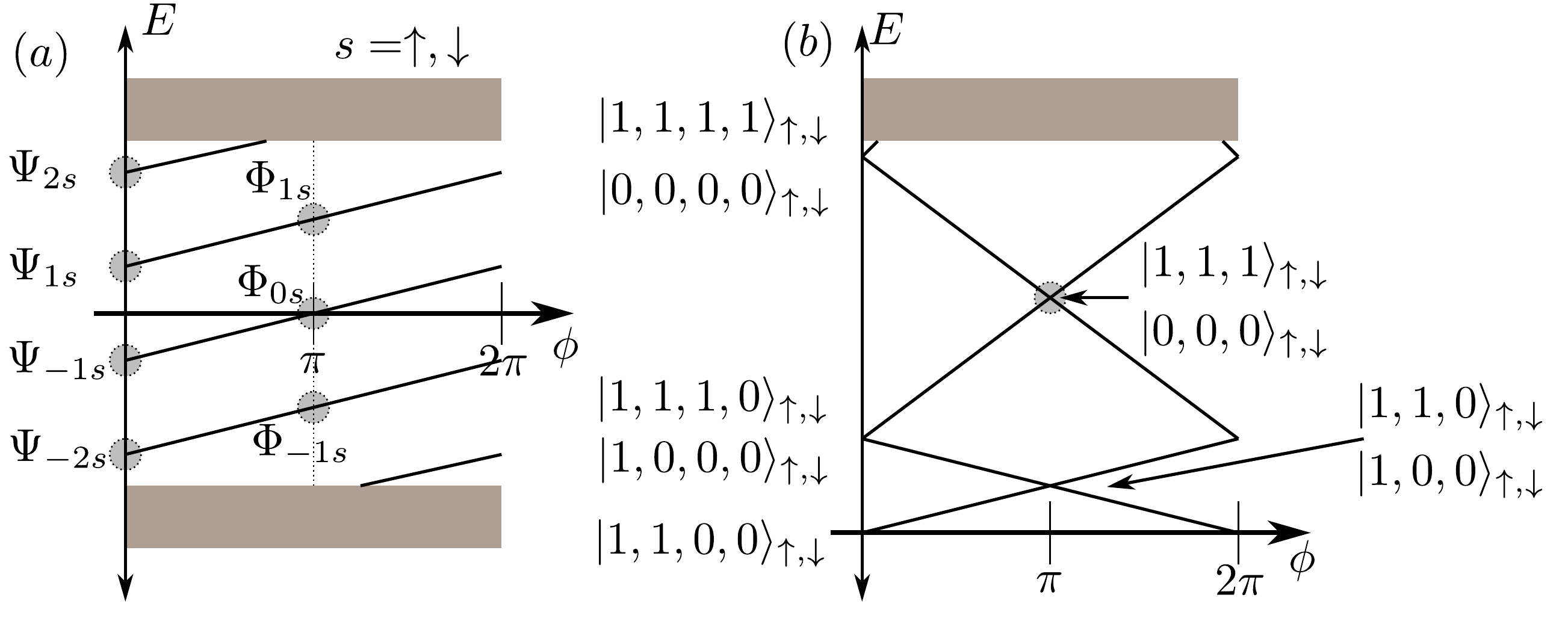}
    \caption{Single-particle and multi-particle spectra of the fractional Josephson effect in TBG-TSC.  (a) Single-particle spectra.  Multiple in-gap states have been created by modifying the potential.  The spectrum is doubled for both spins.  If $\PH$ is preserved, the spectrum is quadrupled.  (b) Many-body spectrum, $\PH$ broken.  The many-body states at $\phi = 0$ are labeled by occupation of $\Psi_{-2s},\Psi_{-1s},\Psi_{1s},\Psi_{2s}$ respectively.  The subscript $\uparrow\downarrow$ indicates the exact same occupation in both spin sectors.  At $\phi = \pi$, the occupation is labeled by $\Phi_{-1s},\Phi_{0s},\Phi_{1s}$, respectively.  The crossing in the shaded circle can be avoided by the term $\Phi_{-1\uparrow}\Phi_{0\uparrow}\Phi_{1\uparrow}\Phi_{-1\downarrow}\Phi_{0\downarrow}\Phi_{1\downarrow} + H.c.$.  There are many other states in the multi-particle spectra, but their presence does not affect the pumping cycle qualitatively.}
    \label{fig:josephson_final}
\end{figure}

Consider now a pumping cycle that winds $\phi$ by $2\pi$.  Beginning in the ground state at $\phi = 0$, what is the fate of the system after $\phi$ is adiabatically changed by $2\pi$?
First, let us pull down additional in-gap states so that at $\phi = 0$, there are $4$ in-gap fermion states per spin sector, or $8$ total; there are an additional $8$ modes at the $C_{2x}$-symmetric position that we do not consider.  The eight fermions at $\phi = 0$ are $\Psi_{b\uparrow}, \Psi_{b\downarrow}$, where $b$ labels the flavor of $\Psi$ (\emph{not} the corners separated in space!).  At $\phi = \pi$ the six in-gap states are $\Phi_{b\uparrow}, \Phi_{b\downarrow}$; again, $b$ refers to the flavor of in-gap state, not the different $C_{2x}$-related corners.
This is depicted in Fig.~\ref{fig:josephson_final}.
Without interactions, we take the non-interacting Hamiltonian
\begin{align}
    &H_\text{in-gap}[\phi = 0] = \sum_s E_1 (\Psi_{1s}^\dagger \Psi_{1s} -\frac{1}{2}) + E_2 (\Psi_{2s}^\dagger \Psi_{2s} - \frac{1}{2}) \nonumber \\
    &\qquad - E_1 (\Psi_{-1s}^\dagger \Psi_{-1s} -\frac{1}{2}) - E_2 (\Psi_{-2s}^\dagger \Psi_{-2s} - \frac{1}{2}), \\
    &H_\text{in-gap}[\phi = \pi] = \sum_s E_1' (\Phi_{1s}^\dagger \Phi_{1s} -\frac{1}{2}) - E_1' (\Phi_{-1s}^\dagger \Phi_{-1s} -\frac{1}{2}).
\end{align} The hybridization energies satisfy $E_1,E_2, E_1' < |\Delta|$.  This spectrum is depicted in Fig.~\ref{fig:josephson_final}.
Following the single-particle spectrum, the ground state initially consists of all negative energy states unoccupied.  After evolution, the states $\Psi_{1s}$ go from unoccupied to occupied, and the ground state is not mapped back to its original form.  An additional evolution of $\phi$ by $2\pi$ will then turn $\Psi_{2s}$ from unoccupied to occupied.  Each evolution of $2\pi$, the valley number of the state changes by $+2$, as states are dragged from the negative energy continuum and pumped into the positive energy continuum.  So long as valley-U(1) is preserved, there is no way for the ground state to return to its original form.  The Josephson cycle is aperiodic.  To remedy this issue, we must consider interactions, which will serve to reduce the valley-U(1) symmetry down to $Z_3$.

The valley-U(1) symmetry originates from the fact that the twist angle between the bilayers of graphene is small.  This means that the Moir\'e potential is smooth on the scale of the graphene lattice vectors, and the graphene translation operators $T_{a_i}$ are approximately good operators.  In momentum space, the low-energy excitations are near the Dirac points $K, K'$, meaning that one requires a large number of Moir\'e BZ vectors to scatter a state in one valley to the other.  In other words,
\begin{align}
    T_{a_i} \psi_+ \approx e^{-iK \cdot a_i} \psi_+, \\
    T_{a_i} \psi_- \approx e^{iK \cdot a_i} \psi_-.
\end{align}
The states around the positive valley $\psi_+$ are located at $K$, while the opposite valley at $K' = -K$.  Thus hoppings between the two valleys are forbidden.  At the free fermion level, we thus have independent number conservation of both the $+$ valley fermions and $-$ valley fermions.  Note that pairing between opposite valleys is allowed, and is the form of the pairing Hamiltonian we take.

Once we turn to the multi-particle language, more terms are allowed.  Notice that $3K = 0$, that is, three times the $K$ point yields a reciprocal lattice vector in the graphene BZ.  This means that, despite not conserving valley-U(1), terms of the form $\psi_{1+} \psi_{2+} \psi_{3+} \psi_{4+} \psi_{5+} \psi_{6+}$ preserve translation symmetry in the graphene lattice.  We thus claim that in the many-body formalism, valley-U(1) symmetry reduces to $Z_3$, with an extra $Z_2$ from fermion parity yielding $Z_6$.

With valley-U(1) broken to $Z_3$, the Josephson effect is now $6\pi$ periodic.  The multi-particle ground state evolution is depicted in Fig.~\ref{fig:josephson_final}(b).
In the multi-particle spectrum, all crossings are protected by $Z_3$ symmetry, except for the shaded circle, which can be gapped by
\begin{align}
    u\Phi_{-1\uparrow}\Phi_{0\uparrow}\Phi_{1\uparrow}\Phi_{-1\downarrow}\Phi_{0\downarrow}\Phi_{1\downarrow} + H.c.,
\end{align} which preserves $Z_3$ symmetry.  The $\Phi_{0s}$ operator is the zero mode at $\phi = \pi$ and the $\Phi_{1s},\Phi_{-1s}$ are two in-gap states, as depicted in Fig.~\ref{fig:josephson_final}(a). Tracing out the evolution of the ground state reveals a $6\pi$ periodicity, or $Z_3$ fractional Josephson effect.

We explain the evolution of the ground state explicitly.  Consider the two spin sectors to be identical copies that behave identically under evolution of $\phi$:  in the spin up sector, $\ket{1,1,0,0}_\uparrow$ is the ground state, the numbers denoting the occupation of $\Psi_{-2\uparrow},\Psi_{-1\uparrow},\Psi_{1\uparrow},\Psi_{2\uparrow}$.  After $2\pi$ evolution, this state flows to $\ket{1,1,1,0}_\uparrow$, and then to $\ket{1,1,1,1}_\uparrow$, and then into the continuum.  In order to prevent this, we must open a gap between two states of different valley number, which we have argued can only occur when the valley number differs by a multiple of $3$.

To see the cycle explicitly, define the ground state at $\phi = 0$ as $\ket{\Omega(\phi = 0)}$, which satisfies the following:
\begin{align}
    \Psi_{-1s}^\dagger \ket{\Omega(\phi = 0)} = 0\\
    \Psi_{-2s}^\dagger \ket{\Omega(\phi = 0)} = 0\\
    \Psi_{1s} \ket{\Omega(\phi = 0)} = 0\\
    \Psi_{2s} \ket{\Omega(\phi = 0)} = 0,
\end{align} and at $\phi = \pi$, the two ground states $\ket{\Omega(\phi = \pi)}, \Phi_{0s}^\dagger \ket{\Omega(\phi = \pi)} $ satisfy
\begin{align}
    \Phi_{0s} \ket{\Omega(\phi = \pi)} = 0 \\
    \Phi_{1s} \ket{\Omega(\phi = \pi)} = 0 \\
    \Phi_{-1s}^\dagger \ket{\Omega(\phi = \pi)} = 0.
\end{align}
We now study the flow of the ground state under winding $\phi$.  The ground state evolves as
\begin{align}
\ket{\Omega(\phi = 0)} &\rightarrow \Phi_{0\uparrow}^\dagger \Phi_{0\downarrow}^\dagger \ket{\Omega(\phi = \pi)} \rightarrow \Psi_{1\uparrow}^\dagger \Psi_{1\downarrow}^\dagger \ket{\Omega(\phi = 0)} \nonumber \\
&\rightarrow \Phi_{1\uparrow}^\dagger \Phi_{0\uparrow}^\dagger \Phi_{1\downarrow}^\dagger \Phi_{0\downarrow}^\dagger \ket{\Omega(\phi = \pi)},
\end{align} or in the occupation number notation
\begin{align}
    \ket{1,1,0,0}_{\uparrow,\downarrow} \rightarrow \ket{1,1,0}_{\uparrow,\downarrow} \rightarrow \ket{1,1,1,0}_{\uparrow,\downarrow} \rightarrow \ket{1,1,1}_{\uparrow,\downarrow}.
\end{align}
But now with interactions, the state $\Phi_{1\uparrow}^\dagger \Phi_{0\uparrow}^\dagger \Phi_{1\downarrow}^\dagger \Phi_{0\downarrow}^\dagger \ket{\Omega(\phi = \pi)}$ can tunnel to $\Phi_{-1\uparrow} \Phi_{-1\downarrow} \ket{\Omega(\phi = \pi)}$, as they differ by $6$ operators: the perturbation $u\Phi_{-1\uparrow}\Phi_{0\uparrow}\Phi_{1\uparrow}\Phi_{-1\downarrow}\Phi_{0\downarrow}\Phi_{1\downarrow} + H.c.$ serves this purpose.  Continuing the evolution gives
\begin{align}
    \Phi_{-1\uparrow} \Phi_{-1\downarrow} \ket{\Omega(\phi = \pi)} &\rightarrow \Psi_{-1\uparrow} \Psi_{-1\downarrow} \ket{\Omega(\phi = 0)} \nonumber \\
    & \rightarrow \ket{\Omega(\phi = \pi)} \rightarrow \ket{\Omega(\phi = 0)},
\end{align} or in the occupation number notation,
\begin{align}
    \ket{0,0,0}_{\uparrow,\downarrow} \rightarrow \ket{1,0,0,0}_{\uparrow,\downarrow} \rightarrow \ket{1,0,0}_{\uparrow,\downarrow} \rightarrow \ket{1,1,0,0}_{\uparrow,\downarrow}.
\end{align}
This is the $Z_3$ fractional Josephson effect.

Two important considerations needed to protect the $Z_3$ fractional Josephson effect are that the states $\ket{1,1,1,0}_{\uparrow\downarrow}, \ket{1,0,0,0}_{\uparrow\downarrow}$ have a protected crossing, as well as the states $\ket{1,1,0}_{\uparrow \downarrow}, \ket{1,0,0}_{\uparrow\downarrow}$.  Both crossings are protected by valley symmetry -- their total valley numbers do not differ by a multiple of $3$.

\end{document}